\documentclass[defaultstyle,11pt]{thesis}

\usepackage{amssymb}		
\usepackage{amsmath}
\usepackage{graphicx}		
\usepackage{hyperref}               
\usepackage{multirow}
\usepackage{adjustbox}		

\title{Strongly Interacting Fermi Gases: Hydrodynamics and Beyond}

\author{William E.}{Lewis}

\otherdegrees{B.S., University of Arkansas, 2012 \\
	      M.S., University of Colorado at Boulder, 2015}

\degree{Doctor of Philosophy}		
	{Ph.D., Physics}		
 
\dept{Department of}			
	{Physics}		

\advisor{Prof.}				
	{Paul Romatschke}			
\reader{Prof.~Murray Holland}		

\abstract{  \OnePageChapter	
	This thesis considers out-of-equilibrium dynamics of strongly interacting non-relativistic Fermi gases in several two and three dimensional geometries. The tools of second-order hydrodynamics and gauge-gravity duality will be utilized to address this system. Many of the themes of this work are motivated by the observed similarities in transport properties between strongly interacting Fermi gases and other very different strongly interacting quantum fluids such as the quark-gluon plasma, high temperature superconductors, and quantum field theories described by gauge-gravity duality. In particular, these systems all nearly saturate the conjectured lower bound on the ratio of shear viscosity to entropy density $\eta/s \geq \hbar/(4 \pi k_B)$ coming from the AdS/CFT correspondence. Among other things, this observation, in conjunction with current experiment and data analysis in atomic, condensed matter, and nuclear physics lends itself to the following questions: How perfect of a fluid is the strongly interacting Fermi gas, and can one find a more stringent constraint on $\eta/s$ in Fermi gases? Do the similarities in transport properties among strongly interacting quantum systems extend beyond dynamics controlled by the hydrodynamical shear viscosity? In regards to the first question, by utilizing second-order hydrodynamics, it will be demonstrated that higher-order collective modes of a harmonically trapped Fermi gas may serve as a more sensitive probe of the shear viscosity. For the second question, both second-order hydrodynamics and a gravity dual theory are used to make predictions about dynamics occurring on short timescales where hydrodynamics is expected to break down. In particular the appearance of a class of ``non-hydrodynamic" collective modes not contained within a Navier-Stokes description of the strongly interacting Fermi gas will be discussed.}

\dedication[Dedication]{	
\noindent To Mom, Dad, Kim, and Rex:\\
\indent You have been my foremost guides in life. Without all of the love, support, and advice you have provided, I wouldn't have had the opportunity and fortitude to achieve my goals. 
	}

\acknowledgements{	\OnePageChapter	
First, I would like to acknowledge my brothers James, Thomas, and Paul. I am continually inspired by your intelligence, independence, kindness, and willingness to delve into the next stage of your lives despite how daunting that may be. Thomas, I owe you a special debt of gratitude. You have shared stories of your own difficulties in graduate school, and how you overcame them. This connection has provided me with confidence that I could get through similar difficulties. Since starting my higher-education you have also consistently shown me the value of listening to those around me. I wouldn't be where I am without this advice. I also want to thank Jasmine Brewer. She helped me trust myself and find the courage to switch research groups in my third year. She has been supportive ever since. Along the same lines, I want to express my appreciation of Professor Paul Romatschke for giving me the opportunity to continue my graduate career after this switch. Finally, I'd like to express gratitude to my closest friends Ben Galloway, Mathis Habich, Carrie Wiedner, John Bartolotta, Sam Johnson, and Zack Joseph (and his family). You helped me maintain my day to day sanity and have plenty of fun during my stint at CU.}

\ToCisShort	

\LoFisShort	

\LoTisShort	

\begin{document}

\newtheorem{theorem}{Theorem}

\newcommand{\diff}[2]{\frac{\partial #1}{\partial #2}}
\newcommand{\diffr}[1]{\diff{#1}{r}}
\newcommand{\diffth}[1]{\diff{#1}{\theta}}
\newcommand{\diffz}[1]{\diff{#1}{z}}

\newcommand{\vth}{V_{\theta}}

\newcommand{\twochoices}[2]{\left\{ \begin{array}{lcc}
        \displaystyle #1 \\ \vspace{-10pt} \\
        \displaystyle #2 \end{array} \right. } 

\newcommand{\threechoices}[3]{\left\{ \begin{array}{lcc}
        #1 \\ #2 \\ #3 \end{array} \right. }    

\newcommand{\fourchoices}[4]{\left\{ \begin{array}{lcc}
        #1 \\ #2 \\ #3 \\ #4 \end{array} \right. }      

\newcommand{\twovec}[2]{\left(\begin{array}{c} #1 \\ #2 \end{array}\right)}
\newcommand{\threevec}[3]{\left(\begin{array}{c} #1 \\ #2 \\ #3 \end{array}\right)}
\newcommand{\twomatrix}[4]{\left(\begin{array}{cc} #1 & #2 \\ #3 & #4 \end{array}\right)}

\chapter{Introduction}
\label{introchap}
\section{A Brief History of Cold Atoms}
Broadly, this thesis aims to contribute novel results to the ever growing field of cold atomic physics. This field has a rich and fascinating progression. No attempt is made to provide comprehensive discussion of that history here. However, a few key developments are highlighted. The theoretical underpinnings of cold quantum gases can be traced at least as far back as the early $1920$s with Satyendranath  Bose and Albert Einstein's work on the statistics of bosons \cite{Bose1924,Einstein1924}. Bose first applied Bose-Einstein statistics to photons in the case of thermal blackbody radiation. Soon after, Einstein extended these ideas to non-interacting bosonic atoms. This work quickly culminated in the theoretical discovery of a condensed phase termed the Bose-Einstein condensate (BEC) characterized by all of the atoms simultaneously occupying the quantum mechanical ground state of the system \cite{Einstein1925}. This phase of matter could theoretically be achieved in a gas of non-interacting bosonic atoms cooled to very low temperatures such that $n^{1/3} \lambda_{dB}  \gtrsim 1$, where $n$ is the atomic density and $\lambda_{dB}= h/\sqrt{2 \pi m k_B T}$ is the thermal de Broglie wavelength. Just over a decade later, in 1938, Bose-Einstein condensation was proposed by Fritz London as an explanation for the observed superfluidity in liquid $^4$He \cite{Fritz1938}.

\begin{figure}[ht!]
 \centering \includegraphics[width=0.75\textwidth]{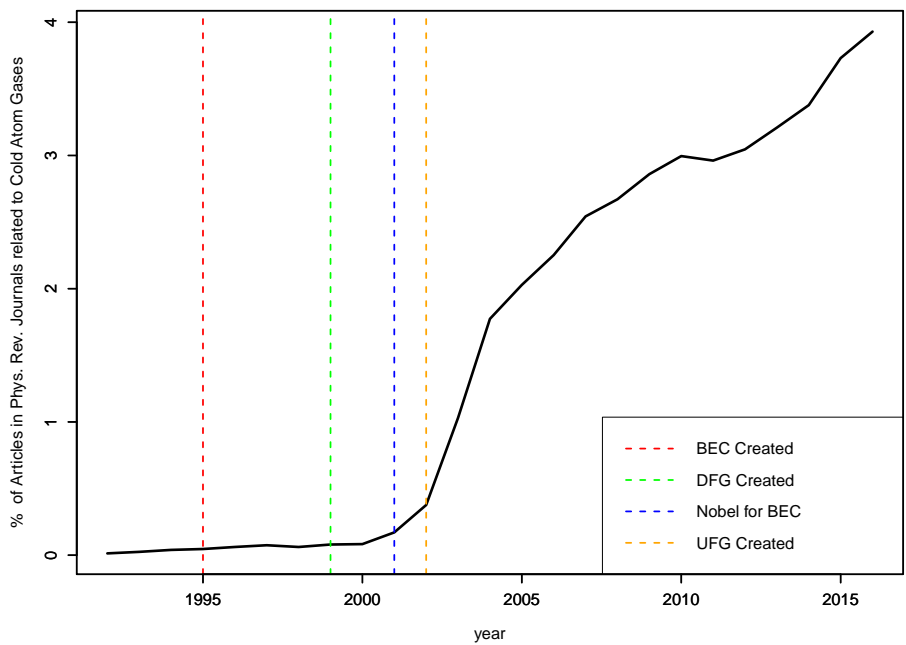}%
  \caption{Fraction of articles published by Physical Review each year relating to cold atom gases. The number of articles relating to cold atom gases was naively measured by the value returned from  a search "Cold Atomic Gas" specifying the year start and end dates at the physical review article database available at \url{https://journals.aps.org/archive/}.
}
  \label{fig:pop}
\end{figure}

In 1926, shortly after the theoretical introduction of the Bose-Einstein condensate, there were two papers, one by Enrico Fermi \cite{Fermi1926} (translated in Ref.~\cite{Zannoni1999}) and another by Paul Dirac \cite{Dirac1926}, discussing Fermi-Dirac statistics in a monoatomic ideal gas. Their results found near immediate application to a range of problems including the collapse of a star to a white dwarf \cite{Fowler1926} and the resolution of a number of condensed matter conundrums such as the electron theory of metals \cite{Sommerfeld1927} and emission of electrons from metals in intense electric fields \cite{Fowler1928}.  The ability of even simple models of non-interacting atoms incorporating quantum statistics to span a range of physics from condensed matter to astrophysical objects gives a first indication at a reason for the rise in prevalence of cold atomic gas physics observed around the turn of the twentieth century (see Fig.~\ref{fig:pop} demonstrating the increased volume of journal articles relating to cold atom physics). Namely, these models of non-interacting atoms with quantum statistics provide a wealth of physical insight applicable across many subfields.

Several decades later, in the $1990$s, there was a series of events that triggered a rapid and vast growth in popularity of cold atomic physics. It is important to note that these events were possible largely due to advances in the field of atom trapping and cooling. However, the reader is referred to other discussions of those developments (see \textit{e.g.} Ref.~\cite{2001Nobel}). Here only a handful of major results coming from applying those techniques to cold atom systems will be discussed. To begin this part of the epoch, in 1995, the group of Eric Cornell and Carl Wieman succeeded in producing the first atomic BEC in a gas of weakly interacting rubidium$-87$ atoms \cite{Anderson198}. This achievement was followed shortly by Wolfgang Ketterle who demonstrated a variety of interesting features of BECs such as matter wave interferometry \cite{Andrews637}.

In 1999, the group of Deborah Jin succeeded in cooling a gas of weakly interacting fermionic potassium$-40$ atoms to degeneracy \cite{DeMarco1703}. This was the first demonstration of a degenerate atomic Fermi gas created in the lab. Degeneracy in a Fermi gas is achieved when the pressure in the gas is primarily a result of fermionic quantum statistics which do not allow two identical particles to occupy the same state. Similarly to the case of bosonic atoms, these quantum effects become significant when $n^{1/3} \lambda_{dB} \gtrsim 1$. Finally, in 2002, the group of John Thomas succeeded in observing a strongly interacting degenerate Fermi gas of Lithium$-6$ atoms later termed a ``unitary Fermi gas" (UFG) \cite{OHara2179}. This experiment demonstrated hydrodynamic evolution of the gas, and motivates the foundational goals of this work. Namely, a theoretical treatment of hydrodynamic as well as ``non-hydrodynamic" modes in strongly interacting Fermi gases. However, before moving on to those details, it is interesting to attempt to quantify the change in perception of cold atom physics over time. Fig.~\ref{fig:pop} indicates that there was a dramatic rise in appearance of the topic of cold atomic gases in the literature almost immediately following the series of events starting with the first observation of a BEC in 1995 and culminating in the creation of a UFG in 2002. 
  
The growth in popularity of cold atom physics demonstrated by Fig.~\ref{fig:pop} is not due solely to an increase in the amount of research explicitly in the field. In fact, the broad and rather naive search phrase ``cold atom gas" used in the physical review article database for generating Fig.~\ref{fig:pop} was intentionally vague. A cursory survey of the resulting papers indicates a large number contributions from other sub-fields including condensed matter and nuclear physics. Perhaps the key reason is that, as of yet, cold atom gases provide one of the cleanest experimental realizations of a quantum mechanical system, offering exquisite control over a large number of features including dimensionality, lattice structure, interaction strength and type, mass imbalance, spin imbalance, and disorder. This enables cold atoms to play a role as a testbed for studying some of the most cutting edge physics across sub-disciplines ranging from topological matter and strongly interacting many body systems to far-from-equilibrium physics. The next section follows this trend of taking inspiration from other sub-fields, particularly nuclear physics and gauge-gravity duality, to introduce the questions that will be considered in this work.  


\section{Strongly Interacting Fermi Gases: Taking Inspiration from Nuclear Physics and Gauge-String Duality}

Strongly interacting quantum systems may be observed in a variety of settings, \textit{e.g.} high $T_c$ superconductors \cite{Rameau2014},  clean graphene \cite{Muller2009}, the quark-gluon plasma \cite{RHICWhite}, and Fermi gases tuned to a Feshbach resonance \cite{Book}. However, the theoretical treatment of these systems often precludes or at best obfuscates the use of standard perturbation theory and/or kinetic theory techniques. In the former, there is not a small parameter with which to organize a perturbative calculation, and in the latter, the possible lack of a quasiparticle description in the regime of strong coupling means the kinetic theory treatment is ill-defined and not well controlled. Thus it is of interest to develop techniques that do not rely on these approaches. 

Various extended formulations of hydrodynamics and the conjectured gauge-string duality are two intimately linked non-perturbative approaches that do not rely on the validity of a quasi-particle picture. These tools are quite general and may be adapted to treat a variety of interesting systems. Due to the experimental flexibility of cold quantum gases, this work applies these techniques to strongly interacting Fermi gases. The hope is that in addition to gaining a better understanding of strongly interacting atomic fermi gases, this work may aid in developing general insight into the nature of strongly interacting quantum systems. 

As indicated by the section title, the particular questions addressed in this work are best framed by appealing to some interesting properties of the now well-known conjectured duality between gravity in anti-de-Sitter space and conformal field theory (AdS/CFT). Additional insight and inspiration is taken from experimental measures of transport properties in the quark gluon plasma created in heavy-ion collisions of nuclei. 

\begin{figure}[ht!]
  \includegraphics[width=\textwidth]{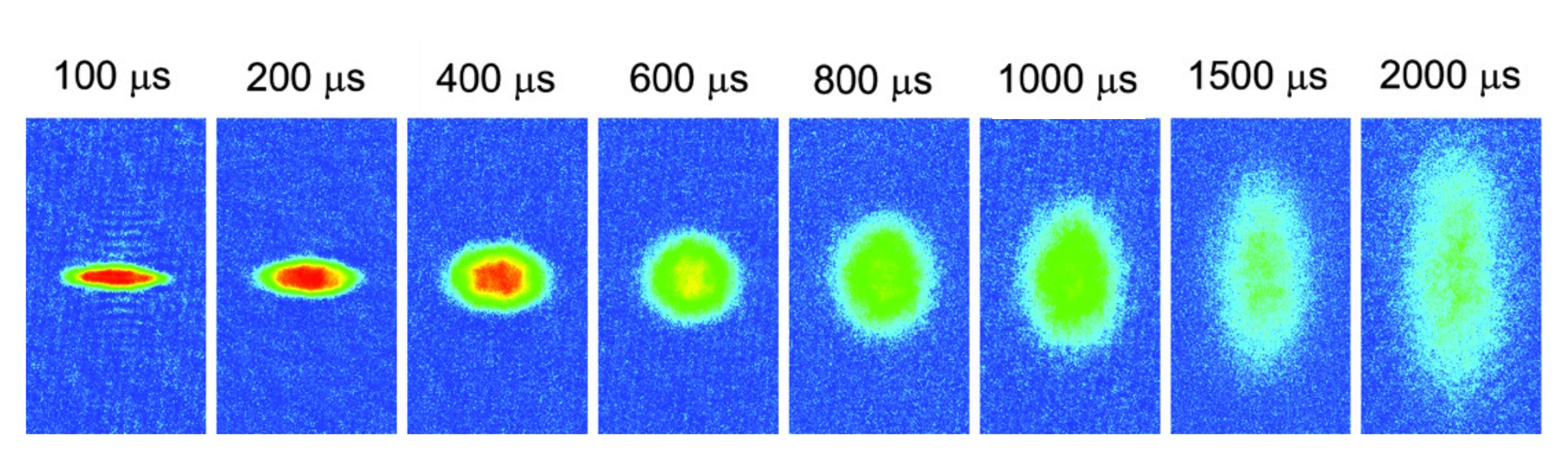}%
  \caption{Demonstration of hydrodynamic expansion of a unitary Fermi gas from an initially elliptical configuration. Note that the aspect ratio of the ellipse $A.R.=a_y/a_x$, where $a_x (a_y)$ the cloud radius along the $x-(y-)$axis, is initially $A.R.<1$ but grows to $A.R.>1$. This is driven by pressure anisotropy in the fluid and is distinct from the case of ballistic expansion where $A.R. \leq 1$. Figure modified with permission from Ref.~\cite{OHara2179}. 
}
  \label{fig:eflow}
\end{figure}

In experiments on a two-spin Fermi gas of $^6$Li in Ref.~ \cite{OHara2179} the interaction strength between the spin species was controlled with an external magnetic field by tuning through a Feshbach resonance (to be discussed in more detail later). This allowed the system to be tuned from weakly to strongly interacting, with expansion of an initially elliptically shaped cloud exhibiting ballistic and hydrodynamic evolution respectively. In the case of hydrodynamic expansion, the subsequent evolution is referred to as elliptic flow, and is shown in Fig.~\ref{fig:eflow}. This type of flow is a signature of hydrodynamics as will be discussed in Chap.~\ref{ifgintro}. Follow up experiments on elliptic flow as well as a collective oscillation known as the breathing mode of the gas in a cigar shaped trap (see Ch.~\ref{sohiso} for details on this and other hydrodynamic modes) were used to extract a the ratio of shear viscosity to entropy density in the gas ($\eta/s$), with a minimum value in the range of $(\eta/s)_{\text{UFG}} \sim 0.2 - 0.4$ in units where $\hbar=1=k_B$ \cite{CaoSci2011}.

\begin{figure}[ht!]
  \includegraphics[width=\textwidth]{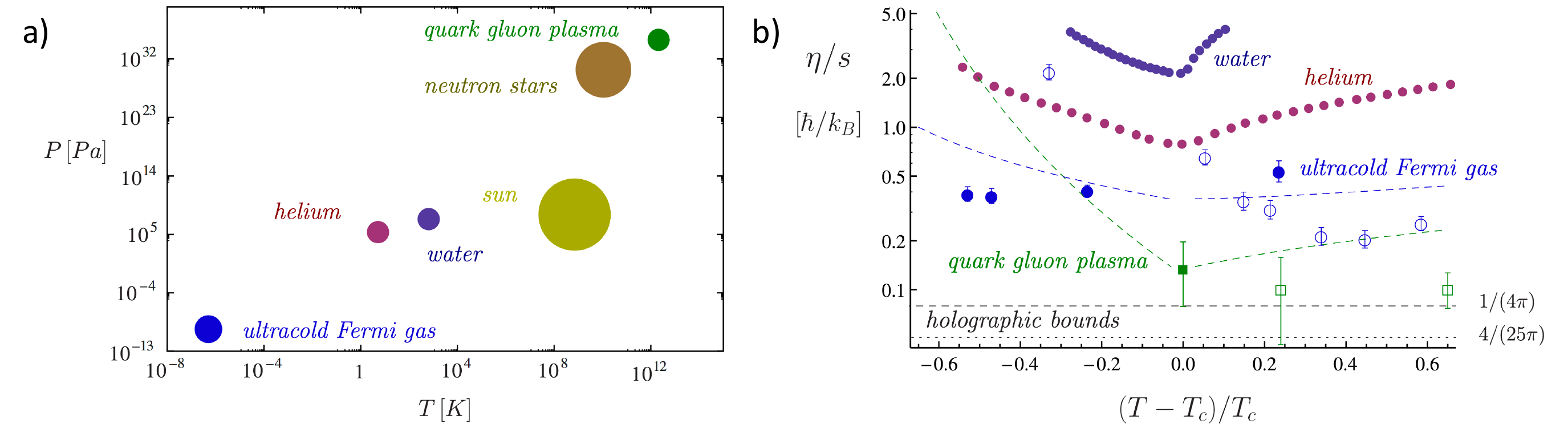}%
  \caption{Left: Plot indicating pressure and temperature of a number of familiar systems. Of particular interest are the ultracold Fermi gas, superfluid $^4$He, and the quark gluon plasma which share quantitatively similar values of $\eta/s$ as shown in b). Reproduced from Ref.~\cite{SchaferPR2009} in accordance with the Attribution-NonCommercial-ShareAlike 3.0 Creative Commons License. Right: Plot of $\eta/s$ vs $T-T_c/T_c$ for ultracold Fermi gas, superfluid $^4$He, the quark gluon plasma, and water for reference. $T_c$ is the superfluid transition temperature for the Fermi gas, the deconfinement temperature for the QGP and the endpoint temperature of the liquid-gas transition for water and helium. Open symbols are from lattice calculations. Reproduced from Ref.~\cite{Adams2012} in accordance with the Attribution-NonCommercial-ShareAlike 3.0 Creative Commons License.
  }
  \label{fig:SIQFs}
\end{figure}

Beyond serving the purpose of characterizing the transport properties of strongly interacting atomic fermi gases, the measured value of $(\eta/s)_{\text{UFG}}$ is surprisingly similar that of the quark-gluon plasma (QGP) created in relativistic heavy ion collisions $(\eta/s)_{\text{QGP}} \sim 0.1-0.2$ extracted from elliptic flow measurements\cite{Song2011}. This fact is all the more striking when considering the fact that these fluids differ by no less than $17$ orders of magnitude in temperature and more than $40$ orders of magnitude in pressure (see Fig.~\ref{fig:SIQFs}a) \cite{SchaferPR2009}. Furthermore, a number of other systems have been calculated or measured to have similar values of $\eta/s$. For example calculations for electron transport in clean graphene (CG) near room temperature give $(\eta/s)_{\text{CG}} \sim 0.2$ for electron transport \cite{Muller2009}, experimental measurements of electron transport in the high$-$T$_c$ superconductor (HTSC) Bi2212 give (an albeit controversial estimate) $(\eta/s)_{\text{HTSC}}\sim0.2$ \cite{Rameau2014}, and for strongly coupled electromagnetic plasma (EMP) values were extracted to be of order $(\eta/s)_{\text{EMP}} \sim 0.1$ \cite{Fortov2013}. One may then ask whether there is some unifying framework with which to think about the quantitative similarity in $\eta/s$ for these vastly different systems. 

For a first pass at understanding this similarity, one may consider an approach using kinetic theory and the Heisenberg uncertainty principle (this argument follows that given for example in Ref.~\cite{Schaferexpose}). Kinetic theory for a dilute gas gives the shear viscosity as $\eta=n p \ell/3$ where n is the quasiparticle density, p the average momentum of a particle, and $\ell$ the mean free path. At the same time, the entropy density is expected to be proportional to density ($s\propto n$). Combining these two expressions with the uncertainty principle $p \ell \geq \hbar$ gives $\eta/s \gtrsim 1$ up to a numerical prefactor. This argument relies on kinetic theory and cannot be expected to hold quantitatively at large coupling and in systems that lack quasiparticles. However, it does give insight that a lower bound on $\eta/s$ in quantum fluids may be expected on fairly general grounds.

A more precise calculation of this type of bound on $\eta/s$ derives from the conjectured duality between certain four-dimensional field theories and string theory on a higher dimensional curved space. For example, an AdS/CFT calculation gives $(\eta/s)_{\text{AdS/CFT}} \geq 1/(4 \pi)$ (known as the KSS bound), while another theory by the name of Gauss-Bonnet gives $(\eta/s)_{\text{GB}} \geq 4/(25 \pi)$. A detailed discussion of the exact nature of these bounds is beyond the scope of this work. However, it is clear that there are some grounds for the expectation of a lower bound on $\eta/s$ in quantum fluids dependent for example on symmetry properties of the system under consideration \cite{Bantilan:2016qos}. Thus it is of interest to make precision studies of transport properties in strongly coupled quantum fluids to address ``big picture" questions like the nature of these lower bounds on transport coefficients. This leads to one of the first major items addressed by this thesis. Namely, are there unexplored methods which might aid in extracting a more precise value for $\eta/s$ in the unitary fermi gas? This question will be addressed by taking inspiration from the physics of the quark gluon plasma. There, $\eta/s$ was first extracted using elliptic flow measurements, but later it was realized that higher order flows (\textit{e.g.} triangular shape expansion) exhibited higher sensitivity to $\eta/s$ and could be used to obtain a better constraint on this transport coefficient \cite{Schenke2012}. It will be shown that collective modes in a harmonically trapped strongly interacting Fermi gas (SIFG) obtained from a modified non-relativistic hydrodynamics  formalism (here referred to as second-order hydrodynamics) demonstrate an analogous feature. This leads to the proposition that higher-order collective oscillations should be used as a probe of transport properties in SIFGs.

In addition to  working towards a more quantitative understanding of how close the SIFG comes to saturating the KSS bound $(\eta/s)_{\text{AdS/CFT}} \geq 1/(4 \pi)$, one may look for other ways in which the SIFGs exhibit quantitatively similar features to other strongly interacting fluids. To this end inspiration is taken yet again from AdS/CFT correspondence. As will be discussed later, the ringdown modes of a black hole in a gravity dual description contain both hydrodynamic modes as well as modes not contained in the standard hydrodynamic description of a fluid (\textit{i.e.} ``non-hydrodynamic modes"). This naturally leads to the question: Does the similarity in transport properties of different strongly interacting quantum fluids extend beyond $\eta/s$? If so, one would expect to find non-hydrodynamic modes when studying the early time dynamics of SIFGs, as such modes are a general feature in the quasi-normal mode spectra of black hole duals. This thesis aims to contribute to the understanding of this topic by working with both second-order hydrodynamics and an approximate gravity dual to the UFG to explore the possibility for and expected properties of such non-hydrodynamic modes in SIFGs.

\section{Overview}

This thesis is organized as follows. Chap.~\ref{ifgintro} reviews the necessary prerequisites needed to understand how hydrodynamic behavior emerges in Fermi gases. This will rely on the existence of a Feshbach resonant interaction. Therefore, the chapter proceeds from the basics of scattering theory to a two-channel model of Feshbach resonances. Although this is a toy model for Feshbach resonance, it will demonstrate all of the important features. In Chap.~\ref{hydrointro}, the kinetic theory approach to deriving hydrodynamics is discussed, and subsequently existing experimental techniques for extracting transport properties are outlined. Chap.~\ref{hydrointro} ends with a discussion of some challenges that arise in applying a typical hydrodynamic framework to extract $\eta/s$ in these experiments. This will  motivate the use of second-order hydrodynamics in this work to treat transport in SIFGs. Chap.~\ref{chap:nhintro} introduces various forms of modified hydrodynamics. The physical interpretation of the additional fluid degrees of freedom introduced by these techniques is discussed. Particularly, the existence of ``non-hydrodynamic" modes is introduced. In Chap.~\ref{sohiso}, collective modes up to the decapole mode of an isotropically harmonically trapped gas are derived from linearized second-order hydrodynamics. After the derivation, properties of higher-harmonic modes lead to a proposal for their use in an alternate technique for constraining $\eta/s$. The structure of non-hydrodynamic modes, which are collective oscillations that are not described by the standard Navier-Stokes formalism is then discussed. Properties of these modes arising from the two models of second-order hydrodynamics as well as an approximate gravity dual calculation will be one of the primary focuses of this work. Subsequently, non-hydrodynamic mode excitation amplitudes are calculated within this framework. It is found that these modes should be experimentally observable. Chap.~\ref{sohaniso} addresses the same problem in an anisotropically trapped gas. In this case, the two transverse trapping frequencies are allowed to differ. Here the properties of a smaller set of modes, which will include the experimentally relevant scissors mode are derived. In Chap.~\ref{sohunif} frequencies and damping rates of the shear and sound modes in second-order hydrodynamics with a uniform density and temperature background are calculated. These results may find application to model non-hydrodynamic behavior in experiments currently being developed by M. Zwierlein's group at MIT \cite{Zwierlein2017}. Chap.~\ref{ggintro} will provide a brief overview of the conjectured gauge-gravity duality needed to understand the results of Chap.~\ref{fgd}. In Chap.~\ref{fgd}, an approximate gravity dual for UFGs is used as a model to make predictions for the non-hydrodynamic mode spectrum of a harmonically trapped UFG. Chap.~\ref{conclusion} provides a summary of key results and future outlook for this area of research.

\chapter{Introduction to Strongly Interacting Fermi Gases}
\label{ifgintro}

This chapter reviews several aspects of Fermi gases relevant for this work. As will be discussed, the presence of a Feshbach resonance allows tuning from weak to strong interactions between atoms in the gas. A key step in understanding Feshbach resonances is the characterization of the interaction potential by a single parameter, the scattering length, particularly in the low-energy scattering limit. The chapter begins with a review of the scattering theory needed to understand the scattering length. A small detour is taken to shown how the scattering cross section is related to the s-wave scattering length with a brief discussion of the relevant features of this relationship for the purposes of this thesis. The chapter then closes with a simple model of a Feshbach resonance. Despite its simplicity, this model will give a realistic picture of how Feshbach resonances arise in atomic Fermi gases. 

\section{Tunable Interactions: Feshbach Resonances}

Perhaps the most straightforward way to reach an understanding of Feshbach resonances is to consider simple two atom scattering where the structure of the atoms is ignored. In doing so the concepts of scattering amplitude, the partial wave expansion, and subsequently scattering length for the low-energy s-wave scattering case are reviewed. After doing this, a simple two-channel scattering model accounting for the atomic structure is introduced to explain the existence of Feshbach resonances. It should be noted that much of this section on Feshbach Resonances is distilled from the review article Ref.~\cite{DuinePhysRep2004}. In addition, some of the discussion in this section is emphasized and clarified by including material which may be found in quantum mechanics textbooks and notes such as Refs.~\cite{Sakurai,DeGrand}. On a final note, scattering in $d=2$ spatial dimensions exhibits a number of subtleties, the details of which are omitted. However, the result for the scattering amplitude in $d=2$ in terms of the s-wave scattering length will be quoted.

\subsection{Scattering Theory: Lippmann-Schwinger Equation}

\begin{figure*}[ht!]
\begin{center}
\includegraphics[width=0.5\textwidth]{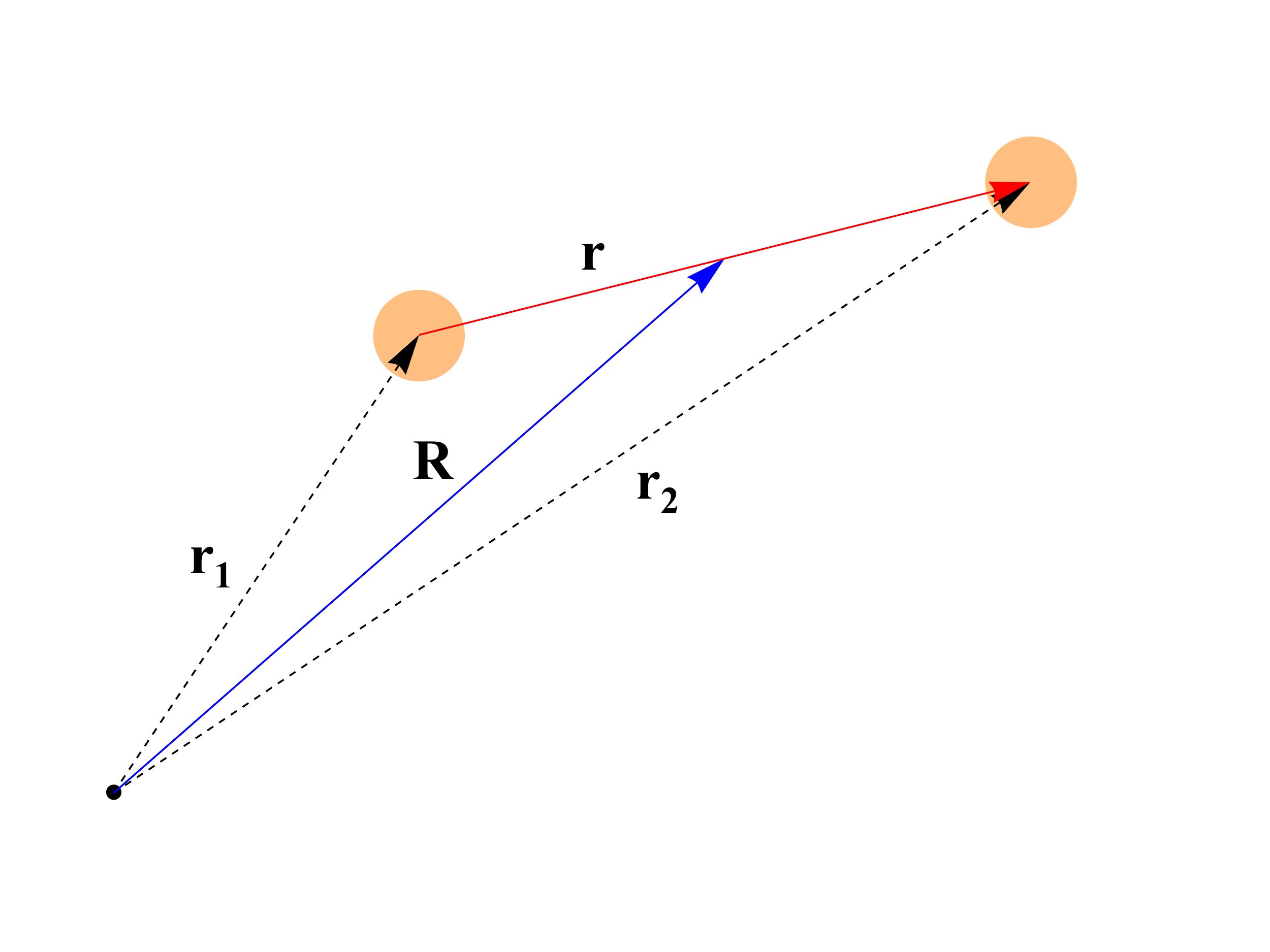}
\end{center}
  \caption{Coordinate system for evaluating the two-particle Schr\"odinger equation.}
  \label{fig:CMcoords}
\end{figure*}

The starting point of quantum mechanical scattering theory of two particles of equal mass in a non-relativistic setting is the two-particle (time-independent) Schr\"odinger equation
	\begin{equation}
		\Big(\dfrac{-   \nabla_{\mathbf{r}_1}^2}{2m} + \dfrac{-   \nabla_{\mathbf{r}_2}^2}{2m} +V(|\mathbf{r}_1-\mathbf{r}_2|)\Big) \psi= E \psi,
	\end{equation}
where E is the total energy of the two particle system. Working in relative and center-of-mass coordinates, $\mathbf{r} = \mathbf{r}_1 - \mathbf{r}_2$ and  $\mathbf{R} = (\mathbf{r}_1 + \mathbf{r}_2)/2$ respectively, one can separate the wave function into a product of functions depending only on the center-of-mass (CM) and relative coordinates as $\psi = \psi_{CM}(\mathbf{r}) \psi_{rel}(\mathbf{R})$ (see Fig.~\ref{fig:CMcoords}) where the two functions satisfy
	\begin{align}
		\label{eq:CMSE}
		&\dfrac{- \nabla^2_{\mathbf{R}}}{2 (2m)} \psi_{CM} = E_{CM} \psi_{CM},\\
		\label{eq:relSE}
		&\Big(\dfrac{- \nabla^2_{\mathbf{r}}}{2 (m/2)} + V(r) \Big) \psi_{rel} = E_{rel} \psi_{rel},
	\end{align} 
and $E = E_{CM} + E_{rel}$. Eq.~\eqref{eq:CMSE} is the Schr\"odinger equation for a free particle of mass $2m$ and its solution is simple a plane-wave $\psi_{CM} \propto e^{-i \mathbf{K} \cdot \mathbf{R}}$ with dispersion relation $E_{CM} =   \mathbf{K}^2/(4 m)$, and does not contain any physics of interest here. Thus Eq.~\eqref{eq:relSE} contains all of the relevant physics of scattering needed here. In order to simplify notation the substitutions $\psi_{rel}\to\psi$ and $E_{rel}\to E$ will be made in the following.

It will be useful to recast Eq.~\eqref{eq:relSE} in a more formal language as
	\begin{equation}	
		\label{eq:formSE}
		\Big(H_0 + V\Big) |\psi \rangle = E |\psi \rangle,
	\end{equation}
where $H_0$ is the free part of the Hamiltonian in Eq.~\eqref{eq:relSE}. Formal manipulation of Eq.~\eqref{eq:formSE} gives the Lippmann-Schwinger equation:
	\begin{equation}
		\label{eq:LipSchE}
		|\psi^{(\pm)}\rangle = \dfrac{1}{E-H_0 \pm i \epsilon}V|\psi^{(\pm)}\rangle+ |\phi\rangle,
	\end{equation}
where the $\pm i \epsilon$ is required to regulate the divergence caused by the continuous nature of the spectrum of $H_0$, and $|\phi\rangle$ is the (plane-wave) solution of Eq.~\eqref{eq:formSE} with vanishing potential, \textit{i.e.} $V=0$.

The Lippmann-Schwinger equation encodes an interesting feature of the solution $|\psi^{(\pm)}\rangle$ far from the center of the scattering potential. Namely, the wave function may be written as a sum of an incoming plane wave plus an outgoing spherical wave. This fact will be useful to define a scattering amplitude which, in the low-energy scattering limit, may be characterized to lowest order by a single value, the scattering length.

\subsection{Scattering Theory: Scattering Amplitudes and the T-Matrix}
To see that the wave function may be written as a sum of an incoming plane wave plus an outgoing spherical wave, consider $\psi^{(\pm)}(\mathbf{r})$ given by
	\begin{equation}
		\psi^{(\pm)}(\mathbf{r}) = \langle \mathbf{r} |\dfrac{1}{E-H_0 \pm i \epsilon}V|\psi^{(\pm)}\rangle+ \langle \mathbf{r} |\phi\rangle.
	\end{equation}
Working in $d=3$, insertion of the identity gives
	\begin{equation}
		\psi^{(\pm)}(\mathbf{r}) = \int d^3r' \phantom{.}\langle \mathbf{r} |\dfrac{1}{E-H_0 \pm i \epsilon}|\mathbf{r'}\rangle\langle\mathbf{r'}|V|\psi^{(\pm)}\rangle+ \langle \mathbf{r} |\phi\rangle.
	\end{equation}
	
Now, recall that $G^{(\pm)}(\mathbf{r},\mathbf{r'})=\langle \mathbf{r} |\dfrac{1}{E-H_0 \pm i \epsilon}|\mathbf{r'}\rangle$ is the free-particle Green's function which up to a proportionality constant is given by
	\begin{equation}
	\label{GF}
		G^{(\pm)}(\mathbf{r},\mathbf{r'}) \propto \dfrac{e^{\pm i k |\mathbf{r}-\mathbf{r'}|}}{ |\mathbf{r}-\mathbf{r'}|},
	\end{equation}
so that
	\begin{equation}
		\label{eq:projectedpsi}
		\psi^{(\pm)}(\mathbf{r}) = \int d^3r' \phantom{.} \dfrac{e^{\pm i k |\mathbf{r}-\mathbf{r'}|}}{ |\mathbf{r}-\mathbf{r'}|}V(r')\langle\mathbf{r'}|\psi^{(\pm)}\rangle+ \langle \mathbf{r} |\phi\rangle,
	\end{equation}
where for now, the constant of proportionality in Eq.~\eqref{GF} is dropped as it is unimportant for the purposes of this chapter. If the potential $V$ is localized within some region of width $\Delta_V$ around $\mathbf{r'}=\mathbf{r'}_0$, the integrand of Eq.~\eqref{eq:projectedpsi} may be expanded for $|\mathbf{r}-\mathbf{r'}_0|\gg \Delta_V$. Without loss of generality, let $\mathbf{r'}_0 = 0$ in which case one may use $|\mathbf{r}-\mathbf{r'}|\approx r - \mathbf{\hat{r}}\cdot\mathbf{r'}$ in Eq.~\eqref{eq:projectedpsi} to obtain
	\begin{equation}
		\label{eq:projectedpsiapprox}
		\psi^{(\pm)}(\mathbf{r}) \approx \frac{e^{\pm i k r}}{r}\int d^3r' \phantom{.} e^{\mp i\mathbf{k'} \cdot \mathbf{r'}} V(r')\langle\mathbf{r'}|\psi^{(\pm)}\rangle+ e^{i \mathbf{k}\cdot\mathbf{r}},
	\end{equation}
where $\mathbf{k'}=k \mathbf{\hat{r}}$, and the plane wave solution with $V=0$ was used.

Upon inspection, one notices that Eq.~\eqref{eq:projectedpsiapprox} takes the form of a linear combination of an outgoing (incoming) spherical wave for the plus (minus) choice of sign in $\psi^{(\pm)}$ and a plane wave. Since a scattered state is considered here, the outgoing spherical wave is chosen giving the scattering amplitude as
	\begin{equation}
		\label{eq:scattamp}
		f(\mathbf{k'},\mathbf{k}) \equiv \int   d^3r' \phantom{.} e^{- i\mathbf{k'} \cdot \mathbf{r'}} V(r')\langle\mathbf{r'}|\psi^{(+)}\rangle.
	\end{equation}
One may define the $T-$Matrix operator by $T|\phi\rangle = V |\psi^{(+)}\rangle$ such that Eq.~\eqref{eq:scattamp} may be manipulated to give
	
	\begin{align}
	        \nonumber  f(\mathbf{k'},\mathbf{k}) &= \int   d^3r' \phantom{.} e^{- i\mathbf{k'} \cdot \mathbf{r'}} V(r')\langle\mathbf{r'}|\psi^{(+)}\rangle\\
	      \nonumber    &=  \int   d^3r' \phantom{.} e^{- i\mathbf{k'} \cdot \mathbf{r'}} \langle\mathbf{r'} | T | \phi \rangle\\
	         \nonumber &=  \langle\mathbf{k'}| T | \mathbf{k}\rangle.         
	\end{align}
This result indicates that for $r \gg \Delta_V$ the scattering problem is fully specified by the $T-$Matrix. However, finding the $T-$Matrix is a difficult problem in its own right, and this form is not very useful for the purposes of this work. To arrive at a more useful formulation of the scattering amplitude, additional geometric and physical features of the problem must be utilized.

\subsection{Scattering Theory: Partial Wave Expansion and Scattering Phase Shift}

\begin{figure*}[t!]
\includegraphics[width=\textwidth]{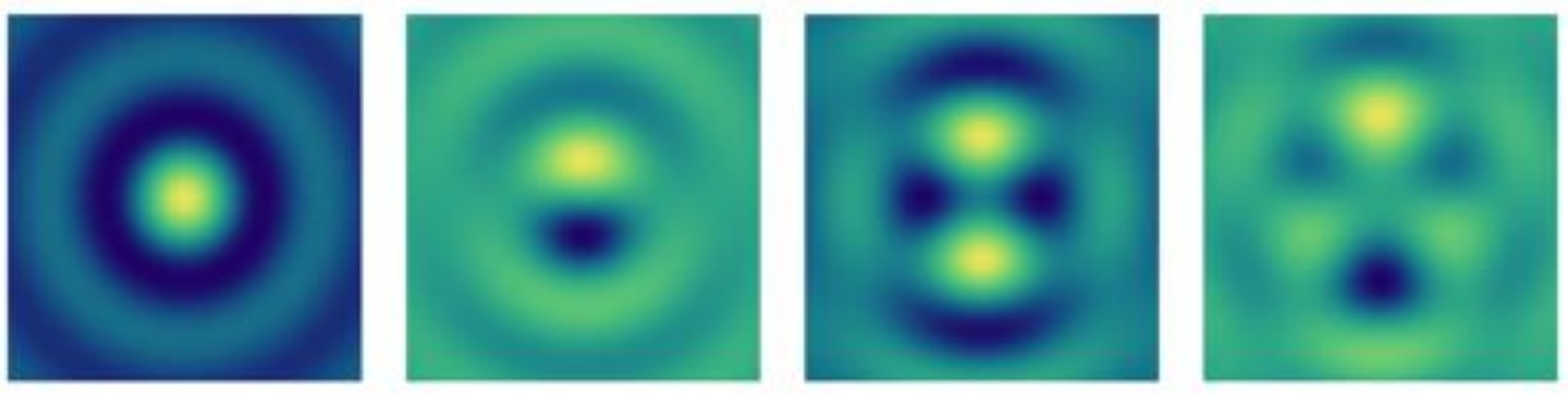}
  \caption{Cross sectional geometry of the partial wave scattering states with the vertical axis being the axis of rotation symmetry. Notice in particular that the s-wave has continuous rotation symmetry.}
  \label{fig:PWE} 
\end{figure*}

Notice that as long as the scattering potential $V$ is spherically symmetric, so is the $T-$Matrix. Hence, $f(\mathbf{k'},\mathbf{k})$ depends only on the magnitude $k$ and the angle $\theta$ between $\mathbf{k}$ and $\mathbf{k'}$. As a result, one may expand $f(\mathbf{k'},\mathbf{k})$ in terms of Legendre polynomials as
	\begin{equation}
		\label{eq:partialwaveexp}
		f(\mathbf{k'},\mathbf{k}) = \sum_{\ell=0}^{\infty} f_\ell(k) P_\ell(\cos \theta).
	\end{equation}
This result will be combined with the expansion of of plane wave in Legendre polynomials	   	\begin{equation}
		\label{eq:planewaveexp}
		e^{i \mathbf{k} \cdot \mathbf{r}} = \sum_{\ell=0}^{\infty} (2 \ell + 1) i ^\ell j_\ell(k r) P_\ell(\cos \theta),
	\end{equation}
where $j_\ell(k r)$ denotes the spherical Bessel function of the first kind. The asymptotic form of $j_\ell(k r)$ for $kr\gg 1$ is
	\begin{equation}
		\label{eq:asypexp}
		j_\ell(k r) \approx \frac{\sin (k r - \ell \pi/2)}{r} = \frac{e^{i (k r - \ell \pi/2)} -e^{-i (k r - \ell \pi/2)}}{2 i r}, 
	\end{equation}
so that Eqs.~\eqref{eq:projectedpsiapprox}-\eqref{eq:asypexp} may be combined to arrive at
	\begin{equation}
		\label{eq:partialwaveapprox}
		\psi(\mathbf{r})=\sum_{\ell=0}^{\infty} (2 \ell + 1) \frac{P_\ell(\cos \theta)}{2 i k} \bigg( \big(1 + \frac{ 2 i k f_\ell (k)}{(2 \ell +1)}\big) \frac{e^{ikr}}{r} - \frac{e^{-i(kr - \ell \pi)}}{r}  \bigg),
	\end{equation}
for $kr\gg 1$. Eq.~\eqref{eq:partialwaveapprox} indicates that $\psi(\mathbf{r})$ is the sum of an incoming and outgoing spherical waves decomposed into a complete angular basis (see Fig.~\ref{fig:PWE}). The amplitudes of these incoming and outgoing waves must match in magnitude if probability is to be conserved. Hence $ f_\ell (k)$ must take the form
	\begin{equation}
		\label{eq:phaseform}
		 f_\ell (k) = \frac{(2 \ell +1)}{2 i k} \big(e^{2 i \delta_\ell(k)} -1\big) = \frac{(2 \ell + 1)}{k \cot\big(\delta_\ell(k)\big) - ik},
	\end{equation}
where $\delta_\ell(k)$ is the energy-dependent scattering phase shift of the $\ell^{th}$ partial wave. To make further progress, the low-energy behavior of Eq.~\eqref{eq:phaseform} is now considered.

\subsection{Scattering Theory: Low-Energy Limit and Scattering Length}
To study the low energy behavior of Eq.~\eqref{eq:phaseform}, it is most convenient to first note a quite general feature of the $\ell^{th}$ scattering phase shift $\delta_\ell$. Namely, it has been shown that for a wide class of short-range potentials $\delta_\ell \sim k^{2 \ell+1}$ for $\ell < \ell_{max}$ with $\ell_{max} > 0$ and $\delta_\ell \sim k^{2 \ell_{max}+1}$ for $\ell>\ell_{max}$ at low-energy (see e.g. review articles Refs. \cite{Sadeghpour2000,Chin2010}). Cursory inspection of Eq.~\eqref{eq:phaseform} with this information clearly indicates that for low enough energy, only the s-wave ($\ell=0$) contributes to the wave-function. 

Focusing on the s-wave phase shift one may expand  $k \cot(\delta_0) \approx -1/a + \mathcal{O}(k^2)$. As a result, the scattering amplitude may be expressed as
	\begin{equation}
		f(k) \equiv f_0(k)  \approx \frac{-1}{\displaystyle \frac{1}{a} + i k}.
	\end{equation}
This is the final form of the scattering amplitude which will be important in describing how the use of Feshbach resonances in cold atom experiments allow for the atoms to strongly interact resulting in the emergence of hydrodynamic behavior. The above derivation relied on the dimensionality $d=3$ of space being considered (see e.g. Ref. \cite{Adhikari1986}), but this thesis considers both two- and three-dimensional gas geometries, and hence the scattering amplitude in both $d=2,3$ is quoted here\cite{Levinsen2015}
	\begin{equation}
		\label{eq:amplfin}
		f(k) = \begin{cases}
			\displaystyle \frac{-4}{\displaystyle \frac{2}{\pi} \log (k a) - i} \text{\phantom{.....}for $d=2$}\\    				\displaystyle \frac{-1}{\displaystyle \frac{1}{a} + i k} \text{\phantom{.....}for $d=3$}\\
		\end{cases}.
	\end{equation}

\subsection{Optical Theorem and Large Scattering Cross Section}

The optical theorem reads \cite{DeGrand,Levinsen2015}
	\begin{equation}
		\sigma(k) = \begin{cases}
			\displaystyle \frac{-1}{k} \text{Im}\lbrack f(k)\rbrack \text{\phantom{.....}for $d=2$}\\    				\displaystyle \frac{4 \pi}{k} \text{Im}\lbrack f(k)\rbrack  \text{\phantom{.....}for $d=3$}\\
		\end{cases},
	\end{equation}
where the form of $f(k)$ appropriate to the dimension of space is taken from Eq.~\eqref{eq:amplfin}. Applying this theorem results in a scattering cross section
	\begin{equation}
		\label{eq:csecfull}
		\sigma(k) = \begin{cases}
			\displaystyle \frac{4}{k \bigg( \displaystyle \frac{4}{\pi^2} \log^2 (k a) + 1\bigg) } \text{\phantom{.....}for $d=2$}\\    
			\displaystyle \frac{4 \pi}{\displaystyle \frac{1}{a^2} + k^2 } \text{\phantom{.....}for $d=3$}\\
		\end{cases}.
	\end{equation}
From this it is obvious to see that if it were possible to tune the scattering length, one could control the scattering cross section. This feature is discussed in more detail below where the concepts of unitarity and scale-invariance, which relate to certain assumptions utilized in later chapters, are introduced. 

\subsubsection{Strong Interactions, Unitarity, and Scale-Invariance}

For a given incident particle momentum $k$, one may find the value of scattering length which maximizes the scattering cross section and hence corresponds the the regime of strongest interactions between the atoms. For $d=2$ the value of $a$ which achieves this is $a = 1/k$ while in $d=3$ one finds $a = \pm \infty$ giving
	\begin{equation}
		\sigma_{max}(k) = \begin{cases}
			\displaystyle \frac{4}{k} \text{\phantom{.....}for $d=2$}\\    
			\displaystyle \frac{4 \pi}{k^2 } \text{\phantom{.....}for $d=3$}\\
		\end{cases}.
	\end{equation}
For a Fermi gas at a temperature not too far from the Fermi temperature, atoms that participate in scattering events have momentum close to the characteristic Fermi momentum $k_F$. Hence $k_F a$ is a good parameter characterizing the interaction strength. Particularly, $k_Fa \sim 1\sim 1$ is the regime of strong interactions in $d=2$, while $| k_F a| \gg 1$ is the regime of strong interactions in $d=3$. 

Now, considering the regime of strong interactions in $d=3$ one has that $a \to \pm \infty$. It turns out that the only length-scale in such a situation is the density \cite{Forbes}. In this case the gas is said to be in the unitary regime, a term which refers to the fact that the cross section takes its maximum value allowed by unitarity of quantum mechanics. Furthermore, $a \to \pm \infty$ also implies that the scattering length drops out as a physically relevant length scale. In this case, the gas is scale-invariant giving rise to the relationship
	\begin{equation*} 
		\text{Strongly Interacting} \equiv \text{Unitary} \equiv \text{Scale Invariant in d=3}.
	\end{equation*}
On the other hand, in $d=2$, the logarithmic momentum dependence of the cross section (see Eq.~\eqref{eq:csecfull}) implies that in the regime of strongest interaction the scattering length is of the same order as the density length scale ($a \sim 1/k_F$) and therefore does not drop out. In this case the gas is not exactly scale invariant, though, as will be discussed more later, collective mode properties behave as though the strongly interacting 2D Fermi gas is scale-invariant \cite{VogtPRL2012,TaylorPRL2012}. Thus in $d=2$
	\begin{equation*} 
		\text{Strongly Interacting} \sim \text{Scale Invariant in d=2 for certain phenomena}.
	\end{equation*}
Notice that since the scattering length is a relevant length scale for strong interactions in $d=2$ the gas is not said to be unitary. For this reason, the present work refers to strongly interacting Fermi gases in both $d=2,3$ to mean gases near a Feshbach resonance where the scattering length is nearly saturated.

To summarize, the regime of strongest interactions is exactly (approximately) scale-invariant in $d=3$ ($d=2$) dimensions. This fact is built into assumptions for modeling hydrodynamic behavior later in this thesis. Furthermore, interaction strength is characterized by the scattering length. Thus if one had experimental control of the scattering length, one could then tune the strength of interactions in the gas. How to achieve this control in practice is the topic of Secs.~\ref{sec:TMat} and \ref{sec:Fesh}.

\subsection{Poles and Branch Cuts of the T-Matrix: Bound and Scattering States }
\label{sec:TMat}
To see the relationship between scattering length and bound state energy, consider the $T-$Matrix. 
From Eq.~\eqref{eq:LipSchE} and the definition $T |\phi \rangle = V |\psi \rangle$ of the $T-$Matrix one has the formal solution
	\begin{equation}
		T = V + V \displaystyle \frac{1}{E-H + i \epsilon} V.
	\end{equation}
Recalling $H=H_0+V$, $V$ in the denominator of the second term may be treated perturbatively by expanding the fraction in a geometric series. In the original problem, the potential will have both bound and scattering states. Letting $n$ denote a discrete index for bound states and $\mathbf{k}$ be a continuous index for scattered state, inserting a complete set of states gives
	\begin{equation}
	\label{aneq}
		T = V + V \sum_{n} \displaystyle \frac{| \psi_n\rangle \langle \psi_n|}{E-\epsilon_n + i \epsilon} V + \int \frac{d^dk}{(2 \pi)^d} V  \displaystyle \frac{| \psi_\mathbf{k} \rangle \langle\psi_\mathbf{k}|}{E-\epsilon_\mathbf{k} + i \epsilon} V.
	\end{equation}
Thus, the analytic structure of the $T-$Matrix in the complex energy plane is such that it has poles at (negative) bound state energies (second term in Eq.~\eqref{aneq}) and a branch-cut along the positive energies corresponding to scattered states (third term in Eq.~\eqref{aneq}) . Now, utilizing Eq.~\eqref{eq:amplfin} one may see that for small positive energies $E =k^2/m$ the $T$ matrix is given by
	\begin{equation}
		T \approx \displaystyle \frac{1}{1-i a \sqrt{mE}}.
	\end{equation}
Analytic continuation to negative energies gives
	\begin{equation}
		T \approx \displaystyle \frac{1}{1-a \sqrt{-mE}},
	\end{equation}
which has a pole at energy $E = -1/(m a^2)$ where the scattering length $a$ is taken large and positive. Thus control over the energy of a bound state of the potential with small negative energy, would allow tuning of the scattering length. This is achieved in practice through the use of Feshbach resonances. 

\subsection{Two-Channel Model of Feshbach Resonances}
\label{sec:Fesh}
A variety of Fermonic alkali atoms/isotopes are typically utilized in experiments, perhaps the most relevant being potassium-40 ($^{40}$K) and lithium-6 ($^6$Li). Internal atomic structure will lead to many hyperfine states of slightly differing energy when an external magnetic field is applied. Experimentally, two hyperfine states are selected and the system is prepared in an $50:50$ incoherent mixture of these two states. Hence each atom may take on one of two internal states labeled $|\uparrow\rangle$ and $|\downarrow\rangle$. Now, for two colliding atoms, the internal states may be in a singlet or triplet configuration which are up to normalization are
	\begin{align}
		\nonumber \text{Singlet}&\propto |\uparrow \downarrow\rangle-| \downarrow\uparrow\rangle,\\
		\text{Triplet} &\propto \begin{cases}
			 |\uparrow \uparrow \rangle\\
			 |\uparrow \downarrow\rangle+| \downarrow\uparrow\rangle\\    
			 | \downarrow\downarrow\rangle.
		\end{cases}.
	\end{align}
Recalling that quantum statistics governs that Fermions (Bosons) in the same state can only scatter via odd (even) partial waves and that p-wave scattering will be negligible when there is s-wave scattering\cite{DeGrand}, the triplet configuration may be effectively reduced to one state for consideration in the problem of s-wave scattering. Labelling the singlet state $|S\rangle$ and triplet state $|T\rangle$ gives
	\begin{align}
		\nonumber |S\rangle &\propto |\uparrow \downarrow\rangle-| \downarrow\uparrow\rangle\\
		|T\rangle &\propto |\uparrow \downarrow\rangle+| \downarrow\uparrow\rangle.
	\end{align}
Quantum statistics ensures that the electrons may get closer together in the singlet configuration. Hence one expects that the interatomic potential for atoms in the singlet state is deeper than that in the triplet state. 

Working in $d=3$ and following the treatment of Ref.~\cite{DuinePhysRep2004}, one may approximate the interatomic potential as a simple square-well of range $ R$ taken to satisfy $R\ll k_F^{-1/3}$ (\textit{i.e.} the range is much smaller than the average inter-particle particle spacing) for both the singlet and triplet channels. The Schr\"odinger equation will then be of the form
	\begin{equation}
		\label{eq:TwochSE}
		\begin{pmatrix} -\displaystyle \frac{\nabla^2}{m} + V_T(\mathbf{r}) - E & V_{hf}\\
			V_{hf} & -\displaystyle \frac{\nabla^2}{m} + V_S(\mathbf{r}) + \Delta\mu B - E 
		 \end{pmatrix} \begin{pmatrix}  \psi_T(\mathbf{r}) \\ \psi_S(\mathbf{r})    \end{pmatrix} = 0,
	\end{equation}
where
	\begin{equation}
		V_{S,T}(\mathbf{r}) = \begin{cases}
			-V_{S,T} \text{\phantom{.....}for $r<R$}\\    
			0 \text{\phantom{.....}for $r>R$}
		\end{cases}.
	\end{equation}
The off diagonal terms $V_{hf}$ in Eq.~\eqref{eq:TwochSE} are the hyperfine terms which couple the internal states, and $\Delta\mu B$ is the Zeeman energy splitting caused by the difference in total magnetic moment of the singlet and triplet configurations. For simplicity, it will be assumed that $V_S$ is deep enough to have one, but only one bound state. Furthermore, it is assumed that  $0<V_{hf} \ll V_T,V_S,\Delta \mu B$. This model has three key features allowing for Feshbach resonant behavior. They are:
	\begin{itemize}
	\item Two channels with one of the channels having an interatomic potential with a bound state (here the singlet channel).
	\item The ability to tune the location of the bound state within that channel (here through the Zeeman splitting).
	\item Coupling of the bound state (singlet) to the continuum channel (triplet). Here the hyperfine interaction couples the channels.
	\end{itemize}
	\begin{figure}[ht!]
\includegraphics[width=\textwidth]{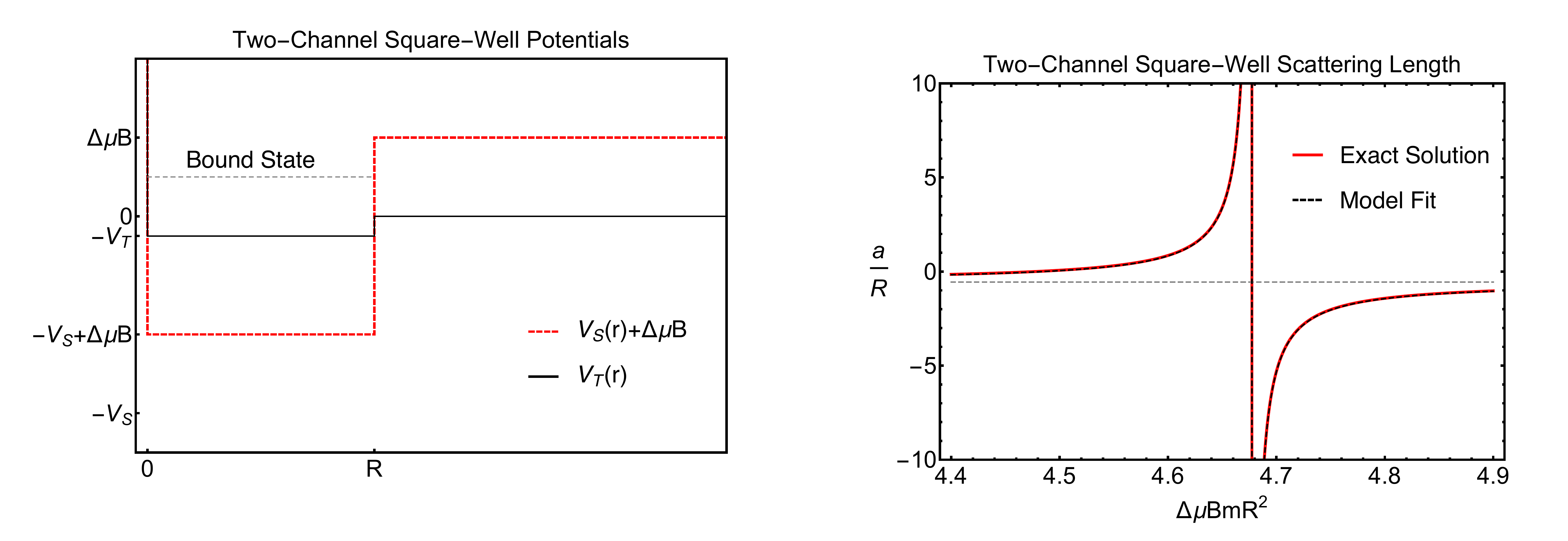}
  \caption{Left: Potential functions including Zeeman shift for the square-well model. Right: Resulting scattering length versus applied magnetic field demonstrating divergence in scattering length. Here we used $V_S=10$, $V_T=1$, and $V_{hf}=0.1$ in units of $(m R^2)^{-1}$.}
  \label{fig:feshscattlen}
\end{figure}
The detailed solution of this model is not particularly enlightening for the purposes of this thesis and is not presented here (see Ref.~\cite{DuinePhysRep2004} for the full solution). However, after finding the solution one may calculate the resulting s-wave scattering length which is shown versus magnetic field in Fig.~\ref{fig:feshscattlen}. The resulting scattering length is well fit by the functional form
	\begin{equation}
		a(B) = a_{bg} \bigg( 1 - \displaystyle \frac{\Delta B}{B-B_0}\bigg),
	\end{equation}
where $a_{bg}$ is a background scattering length at large magnetic field, $B_0$ gives the location of the resonance, and $\Delta B$ the width of the resonance. Clearly by tuning to the resonance, one obtains unitarity in $d=3$. This will be the regime of interest when discussing the applicability of hydrodynamics to cold atomic gases in the next chapter.
\chapter{Emergent Hydrodynamic Behavior}
\label{hydrointro}

Hydrodynamics is a set of partial differential equations governing the dynamics of macroscopic fields of density, temperature, and fluid velocity which arise from conservation of mass, momentum, and energy. As will be discussed, the equations of hydrodynamics as the governing theory of a strongly interacting Fermi gas may be arrived at through at least two approaches. The approach considered in this chapter is one in which hydrodynamics arises from taking a strong-coupling limit of a microscopic kinetic theory. In doing so, one may take moments of the kinetic theory phase-space distribution function with respect to so called collision invariants. In order to close the conservation equations that arise from this process, one may take two different avenues. First one may arrive at a controlled expansion with an appropriate small parameter, namely the time between collisions of atoms in the gas $\tau_0$. This collision time is inversely proportional to the scattering cross section $\tau \propto 1/\sigma$ so that in the regime of strong interactions one expects $\tau_0$ to be small. The second approach treats hydrodynamics as a gradient expansion supplemented by symmetry considerations, and will be further discussed in Chap~\ref{chap:nhintro}.

In the remainder of this chapter, first the conservation equations of hydrodynamics from kinetic theory are derived. Subsequently, the relaxation time approximation is presented. A short relaxation time provides a small parameter motivating a truncated expression for the non-equilibrium stress tensor in terms of gradients of fluid variables. This expression provides the final equation required for closure of the hydrodynamic conservation equations. Finally, experimental methods which utilize hydrodynamics to extract the shear viscosity coefficient are highlighted. The chapter closes with a discussion of problems that arise in applying the standard hydrodynamic theory to these experiments, setting the stage for the introduction of second-order hydrodynamics in Chap~\ref{chap:nhintro}.

\section{Hydrodynamics from Kinetic Theory}
\label{sec:HydrofromKT}
Boltzmann kinetic theory describes the evolution of a quasi-particle distribution function $f(\mathbf{x},\mathbf{p},t)$ in phase space. Since the purpose of this section is to demonstrate how hydrodynamics arises from a strongly interacting limit of kinetic theory, mean-field interactions and self-energy effects are ignored (see e.g. Ref~\cite{Menotti2002} for inclusion of mean-field interactions). In this case the semi-classical non-relativistic Boltzmann equation for a gas with elastic collisions reads \cite{Uehling1933}
\begin{equation}
	\label{BE}
	D_t f = (\partial_t + \frac{\mathbf{p}}{2m} \cdot \nabla_{\mathbf{x}} -\nabla_{\mathbf{x}} U \cdot \nabla_{\mathbf{p}} ) f = -C[f],
\end{equation}
where $C[f]$ is a functional describing the change in the distribution function due to collisions,
\begin{equation}
\label{eq:ColInt}
C[f] = \int \frac{d^d p_2}{(2 \pi)^2} \frac{2 q}{m} \int d \Omega \frac{d \sigma}{d \Omega} \big[f_3 f_4 (1-\xi f)(1-\xi f_2)-f f_2(1-\xi f_3)(1-\xi f_4)\big].
\end{equation}
In Eq.~\eqref{eq:ColInt} $\xi = 1,0,-1$ for a gas with Fermi-Dirac, Classical, and Bose-Einstein statistics respectively. Furthermore, in Eq.~\eqref{eq:ColInt} $f_i$ is the distribution function evaluated at momentum $\mathbf{p}_i$, $\mathbf{p}$ and $\mathbf{p}_2$ are momenta of the incoming particles, $\mathbf{p}_3$ and $\mathbf{p}_4$ are momenta of the outgoing particles, $q = |\mathbf{p}-\mathbf{p}_2|/2=|\mathbf{p}_3-\mathbf{p}_4|/2$ is the relative momentum for the collision, $\Omega$ is the reflection angle between the incoming and outgoing momenta, and $\sigma(\theta,\phi,q)$ is the scattering cross section. 

Despite the rather complicated form of the collision integral, in a non-relativistic elastic scattering problem, the conservation of particle number, momentum, and energy implies that momentum space moments of the collision integral with weights $1$, $\mathbf{p}$, and $\mathbf{p}^2$ are identically zero. In order to derive conservation laws, one may define the $N^{th}$-order moment operator $N=0,1,2,3...$ by
\begin{equation}
	M^{(N)}_{i_1,i_2,...,i_N}[g] \equiv m^{1-(N+d)}\int d^d p \prod_{j=1}^N p_{i_j} g,
\end{equation}
where $m$ is the particle mass, and if $N=0$ it is understood that 
\begin{equation}
	M^{(0)}[g] \equiv  m^{1-d} \int d^dp\phantom{.}g.
\end{equation}
Note that the distribution function satisfies \cite{BrewerPRA2016}
\begin{align}
M^{(0)}[f] &= \rho(\mathbf{x}),\\
M^{(1)}_i[f] &= \rho(\mathbf{x})u_i,\\
\frac{1}{2}\sum_j M^{(2)}_{j,j}[f] &=\frac{\rho(\mathbf{x})(\mathbf{u}^2 + D \frac{T}{m})}{2}\equiv \epsilon,
\end{align}
where $\rho$ is the mass density, $\mathbf{u}$ is the local fluid velocity, $T$ is the temperature, and $\epsilon$ is the energy density. Also note that
\begin{align}
M^{(0)}[C[f]] &= 0,\\
M^{(1)}_i[C[f]] &= 0,\\
\frac{1}{2}\sum_j M^{(2)}_{j,j}[C[f]] &=0,
\end{align}
encoding that elastic collisions conserve particle number, momentum, and energy. 
Integration by parts for the zeroth-moment leads to
\begin{equation}
	\label{ktmcons}
	M^{(0)}[D_t f] =\partial_t\rho + \partial_i(\rho u_i)=0.
\end{equation}
For the first-order moment one derives
\begin{equation}
\label{ktpcons}
M^{(1)}_i[D_t f] =\partial_t(\rho u_i)+ \partial_j(\rho u_i u_j + P \delta_{ij} + \pi_{ij})=\rho \frac{F_i}{m},
\end{equation}
where P is the pressure. Finally, making use of the second order moment gives
\begin{equation}
\label{ktecons}
	\sum_j M^{(2)}_{j,j}[D_t f] = \partial_t \epsilon + \partial_j\big[u_j\big(\epsilon + P\big)+\pi_{ij}\big ]=\rho \frac{F_k}{m} u_k = 0 = M^{(2)}[C[f]],
\end{equation}

Eqs.~\eqref{ktmcons}-\eqref{ktecons} are a set of conservation laws. Yet, the system of equations is not closed. In the above equations, $ \pi_{ij}$ is related to a second-order moment of the kinetic theory distribution function, but in order to close the system of equations, $\pi_{ij}$ must be expressed in terms of fluid dynamical variables. This is the goal of the next subsection.

\subsection{Relaxation Time Approximation and Shear Viscosity}
\label{sec:RTA}
In order to simplify the analysis of $ \pi_{ij}$, one may make the so called relaxation time approximation
\begin{equation}
	\label{eq:RTA}
	C[f] =\frac{-(f-f_{eq})}{\tau_0},
\end{equation}
where $f_{eq}$ is an equilibrium distribution function satisfying
\begin{equation}
(\partial_t + \frac{\mathbf{p}}{2m} \cdot \nabla_{\mathbf{x}} - \nabla_{\mathbf{x}} U \cdot \nabla_{\mathbf{p}} ) f_{eq} =0.
\end{equation}
Ignoring quantum statistics, $f_{eq}$ is given by the Maxwell-Boltzmann distribution (see e.g. \cite{BrewerPRA2016})
\begin{equation}
	f_{eq} = \dfrac{\rho(\mathbf{x}) N_d}{m^{1-d/2} T^d \pi^{d/2}} exp\bigg(\dfrac{-m^{d/2}(\mathbf{\mathbf{p}/m}-\mathbf{u})^2}{2T^{d/2}}\bigg), 
\end{equation}
where $N_d =1,1/\sqrt{8}$ in $d=2,3$ respectively ensuring integration over momenta gives the density $\rho(\mathbf{x})$. Eq~\ref{eq:RTA} states that collisions cause the system to relax towards equilibrium on a timescale $\tau_0$, capturing the essential physics of collisions. Furthermore, it is interesting to note that this approach has been demonstrated to be quantitatively accurate for small deviations from equilibrium \cite{Cercignani}.

By linearizing Eq.~\eqref{eq:RTA} about equilibrium ( $f = f_{eq} + \epsilon \delta f^{(1)}$) and taking moments as in the previous section one finds \cite{SchaferPRA2014}
\begin{equation}
	\tau_0=\eta/P,
\end{equation}
where $\eta$ is the shear viscosity and $P$ is the pressure. The stress tensor is
\begin{equation}
\label{eq:STen}
\pi_{ij}= -\eta \sigma_{ij},
\end{equation}
with
\begin{equation}
\sigma_{ij} = \big[\partial_i u_j + \partial_j u_i - \frac{2}{d} \delta_{ij} \partial_k u_k \big].
\end{equation}
Eqs.~\eqref{ktmcons}-\eqref{ktecons} along with the closure relation Eq.~\eqref{eq:STen} are the Navier-Stokes equations. It is important to note, however, that Eq.~\eqref{eq:STen} is valid only for $\tau_0 \ll1$ and small spatial and temporal derivatives of fluid variables (not too far from equilibrium). One may systematically take the linearization of the distribution function to higher order. For example let $f = f_{eq} + \epsilon \delta f^{(1)} \to f_{eq} + \epsilon \delta f^{(1)} + \epsilon^2 \delta f^{(2)} $ wherein one obtains terms in Eq.~\eqref{eq:STen} up to second-order in derivatives of fluid variables that are proportional to $\tau_0^2$\cite{SchaferPRA2014}. This process may be continued to even higher orders in the expansion of $\pi_{ij}$ with the behavior of increasing the derivative order by one along with one higher power of $\tau_0$ in the prefactor. Thus one should think of Eq.~\eqref{eq:STen} as the lowest order of gradient expansion for the non-equilibrium stress tensor. Furthermore, the parameter $\tau_0$ which controls this expansion is the relaxation time and should be related to the time between collisions in the gas $\tau_0 \propto \tau_{col} \sim 1/(n \sigma v_F)$. Hence if $\sigma$ is large, which is the case near a Feshbach resonance, one expects $\tau_0$ to be small and thus for the gradient expansion of $\pi_{ij}$ to be well controlled.

\section{Methods for Extracting Transport Properties}

When displaced from equilibrium, the system evolution in hydrodynamics will depend on transport coefficients. Below two types of experiments that have been used to extract values for the shear viscosity coefficient are discussed. Emphasis is placed on the qualitative relationship between the measured quantity and shear viscosity. 

\subsection{Gas Expansion Dynamics
	\label{Expansion}}

One approach for extracting shear viscosity is to release the gas from an elliptically shaped trapping potential. The evolution of the cloud may then be used to extract shear viscosity \cite{CaoSci2011}. Interestingly, this is similar to the method used to estimate shear viscosity of the quark-gluon plasma created in relativistic ion collisions, a fact which will be leveraged later in order to motivate the study of collective modes in Chap~\ref{sohiso}. It is simplest to understand how expansion from an anisotropic initial configuration gives information about shear viscosity by considering two limiting cases of the interaction strength. In particular, it is shown that when the gas is non-interacting (\textit{i.e.} $\eta \to \infty$), the flow is such that the aspect ratio $A$ measured as the ratio of the mean square radii of the short to long axis of the cloud (so that initially $A<1$) satisfies $A \to 1^-$ as $t \to \infty$. When the interactions are so strong that ideal hydrodynamics holds (\textit{i.e.} $\eta=0$), one can show that the aspect ratio actually continues to grow beyond $A=1$ before decreasing again, a phenomenon known as elliptic flow. In essence then, by modeling the hydrodynamic flow for different values of shear viscosity, one can match the measured aspect ratio from expansion dynamics.

\begin{figure*}[ht!]
\includegraphics[width=0.75\textwidth]{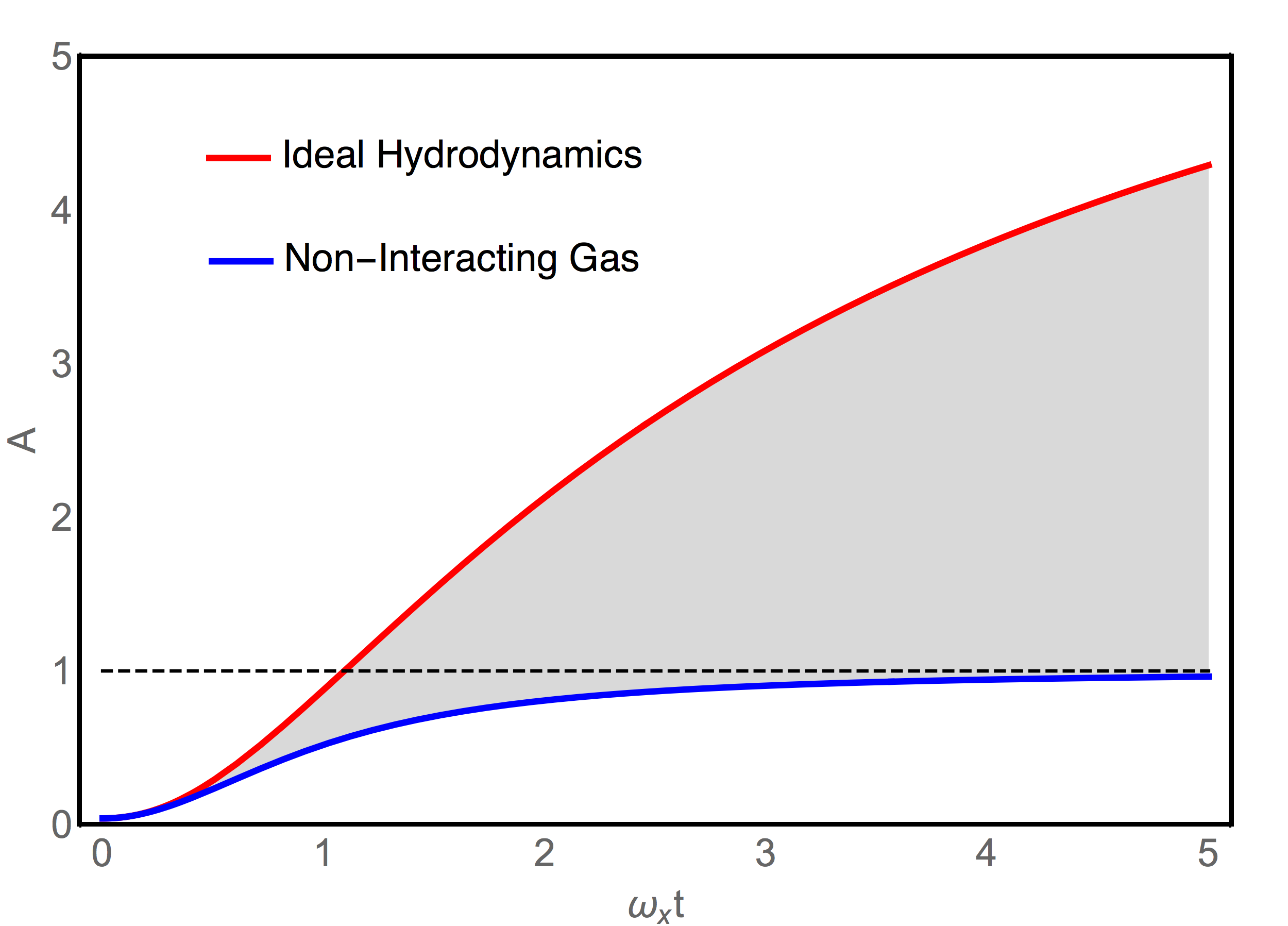}
\centering
  \caption{Aspect ratio for an initial gas configuration having $\omega_y/\omega_x=5$ vs time in the limiting cases of ideal hydro $\eta=0$ (red) and a non-interacting gas $\eta = \infty$ (blue). Here a simple ideal gas equation of state ($P=nT$) is used. See e.g. Chap.~\ref{sohiso} for a discussion of this assumption. Finite non-zero shear viscosity will produce an aspect ratio vs time lying in the gray region assuming $\omega_y/\omega_x=5$. Hence, comparing experimental measurement of the aspect ratio of strongly interacting Fermi gases to hydrodynamic simulations can be used to extract the shear-viscosity.}
  \label{fig:AspRat}
\end{figure*}

Considering first the case of a non-interacting gas, a derivation for the cloud aspect ratio is performed following closely the work of Ref. \cite{Menotti2002}. For a non-interacting gas, the cloud dynamics are described by the collisionless Boltzmann transport equation
\begin{equation}
	(\partial_t -\nabla_{\mathbf{x}}U \cdot\nabla_{\mathbf{v}} + \mathbf{v} \cdot \nabla_{\mathbf{x}})f(t,\mathbf{x},\mathbf{v}) = 0.	
\end{equation}
Denoting the time independent equilibrium distribution by $f_{eq}(\mathbf{x},\mathbf{v})$ one has
\begin{equation}
(-\nabla_{\mathbf{x}}U \cdot\nabla_{\mathbf{v}} + \mathbf{v} \cdot \nabla_{\mathbf{x}})f_{eq}(\mathbf{x},\mathbf{v}) = 0.
\end{equation}
For a harmonically trapped gas then
\begin{equation}
\label{eq:HBE}
(\partial_t-\omega_i^2 \partial_{v_i} + v_i \partial_{x_i})f(t,\mathbf{x},\mathbf{v}) = 0,
\end{equation}
and
\begin{equation}
\label{eq:EqHBE}
(-\omega_i^2 \partial_{v_i} + v_i \partial_{x_i})f_{eq}(\mathbf{x},\mathbf{v}) = 0,
\end{equation}
Multiplying Eq.~\eqref{eq:EqHBE} by $x_j v_j$ and integrating over phase space $\int d^dx d^dv /N$ where N is the total particle number then implies
\begin{equation}
\label{eq:radref}
\omega_j^2 \langle r_j^2\rangle - \langle v_j^2\rangle = 0,
\end{equation}
where it was assumed that $f_{eq}(\mathbf{x},\mathbf{v}) \to 0$ when either $\mathbf{x} \to \infty$ or $\mathbf{v} \to \infty$ sufficiently fast that boundary terms from integrating by parts go to zero. Consider the scaling ansatz
\begin{equation}
	\label{eq:distansatz}
	f(t,\mathbf{x},\mathbf{v}) = f_{eq}(\mathbf{X},\mathbf{V}),
\end{equation}
where $X_i = x_i/\lambda_i(t)$ and $V_i=\lambda_i(t) v_i - \dot{\lambda}_i(t) x_i$. Substituting Eq.~\eqref{eq:distansatz} into Eq.~\eqref{eq:HBE} and utilizing the chain rule relations for partial derivatives one finds for the trapped gas
\begin{equation}
	\label{eq:subBE}
	\bigg(\frac{V_i}{\lambda_i^2}\partial_{X_i}-\lambda_i X_i (\ddot{\lambda}_i-\omega_i^2 \lambda_i)\partial_{V_i}\bigg)f_{eq}(\mathbf{X},\mathbf{V})=0.
\end{equation}
Before moving on, notice that the last term here proportional to $\omega_i^2$ is a result of the trapping force term $-\omega_i^2 \partial_{v_i} f(t,\mathbf{x},\mathbf{v})$ in Eq.~\eqref{eq:HBE}. To study free streaming of a gas released from a harmonic trap with $\omega_x \ll \omega_y$, one should use an equilibrium solution $f_{eq}$ appropriate for the harmonic trap, but remove the term proportional to $\omega_i^2$ in Eq.~\eqref{eq:subBE}. Integrating over phase space $\int X_i V_i d^dX d^dV/N$ one finds
\begin{equation}
      \ddot{\lambda}_i - \frac{\omega_i^2}{\lambda_i^3} =0.
\end{equation}
For free expansion one additionally has the initial conditions $\lambda_i(0)=1$ and $\dot{\lambda}_i(0)=0$, so that the solution is
\begin{equation}
      \lambda_i(t) = \sqrt{1 + \omega_i^2 t^2}.
\end{equation}
For a 2D gas with $\omega_x < \omega_y$ the non-interacting aspect ratio becomes
\begin{equation}
     A_{NI} = \frac{\omega_x^2}{\omega_y^2}\bigg( \frac{1+\omega_y^2 t^2}{1 + \omega_x^2 t^2} \bigg)\xrightarrow[t \to \infty]{} 1.
\end{equation}
A plot of $A_{NI}$ is shown in Fig.~\ref{fig:AspRat}.

\begin{figure*}[ht!]
\includegraphics[width=\textwidth]{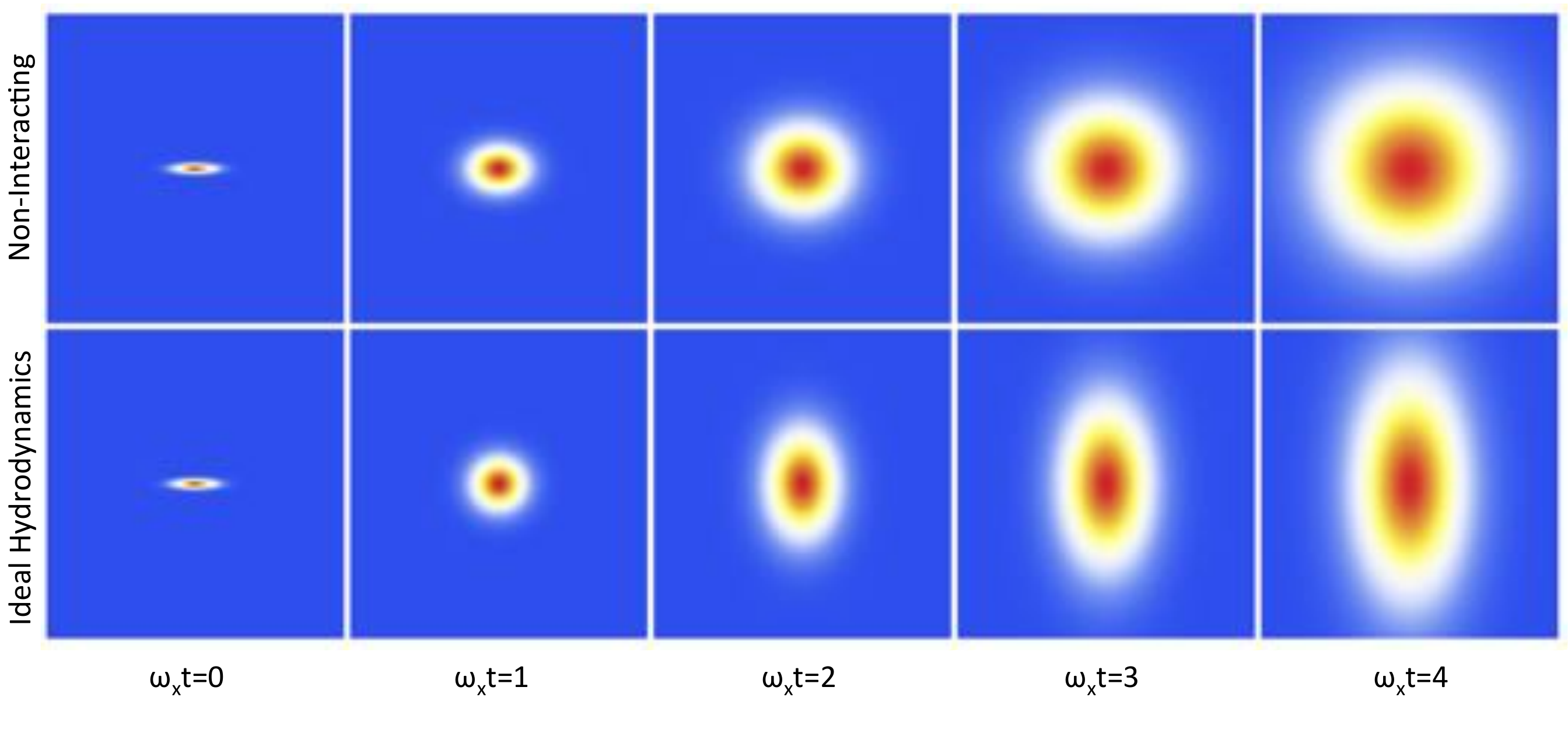}
\centering
  \caption{Time snap shots of density profiles for non-interacting ballistic flow (top) and ideal hydrodynamic elliptic flow (bottom). Notice that for the hydrodynamic case the aspect ratio grows beyond unity while for the non-interacting case it approaches unity.}
  \label{fig:Flowexamp}
\end{figure*}
Considering the case of ideal hydrodynamics with vanishing shear viscosity (and hence vanishing stress tensor), the governing equations of motion are given by
\begin{align}
	\label{idealmasscons}
	&\partial_t\rho + \partial_i(\rho u_i)=0,\\
	\label{idealmomcons}
	&\partial_t(\rho u_i)+ \partial_j(\rho u_i u_j + P \delta_{ij})=\rho \frac{F_i}{m},\\
	\label{idealengcons}
	&\partial_t \epsilon + \partial_j\big[u_j\big(\epsilon + P\big)\big]=\rho \frac{F_k}{m} u_k,
\end{align}
In the above equations, $\epsilon$ is specified in terms of the fluid velocity, mass density and pressure $P$ as 
\begin{align}
	\label{idealeng}
	&\epsilon=\frac{\big(\rho \mathbf{u}^2 + d P\big)}{2}.
\end{align}
Assuming the gas to be either two-dimensional or translationally invariant along the $z-$axis one may make the ansatz
\begin{equation}
\rho(\mathbf{x},t) = \rho_0(\mathbf{X})/(b_x(t) b_y(t)),
\end{equation}
where as before $X_i = x_i/b_i(t)$ $i \in \{1,2\}$. Substituting this ansatz into Eq.~\eqref{idealmasscons} one finds that the fluid velocity must satisfy 
\begin{equation}
 u_i = \frac{x_i \dot{b}_i(t)}{b_i(t)}. 
\end{equation}
Assuming the gas is isothermal $T(x,y,t) = T(t)$, and substituting the initial equilibrium profile of a harmonic trap, $\rho_0(\mathbf{x}) =\rho_0 \exp{[{\frac{-(\omega_x^2 x^2+ \omega_y^2 y^2)m}{2 T_0}}]}$, where $T_0$ is the initial temperature of the gas, into Eqs.~\eqref{idealmomcons} and \eqref{idealengcons} one finds $T(t) = 1/(b_x(t) b_y(t))^{(2/d)}$. The time dependent scale factors when the trapping force is removed are governed by
\begin{equation}
	\ddot{b}_i(t) - \frac{\omega_i^2}{(b_x(t)b_y(t))^{2/d} b_i(t)}=0.
\end{equation}
The aspect ratio obtained from solving these equations for $d=2$ is shown in Fig.~\ref{fig:AspRat}. Of particular interest is the fact that the aspect ratio exceeds unity before reducing again at late times (not shown). Expansion dynamics between the cases of ideal hydrodynamics $\eta =0$ and the non-interacting theory $\eta = \infty$ may be simulated and matched with experiment to extract shear viscosity. For concreteness, Fig.~\ref{fig:Flowexamp} demonstrates the density profile for an initial trap configuration with $\omega_y/\omega_x=5$ in both the non-interacting and ideal hydrodynamic cases. Note the qualitative similarity of the fluid flow in the bottom row of Fig.~\ref{fig:Flowexamp} to the experimentally observed flow of a strongly interacting Fermi gas shown in Fig.~\ref{fig:eflow}. However, in Fig.~\ref{fig:Flowexamp} the peak density is normalized to unity at all times for ease of visualization, while it should scale with the inverse volume of the gas $\rho_{peak}\sim1/(b_x(t) b_y(t))$ in order that particle number is conserved. Finally, while the above derivation was conducted for an ideal gas equation of state, elliptic flow is a ubiquitous signature of hydrodynamics also appearing in different systems such as the quark-gluon plasma created in relativistic heavy-ion collision.

\subsection{Collective Mode Damping \label{Collectivedampshear}}

\begin{figure*}[ht!]
\includegraphics[width=\textwidth]{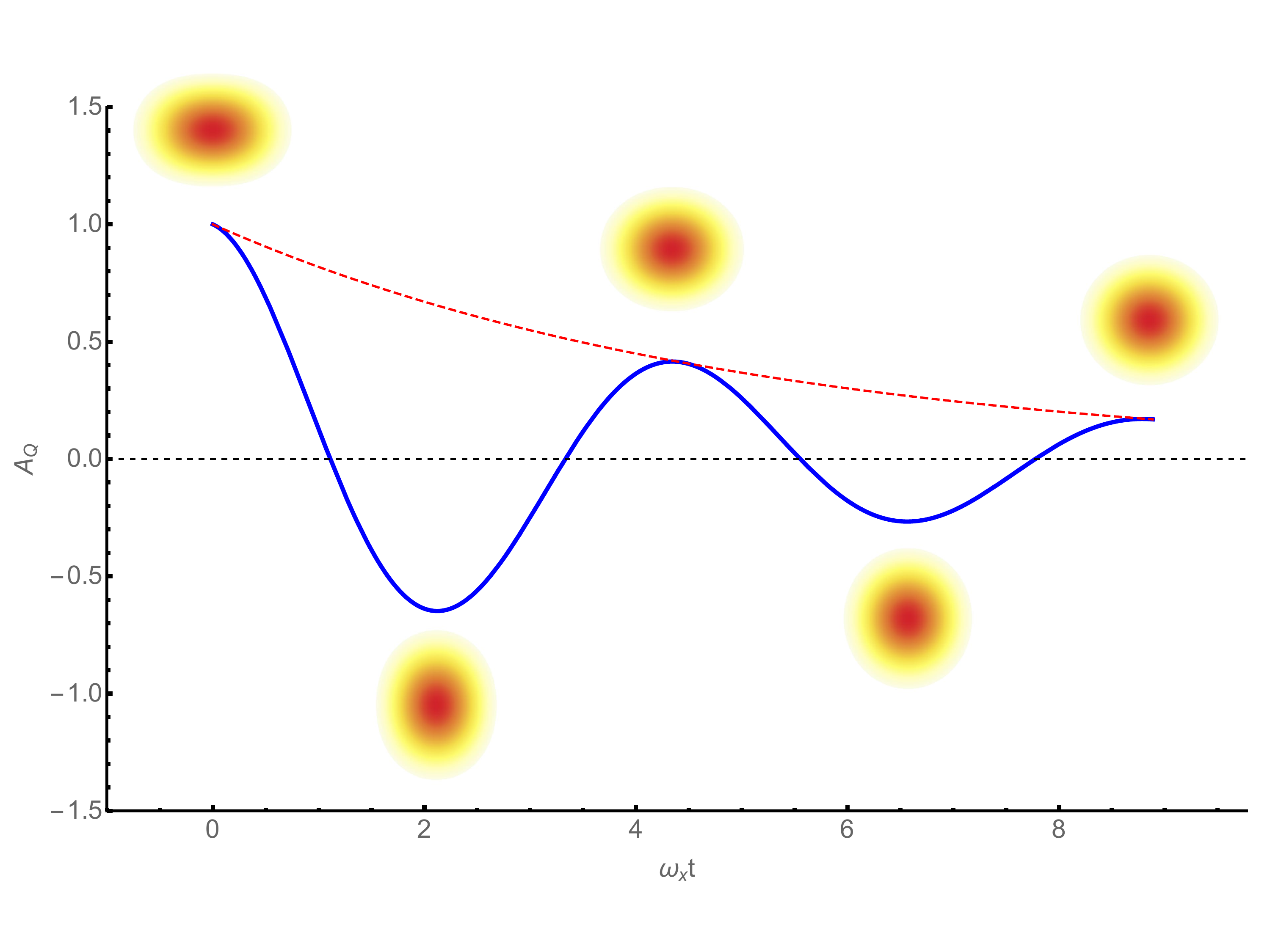}
\centering
  \caption{Time dependence of the quadrupole mode (blue) showing the exponential damping envelope (red) with damping $\Gamma \propto \eta/P$. The orange profiles above the maxima and minima show the density profile of the gas at that time.}
  \label{fig:Quaddampexamp}
\end{figure*}

An additional method for extracting shear viscosity of strongly interacting Fermi gases found in the literature is through the analysis of the damping rates of collective oscillations in nearly harmonic trapping potentials. An example of the time dependence and density profile of the quadrupole mode collective oscillation in an isotropic harmonic trapping potential is shown in Fig.~\ref{fig:Quaddampexamp}. In the hydrodynamic regime (small values of shear viscosity $\eta$), the energy dissipation rate in the fluid is given by \cite{LandauLifshitz} 
\begin{equation}
\dot{E}=- \int d^dx \bigg[\frac{1}{2} \eta(\mathbf{x}) \big(\partial_i u_j + \partial_k u_j + \frac{2}{d} \delta_{ij} \nabla \cdot \mathbf{u} \big)^2 + \zeta(\mathbf{x}) \big(\nabla \cdot \mathbf{u}\big)^2\bigg],
\end{equation}
where $\zeta$ is the bulk viscosity. The damping rate of a collective mode is then $\Gamma=|\langle \dot E\rangle_t / (2 \langle E\rangle_t)|$, where $\langle \phantom{.} \rangle_t$ denotes time averaging \cite{VogtPRL2012}. For the two-dimensional quadrupole mode, parametrizing the shear viscosity as $\eta(\mathbf{x}) = \hbar n_{2D}(\mathbf{x}) \alpha(T/T_F)$, Ref. \cite{VogtPRL2012} obtains $\alpha(T/T_F)= m\langle r^2 \rangle \Gamma_Q/(2 \hbar)$, or $\Gamma_Q \propto \eta/(nT)$ (where the factor of $T$ is from $\langle r^2 \rangle \propto T$). For the case of the breathing mode, the damping rate is proportional to the bulk viscosity which has been demonstrated to be small for the strongly interacting fermi gas experimentally in $d=2$ \cite{VogtPRL2012} and theoretically in $d=3$ \cite{SchaferPRL2013}.

The technique of measuring damping rates to extract shear viscosity was also carried out in Ref.~\cite{CaoSci2011} for the  breathing mode in a cigar shaped trap ($d=3$) where one axis is approximately translationally invariant (see Chap.~\ref{sohiso} for more details about this mode). The measured damping rate near $T_C$ is consistent with measurements from elliptic flow at higher temperatures. Additionally, damping of the scissors mode in $d=3$ (see Chap.~\ref{sohaniso} for more details about this mode) was investigated in Ref.~\cite{Bruun2007}. 

Studying the relationship between collective mode damping and transport coefficients is a central theme of this work. Particularly, beginning in Chap.~\ref{sohiso}, this relationship is investigated within a modified hydrodynamic framework and later in Chap.~\ref{fgd} with an approximate gravity dual theory. However, before moving to the discussion and application of these frameworks, it is important to understand the motivation for approaching the problem using tools beyond the standard Navier-Stokes description. 

\subsection{Breakdown of Hydrodynamics}

The previous sections sketched how to extract transport coefficients from matching hydrodynamic modeling to measurements of cloud dynamics. However, upon closer inspection one finds that these models require the use of a trap averaged shear viscosity which introduces a poorly characterized model dependence through a necessary cutoff radius in the trap averaging procedures \cite{SchaferPRA2014,Bluhm2015,Bluhm2016}. 

The need for this cutoff can be understood from high temperature scaling arguments. In the high temperature regime, kinetic theory arguments show that viscosity scales as $T^{3/2}$ independent of density and produces infinite viscous heating \cite{Bluhm2015,Bluhm2016}. Similar pathologies happen in the low-temperature regime in the low density corona of the cloud. The reason the for the inapplicability of the Navier-Stokes equations in the low density region of the cloud can be understood by appealing to the Knudsen number $Kn = \ell_{mfp}/L_{phys}$ where $\ell_{mfp}$ is the mean free path and $L_{phys}$ is a representative physical length scale (here the cloud radius). If the Knudsen number is large then the continuum assumption of fluid mechanics is a poor approximation, and hence a Navier-Stokes description is inappropriate. In the low density corona, the mean free path will become large due to low densities \cite{SchaferPRA2014}. This then requires the application of a cutoff radius in trap averages of hydrodynamic transport properties to arrive at sensible results. 

Unfortunately, systematic errors are likely to arise from utilizing a cutoff radius in trap averaging the shear viscosity, and the magnitude of these possible errors are poorly understood. Hence it is desirable to develop an approach which circumvents these issues by providing an accurate description of both the hydrodynamic core and low density corona. Furthermore, such a theory would preferably not require matching of kinetic theory calculations in the low density corona to hydrodynamic calculations for the core. Such an approach would likely be computationally intensive. Additionally, one would need to demonstrate that the results were insensitive to the particular matching procedure.
\chapter{Modified Hydrodynamics}
\label{chap:nhintro}

This chapter summarizes two modified theories of hydrodynamics that are capable of addressing the breakdown of the standard Navier-Stokes treatment in the low-density corona of the gas discussed at the end of the last chapter.  The first section introduces second-order hydrodynamics which is the next order in the gradient expansion of the non-equilibrium stress tensor. Particularly, it is possible to use only symmetry properties of the fluid to build the gradient expansion without reference to an underlying kinetic theory.

The second section will discuss a useful phenomenological  theory termed ``resummed" second-order hydrodynamics which actually contains all-order derivatives. This repackaging may be further simplified into a relaxation equation for the non-equilibrium stress tensor which will have a simplified physical interpretation. It is this simplified version of ``resummed" second-order hydrodynamics that will be used in coming chapters.

Next, the closely related theory of anisotropic hydrodynamics is presented. This theory shares many similarities with ``resummed" second-order hydrodynamics. Furthermore, the quantitative applicability of anisotropic hydrodynamics for extracting transport coefficients of Fermi gases has been more extensively studied \cite{Bluhm2015,Bluhm2016}. Thus it is interesting to provide at least a qualitative comparison between second order and anisotropic hydrodynamics pointing out their shared characteristics in order to anticipate novel behavior of the fluid predicted by these equations. 

\section{Second-Order Hydrodynamics}

Ref.~\cite{Chao2012} considers a three-dimensional non-relativistic fluid with conformal symmetry and derives all the allowed terms in a second-order gradient expansion of the non-equilibrium stress tensor. The result is given by \cite{SchaferPRA2014,Chao2012}

\begin{align}
\label{eq:SOHEXP}
\nonumber \pi_{ij} = -\eta \sigma_{ij} &+ \eta \tau_R \big(\dot{\sigma}_{ij} + u^k \nabla_k \sigma_{ij} +\frac{2}{3} \langle\sigma \rangle \sigma_{ij} \big) + \lambda_1 \sigma_{ \langle i}{^k} \sigma_{j \rangle k} + \lambda_2 \sigma_{ \langle i}{^k} \Omega_{j \rangle k}\\ 
\nonumber &+ \lambda_3 \Omega_{ \langle i}{^k} \Omega_{j \rangle k} + \gamma_1 \nabla_{ \langle i}T \nabla_{j \rangle}T + \gamma_2 \nabla_{ \langle i}P \nabla_{j \rangle}P + \gamma_3 \nabla_{ \langle i}T \nabla_{j \rangle}P\\
& \phantom{................}+ \gamma_4 \nabla_{ \langle i} \nabla_{j \rangle}T + \gamma_5 \nabla_{ \langle i} \nabla_{j \rangle}P.
\end{align}
In the above expression $\tau_R, \lambda_\ell$, and $\gamma_m$ for $\ell \in \{1,...,3\}$ and $m \in \{1,...,5\}$ are second-order transport coefficients, $\mathcal{O}_{\langle ij \rangle} = \frac{1}{2}\big(\mathcal{O}_{ij} + \mathcal{O}_{ji} - \frac{2}{3} \delta_{ij} \mathcal{O}^k_k\big)$ is the traceless symmetric part of the tensor $\mathcal{O}_{ij}$, $\langle\sigma \rangle = \partial_k u_k$, and $\Omega_{ij} = \nabla_i u_j - \nabla_j u_i$ is the fluid vorticity. 

\section{Resummed Second-Order Hydrodynamics}

The equations of hydrodynamics with the second-order equation for $\pi_{ij}$ are known as the Burnet equations. The Burnet equations are known to exhibit instabilities \cite{Burnet}. In the setting of relativistic hydrodynamics, these instabilities may be regulated by treating the off-equilibrium stress tensor as an independent dynamical variable in the gradient expansion \cite{ISRAEL1979341}. In the non-relativistic case, one may follow the same procedure to arrive at \cite{SchaferPRA2014}
\begin{equation}
\label{eq:resumhy}
\pi_{ij} = -\eta \sigma_{ij} - \tau_R \big(\dot{\pi}_{ij} + u^k \nabla_k \pi_{ij} +\frac{5}{3} \langle\sigma \rangle \pi_{ij} \big) + ...,
\end{equation}
where other terms from Eq.~\eqref{eq:SOHEXP} are denoted by ``...".

It is not difficult to see that solving iteratively, Eq.~\eqref{eq:resumhy} matches Eq.~\eqref{eq:SOHEXP} up to second order but contains derivatives of all orders. While with this approach it is possible to correctly capture the gradient expansion up to arbitrary order, a useful phenomenological approach is to reduce this to a relaxation equation for $\pi_{ij}$ \cite{Book}
\begin{equation}
\label{eq:MIS}
\pi_{ij} + \tau_\pi \dot{\pi}_{ij} = -\eta \sigma_{ij}, 
\end{equation}
where $\tau_\pi$ is the timescale on which the non-equilibrium stress tensor relaxes to its Navier-Stokes form. Eq.~\eqref{eq:MIS} in conjunction with the conservation equations Eqs.~\eqref{ktmcons}-\eqref{ktecons} are henceforth referred to as the equations of second-order hydrodynamics, but it is important to note that Eq.~\eqref{eq:MIS} contains gradients beyond this order, which will not exactly match all the gradients obtained from symmetry arguments.

\section{Anisotropic Hydrodynamics}

One way to view the equations of second-order hydrodynamics is that the Navier-Stokes description is supplemented by additional ``non-hydrodynamic" degrees of freedom allowing a smooth crossover to the ballistic regime. In Eq.~\eqref{eq:MIS} these degrees of freedom are contained in the components of the off-equilibrium stress tensor. Another formalism discussed in Refs.~\cite{Bluhm2015,Bluhm2016} termed anisotropic hydrodynamics takes exactly this approach arriving at a slightly different set of dynamical equations. In Refs.~\cite{Bluhm2015,Bluhm2016}, the modified fluid equations are motivated by appealing to an underlying kinetic theory description and following the moment procedure outlined in Ch.~\ref{sec:HydrofromKT}-\ref{sec:RTA}. However, instead of expanding around the equilibrium distribution function $f_0$ in Ch.~\ref{sec:RTA}, an expansion around an exact solution for ballistic expansion is used. The resulting equations of motion for the fluid (without a trapping potential) are given by\cite{Bluhm2015,Bluhm2016}
\begin{align}
	\label{eq:AH1}
	&D_0\rho = -\rho \partial_i u_i,\\
	&D_0 u_i= -\frac{1}{\rho} ( \partial_i P + \partial_j \pi_{ij}), \\
	&D_0 \mathcal{E} = -\frac{1}{\rho} \partial_i (u_i P + u_j \pi_{ij}),\\
	\label{eq:AH4}
	&D_0 \mathcal{E}_a = -\frac{1}{\rho} \partial_i (\delta_{ia}u_i P + \delta_{ia} u_j \pi_{ij}) - \frac{P}{2 \eta \rho}(P_a-P),
\end{align}
where $D_0 = \partial_t + u_k \partial_k$, $\mathcal{E} = \epsilon/\rho$ is the energy per mass, $P = 2/3 \mathcal{E}^0$, $\mathcal{E}^0 = \mathcal{E}-1/2\rho \mathbf{u}^2$, $P_a = 2 \mathcal{E}^0_a$, $\mathcal{E}^0_a = \mathcal{E}_a - 1/2\rho u_a^2$, $P=1/3 \sum_a P_a$, and $\pi_{ij} = \sum_a \delta_{ia} \delta_{ja} (P_a-P)$ where $a \in \{1,2,3\}$. Notice from these relations that Eqs.~\eqref{eq:AH1}-\eqref{eq:AH4} may be rewritten in terms of a set of standard hydrodynamic degrees of freedom $\rho$, $\mathbf{u}$, and $\mathcal{E}$ along with the non-hydrodynamic degrees of freedom $\mathcal{E}_a$.
\section{Additional Modes in Modified Hydrodynamics}
As previously mentioned, both second-order and anisotropic hydrodynamics introduce new non-hydrodynamic degrees of freedom. Hence one would generally expect the presence of additional collective modes and modification of existing hydrodynamic modes arising from the inclusion of these new degrees of freedom. Furthermore, from the form of Eq.~\eqref{eq:MIS} as a simple relaxation equation, one might expect new modes decaying on the timescale $\tau_\pi$. As will be seen in the next chapter this is indeed the case. These modes are termed ``non-hydrodynamic" modes as they are not present in the standard Navier-Stokes framework. Non-hydrodynamic modes along with their hydrodynamic counterparts are discussed in detail in various experimentally relevant circumstances in coming chapters. In particular, questions of how these modes alter the dynamical behavior of the fluid, as well as whether or not these modes should be experimentally observable will be addressed.

\chapter{Isotropically Trapped Fermi Gas in Second-Order Hydrodynamics}
\label{sohiso}
This chapter provides a study of collective modes of a harmonically trapped gas in second-order hydrodynamics (SOH) following very closely the presentation by the author of this thesis and P. Romatschke in Ref.~\cite{Lewis2017}. Particularly, the existence of a set of purely damped non-hydro modes as well as higher-than-quadrupole-order hydrodynamic modes whose dispersion is modified from the Navier-Stokes form are discussed. It is found that the damping of these higher-order modes exhibits increased dependence of the damping rate on shear viscosity. Thus these modes may be of use in extracting transport properties of strongly interacting Fermi gases. 

The results of this chapter are closely related to Ref.\cite{Rosi2015}, and in many aspects are complementary to the results therein. In this chapter, the focus is on effects arising from a non-vanishing shear viscosity. Consideration is limited to ideal equation of state. In contrast, Ref.~\cite{Rosi2015} considered collective modes for polytropic equations of state and zero shear viscosity.

The chapter begins by presenting the equations of second-order hydrodynamics along with the assumptions to be made in this chapter. The equations are then linearized for small deviation from equilibrium. Next, the configuration space expansion solution method for these equations is presented.Thee configuration space expansion is used to derive the frequencies, damping rates, and spatial structures for the collective modes of harmonically trapped gases in both two and three dimensions in Secs.~\ref{har2dmodes} and \ref{har3dmodes}. Finally, amplitudes for the breathing and quadrupole modes are calculated for experimentally relevant excitation conditions in Sec.~\ref{ampcalcs}.

\section{Solving Second-Order Hydrodynamics}\label{SOH}

\subsection{Equations of Motion}
 
The equations of second-order hydrodynamics for a fluid in $d$ spatial dimensions with a trapping force $\mathbf{F}$ are given by
\begin{align}
	\label{masscons}
	&\partial_t\rho + \partial_i(\rho u_i)=0,\\
	\label{momcons}
	&\partial_t(\rho u_i)+ \partial_j(\rho u_i u_j + P \delta_{ij} + \pi_{ij})=\rho F_i,\\
	\label{engcons}
	&\partial_t \epsilon + \partial_j\big[u_j\big(\epsilon + P\big)+\pi_{ij} u_i\big]=\rho F_k u_k,\\
	\label{stressten}
	 &\pi_{ij} + \tau_\pi \partial_t \pi_{ij} = -\eta \sigma_{ij},
\end{align}
where $\epsilon$ is the energy density, $\eta$ is the shear viscosity, and $\tau_\pi$ is the relaxation time for the stress tensor. In the above equations, $\epsilon$ and $\sigma_{ij}$ are specified in terms of the fluid velocity, mass density and pressure $P$ as 
\begin{align}
	\label{eng}
	&\epsilon=\frac{\big(\rho \mathbf{u}^2 + d P\big)}{2},\\
	\label{sig}
	&\sigma_{ij} = \big[\partial_i u_j + \partial_j u_i - \frac{2}{d} \delta_{ij} \partial_k u_k \big].
\end{align}
Note that Eq.~(\ref{eng}) corresponds to the equation of state for a scale-invariant system. It is easy see the familiar Navier-Stokes equations are recovered upon taking the limit $\tau_\pi\rightarrow 0$ in Eq.~(\ref{stressten}). 

\subsection{Assumptions}
For simplicity, the bulk viscosity and heat conductivity coefficients are assumed to vanish. The assumption of vanishing bulk viscosity is consistent with measurements in two dimensions \cite{TaylorPRL2012,VogtPRL2012}. Furthermore, calculations of bulk viscosity in $d=3$ imply that the value of bulk viscosity near unitarity in the high temperature limit should be small \cite{SchaferPRL2013}. This chapter will consider a Fermi gas in the normal phase, \textit{i.e.} above the superfluid transition temperature $T_c$. Thus taking the bulk viscosity to vanish should be a good approximation in the case $d=3$ as well. The assumption of vanishing thermal conductivity is justified as it is already a second-order gradient effect as discussed in Ref.~\cite{SchaferPRA2014}. Hence the gas is assumed to be isothermal, \textit{i.e.} the temperature is a function of time only and not of spatial coordinates. However, it is straightforward to see how the procedure below can be extended to treat a non-isothermal gas. Additionally, in order to obtain analytically tractable results, the gas is assumed to be described with an ideal equation of state:
\begin{equation}
	\label{eos}
	P = n T = \frac{\rho T}{m},
\end{equation}
where $n$ is the number density of particles (recall $\hbar=k_B=1$). The effects of a realistic non-ideal equation of state on collective mode behavior in a viscous fluid typically require numerical treatments such as those presented in Ref.~\cite{BrewerPRA2016}. 

Moreover, $\eta/P$ is assumed to be constant. While this assumption is not expected to hold in the low density corona, it will allow analytic access to the spatial structure, frequency and damping rates of collective modes. Numerical studies including temperature and density effects on the shear viscosity are left for future work. 

Finally, in order to access collective mode behavior of the gas, the equations of second-order hydrodynamics are linearized for small perturbations around a time independent equilibrium state characterized by $\rho_0({\bf x})$, $\mathbf{u}_0=0$, and $T_{0}$ which are solutions to Eqs.~\eqref{masscons}-\eqref{eos}. In this case, $\rho = \rho_0 (1+ \delta \rho)$, $\mathbf{u}= \delta \mathbf{u}$, and $T= T_{0}+\delta T$ with $\delta \rho, \delta \mathbf{u}, \delta T$ assumed to be small. Working in the frequency domain one has $\delta \rho(t,\mathbf{x}) = e^{-i \omega t} \delta \rho(\mathbf{x})$, with similar expressions holding for $\delta \mathbf{u}$ and $\delta T$. To simplify notation, from now on perturbations such as $\delta \rho$ denote quantities where the time dependence has been factored out, unless explicitly stated.

\subsection{Linearization}

Expanding Eqs.~\eqref{masscons}-\eqref{stressten} to linear order in perturbations and utilize Eqs.~\eqref{eng}-\eqref{eos} assuming constant $\eta/P$, results in
\begin{align}
	\label{linmasscons}
	&-i \omega \rho_0 \delta \rho + \partial_i(\rho_0 \delta u_i)=0,\\
	\label{linmomcons}
	 &-i \omega \rho_0 \delta u_i + \partial_j(\frac{(T_0 \delta \rho + \delta T \rho_0)}{m} \delta_{ij} + \delta \pi_{ij})-\delta\rho F_i=0,\\
	\label{linengcons}	
	&-i \omega \frac{d}{2} \frac{(T_0 \delta \rho + \delta T \rho_0)}{m} + \partial_j \big( \delta u_j \frac{T_0 \rho_0}{m} \big)-\rho_0 F_k \delta u_k=0,\\
	\label{linstressten}	
	&\big(1-i \tau_\pi \omega\big) \delta \pi_{ij}=-\frac{\eta}{P} \frac{T_0 \rho_0}{m}  \big[\partial_i \delta u_j + \partial_j \delta u_i - \frac{2}{d} \delta_{ij} \partial_k \delta u_k \big].
\end{align}
Eqs.~\eqref{linmasscons}-\eqref{linengcons} with Eq.~\eqref{linstressten} substituted into Eq.~\eqref{linmomcons} are referred to as the linearized second-order hydrodynamics equations. Interestingly, it may be shown that these equations exactly match those arising from linearizing the anisotropic hydrodynamics framework of Ref.~\cite{Bluhm2015}.

\subsection{Configuration Space Expansion}\label{confexp}

The solution for the equilibrium density configuration for a harmonic trapping potential with trapping frequency $\omega_\perp$ is given by $\rho_0(\mathbf{x}) =\rho_0 \exp{[{\frac{-(x^2+y^2)m \omega_\perp^2}{2 T_0}}]}$. For the rest of this work dimensionless units are employed such that distances are measured in units of $(T_0/(m \omega_\perp^2))^{1/2}$, times in units of $\omega_\perp^{-1}$, temperatures in units of $T_0$, and densities in units of $m^{d/2+1}\omega_\perp^d/T_0^{d/2}$. The equilibrium solution is then
\begin{align}
	\label{unitlessequil}
	\nonumber \rho_0(\mathbf{x}) =A_0 &e^{\frac{-(x^2+y^2)}{2 }}, \quad
	\mathbf{u}_0(\mathbf{x})=0,\quad T_0(\mathbf{x})= 1,
\end{align}
where $A_0$ is a dimensionless positive number setting the number of particles (cf. the discussion in App.~\ref{App3}).

In the absence of a trapping potential, a standard approach for obtaining collective modes is the application of a 
spatial Fourier transform of Eqs.~\eqref{linmasscons}-\eqref{linstressten} in order to obtain the shear and sound modes of the system.  However, a harmonic trapping potential (linear trapping force) breaks translation symmetry. Thus, it is more convenient to use a different basis in which to expand the perturbations. Here an expansion in tensor Hermite polynomials is undertaken, though any complete basis of linearly independent polynomials will suffice. The $N^{th}$ order tensor Hermite polynomials in $d$ spatial dimensions may be obtained with the Rodrigues formula\cite{Coelho2014}
	\begin{equation}
		\label{Rodrigues}
		\text{H}^{\phantom{...}(N)}_{i_1 i_2 ... i_N}(\mathbf{x}) = \frac{(-1)^N}{g(\mathbf{x})}\frac{\partial^N g(\mathbf{x})}{\partial x^{i_1}\partial x^{i_2}...\partial x^{i_N}},\quad g(\mathbf{x}) = \frac{1}{(2 \pi)^{\frac{d}{2}}}e^{\frac{-\mathbf{x}^2}{2}},
	\end{equation}
where $i_k \in \{1,2,...d\}$ for $k=\{1,2,...,N\}$.
The tensor Hermite polynomials are orthogonal with respect to a Gaussian weight function which makes them particularly useful for the case of a harmonic trapping potential since in this case the equilibrium solution is itself Gaussian. In particular, the tensor Hermite polynomials satisfy the orthogonality relations
	\begin{equation}
		\int d^d\mathbf{x} \phantom{.}g(\mathbf{x}) \phantom{.}\text{H}^{\phantom{...}(N)}_{i_1 i_2 ... i_N}(\mathbf{x}) \phantom{.}\text{H}^{\phantom{...}(M)}_{j_1 j_2 ... j_M}(\mathbf{x}) = \delta^{NM} (\delta^{i_1j_1}\delta^{i_2j_2}...\delta^{i_Nj_N}+\text{all permutations of $j$'s}).
	\end{equation}
In this work, only cigar shaped traps in $d=3$ are considered. Hence to a good approximation, translational invariance along the z-axis in $d=3$ spatial dimensions may be assumed. In this case, the expansion in both $d=2$ and $d=3$ will involve only the tensor Hermite polynomials for $d=2$. Recalling that the gas is taken to be isothermal, the polynomial expansion of perturbations is
\begin{align}
	\label{perts}
	\nonumber &\delta \rho(\mathbf{x}) =\sum_{M=0}^N \sum_{j=1}^{M+1} a^{(M)}_{\mathbf{m}_j}\text{H}^{(M)}_{\mathbf{m}_j}(\mathbf{x}),\\
	&\delta \mathbf{u}(\mathbf{x})=\sum_{M=0}^N \sum_{j=1}^{M+1} \mathbf{b}^{(M)}_{\mathbf{m}_j}\text{H}^{(M)}_{\mathbf{m}_j}(\mathbf{x}),\\
	\nonumber &\delta T(\mathbf{x})= c,
\end{align}
where, in the sum over $j$, $\mathbf{m}_j$ is understood to run over all combinations of indices unique up to permutations. For example, if $M=2$ the second sum runs over $\mathbf{m}=\{(1,1),(1,2),(2,2)\}$, while $(2,1)$ is excluded. The reason for this restriction is that $\text{H}^{(M)}_{\mathbf{m}}(\mathbf{x})$ is fully symmetric in the indices as can be seen from Eq.~\eqref{Rodrigues}. One should also note that in Eqs.~\eqref{perts}, $\mathbf{b}^{(M)}_{\mathbf{m}}$ is used as shorthand for the polynomial coefficients of all components of $\delta \mathbf{u}$, and for a given $M$ and $\mathbf{m}$ is a column vector with $d$ components. 

Now details of accessing the collective modes whose spatial structure is associated with polynomials of low degree (``low-lying modes'') will be discussed. Substituting Eqs.~\eqref{perts} truncated at polynomial order $N$ into the linearized second-order hydrodynamics equations and taking projections onto different tensor Hermite polynomials of order $K\leq N$ one obtains a matrix equation for the polynomial coefficients in Eqs.~\eqref{perts}. The (complex) collective mode frequencies $\tilde \omega$ are then obtained from requiring a non-trivial null-space of this matrix, and subsequently the spatial structures are obtained from the corresponding null-vectors.

\section{Collective Mode Solutions in $d=2$}\label{har2dmodes}
\begin{figure*}[ht!]
\includegraphics[width=\textwidth]{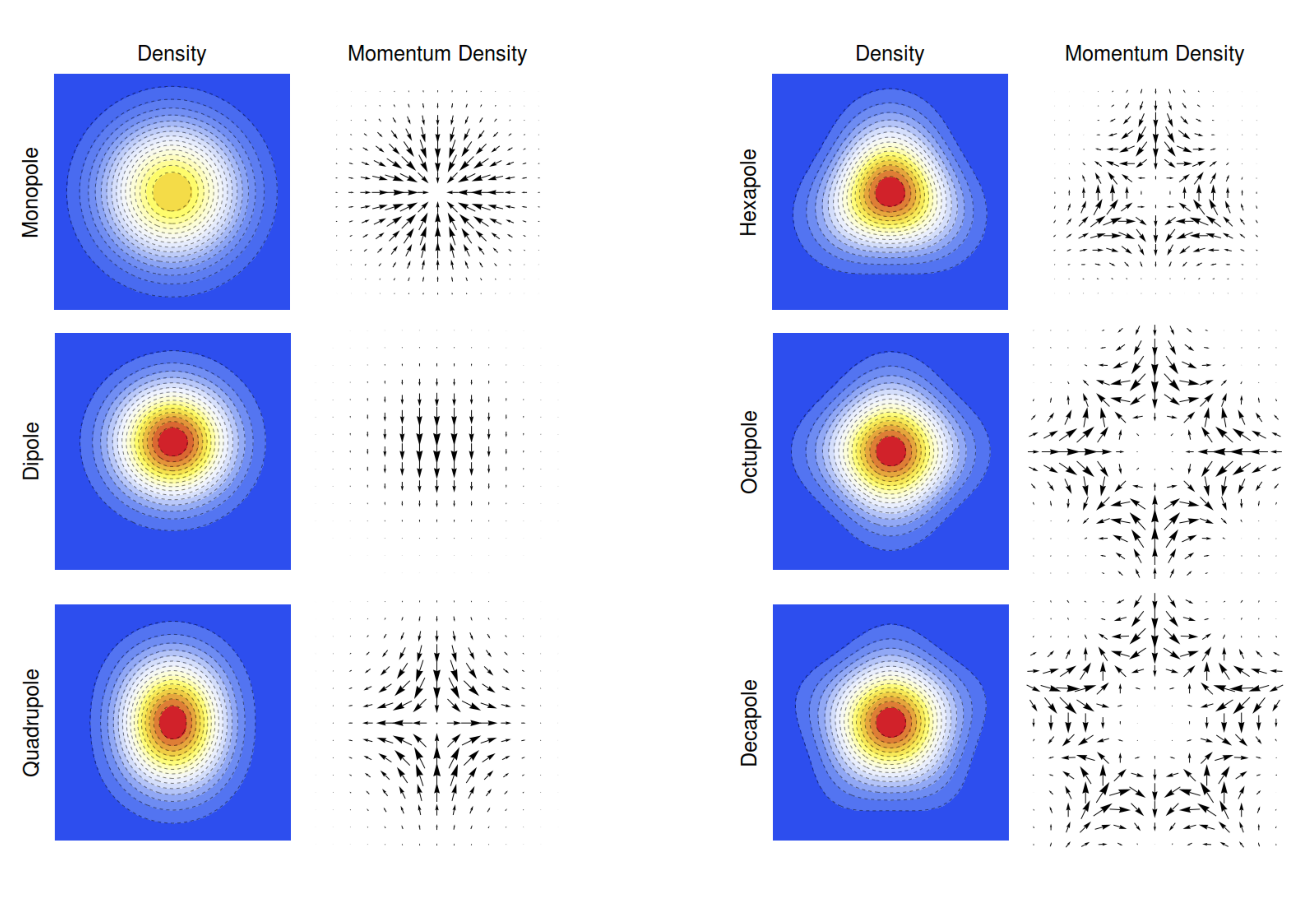}
  \caption{Time snap shots of density profiles and subsequent momentum density ($\rho\mathbf{u}$) for the oscillatory modes in $d=2$. Note that the center of the monopole mode is at a lower density than the centers of the other modes since it is volume changing and has a larger radius than the equilibrium configuration. The damping rate of higher-order modes is more sensitive to $\eta/P$ as discussed in the text. Also note that non-hydrodynamic modes share the same spatial structure as their hydrodynamic counterpart.}
  \label{fig:key}
\end{figure*}

Results for the density and velocity of low-lying collective modes in $d=2$ are shown in Fig.~\ref{fig:key}. In particular, there is a breathing (monopole) mode which corresponds to a cylindrically symmetric oscillatory change in cloud volume, a sloshing (or dipole) mode where the center of mass of the cloud oscillates about the trap center, a quadrupole mode which is elliptical in shape, and higher-order modes corresponding to higher-order geometric shapes. Note that the spatial structure of these collective modes are similar to those reported in Ref.~\cite{Rosi2015}. 
More detailed information about the $d=2$ collective modes can be found in App.~\ref{App1}.

\begin{table}[h!]
\begin{center}
\begin{tabular}{| c | c | c |}
  \hline
  \phantom{...}&$\omega$ & $\Gamma$\\
  \hline			
  Number (Zero Mode) & $0$ & $0$ \\
  \hline			
  Temperature (Zero Mode) & $0$ & $0$\\
  \hline			
  Rotation (Zero Mode) & $0$ & $0$\\
  \hline			
  Breathing (Monopole) & 2 & 0 \\
  \hline
  Sloshing (Dipole) & 1 & 0 \\
  \hline
  Quadrupole & $\sqrt{2}$ & $\frac{\eta}{P}$ \\
  \hline
  Hexapole & $\sqrt{3}$ & $2\frac{\eta}{P}$  \\
  \hline
  Octupole & 2 & $3\frac{\eta}{P}$  \\
  \hline
  Decapole & $\sqrt{5}$ & $4\frac{\eta}{P}$  \\
  \hline  
  Non-hydrodynamic Quadrupole& 0 & $\frac{1}{\tau_\pi} - 2\frac{\eta}{P}$ \\
    \hline  
  Non-hydrodynamic Hexapole & 0 & $\frac{1}{\tau_\pi } - 4\frac{\eta}{P}$ \\
    \hline  
  Non-hydrodynamic Octupole & 0 & $\frac{1}{\tau_\pi } - 6\frac{\eta}{P}$ \\
  \hline  
  Non-hydrodynamic Decapole & 0 & $\frac{1}{\tau_\pi } - 8\frac{\eta}{P}$ \\
  \hline
\end{tabular}
\end{center}
  \caption{Frequencies and damping rates in $d=2$ from linearized second-order hydrodynamics assuming $\frac{\eta}{P},\tau_\pi\ll 1$. The hydrodynamic mode damping rates depend on $\eta/P$ times a prefactor which increases with mode order. Note that for $d=2$ there is no non-hydrodynamic sloshing or breathing mode.  }
\label{2dfreqtab}
\end{table}

The collective mode frequencies $\omega$ and damping rates $\Gamma$ are given as the real and imaginary parts of roots of polynomials, which generally do not admit simple closed form expressions. Hence, in Tab.~\ref{2dfreqtab} expressions for the complex frequencies and spatial mode structure from second-order hydrodynamics for the low-lying modes in the hydrodynamic limit $\eta/P \ll 1$ and $\tau_\pi \ll 1$ (assuming that $\tau_\pi$ and $\eta/P$ are of the same order of magnitude) are reported. In this case simple analytic expressions may be obtained. In addition to the modes shown in Fig.~\ref{fig:key} there are three modes in Tab.~\ref{2dfreqtab} which have zero complex frequency. The first corresponds to a change in total particle number, the second corresponds to a change in temperature and width of the cloud, and the third ``zero mode" is simply a rotation of the fluid about the central axis. While they are required for the mode amplitude analysis (see Sec.~\ref{ampcalcs}), the role of the first two of the zero frequency modes is relatively uninteresting. Hence, detailed discussion of these modes is relegated to App.~\ref{App3}.

\begin{figure*}
  \includegraphics[width=\textwidth]{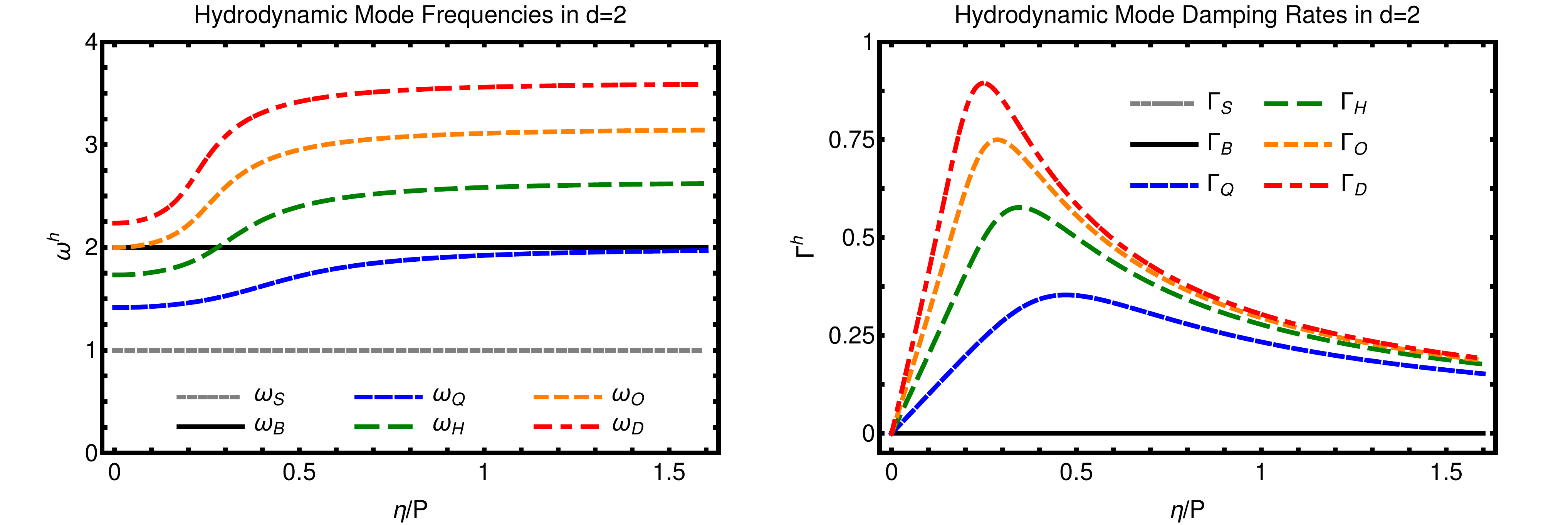}%
  \caption{Two-dimensional hydrodynamic collective mode frequencies $\omega$ (left panel) and damping rates $\Gamma$ (right panel) as a function of $\eta/P$. Subscripts denote mode name (monopole ``B''; dipole ``S''; quadrupole ``Q''; hexapole ``H''; octupole ``O''; decapole ``D''). For the purpose of the figures the kinetic theory relation $\tau_\pi=\eta/P$ has been used.
}
  \label{fig:2dfreq}
\end{figure*}

The rows of Tab.~\ref{2dfreqtab} starting with the number mode and ending with the decapole mode are all hydrodynamic modes. Note that at order $\mathcal{O}(\eta/P)$ the results for these modes match those from an analysis of the mode frequencies with the Navier-Stokes equations at the same order. However, for values of $\eta/P$ where corrections to the hydrodynamic limit become significant, the frequencies found from the Navier-Stokes equations and second-order hydrodynamics disagree. Fig.~\ref{fig:2dfreq} shows the full dependence of the hydrodynamic mode frequencies and damping rates on $\eta/P$ (assuming $\tau_\pi=\eta/P$ based on kinetic theory \cite{Bruun2007,SchaferPRA2014,Kikuchi2016}). Note that the result of second-order hydrodynamics for the quadrupole mode exactly matches the result from kinetic theory when setting $\tau_\pi=\tau_R=\eta/P$ \cite{Baur2013,BrewerPRA2016}.

Furthermore,  results shown in Tab.~\ref{2dfreqtab} demonstrate that the hydrodynamic mode damping rates depend on $\eta/P$ times a prefactor which increases with mode order. This is completely analogous to what has been observed in experiments on relativistic ion collisions, where simultaneous measurement of multiple modes has been used to obtain strong constraints on the value of $\eta/s$ \cite{Schenke2012}. While it appears that higher-order modes have not yet been studied experimentally, it is conceivable that measuring their damping rates could lead to a similarly strong experimental constraint on shear viscosity in the unitary Fermi gas. This approach does not appear to have been suggested elsewhere in the literature. When aiming to use higher-order modes to analyze shear viscosity in Fermi gases one should recall that the present analysis is based on a linear treatment. Quantitative analysis of higher-order flows will, however, require the inclusion of nonlinear effects, especially for analysis of flows beyond hexapolar order due to mode mixing. For this reason, the hexapolar mode is suggested as a prime candidate for the use of higher-order modes to extract shear viscosity. 

\begin{figure}[htbp]
 \begin{center}
      \includegraphics[width=0.5\textwidth]{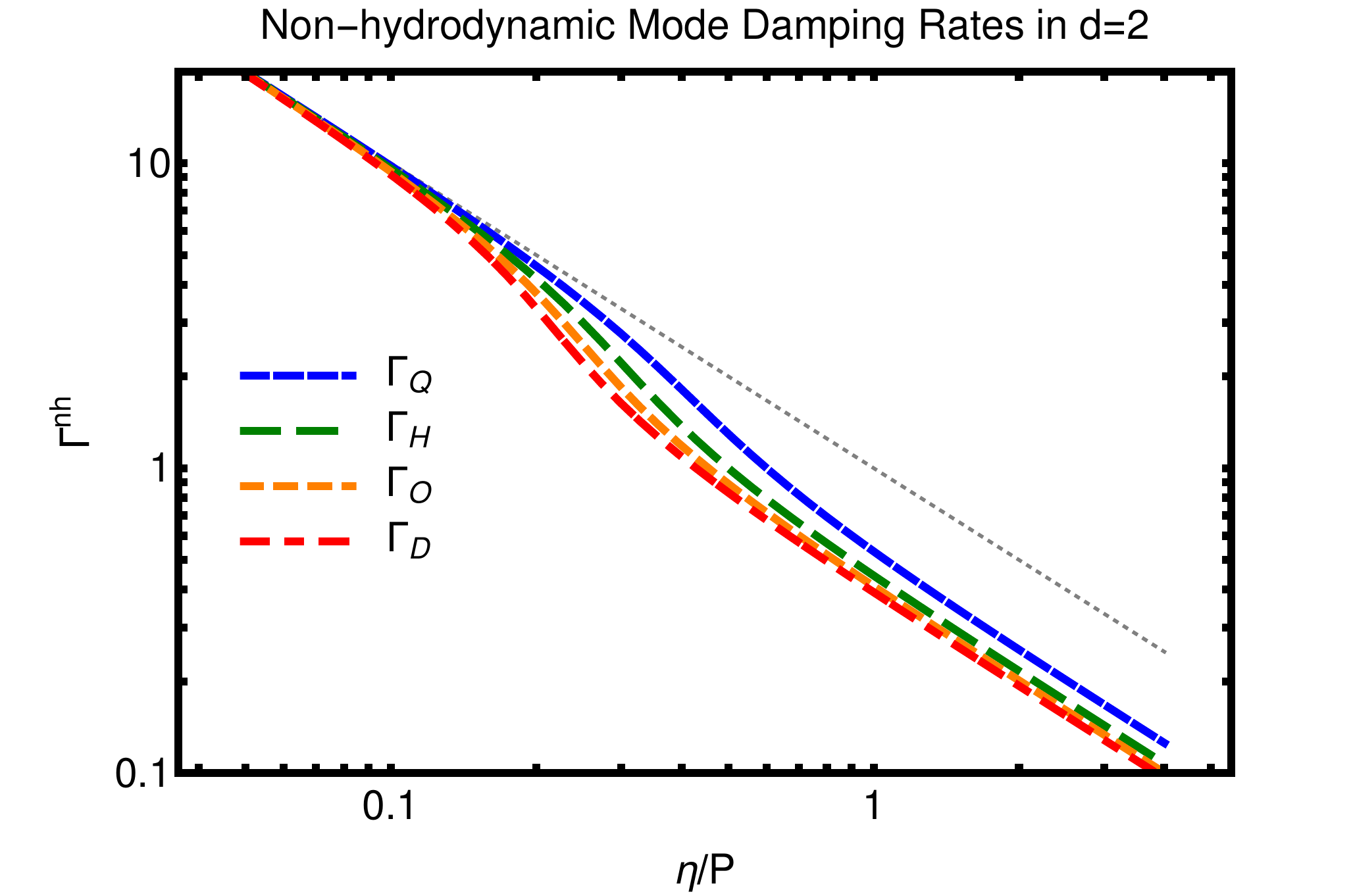}
          \caption{Two-dimensional non-hydrodynamic collective mode damping rates $\Gamma$ as a function of $\eta/P$ (using $\tau_\pi=\eta/P$). Subscripts denote mode name (quadrupole ``Q''; hexapole ``H''; octupole ``O''; decapole ``D''). Dotted line is $\Gamma^{nh}= 1/\tau_\pi$ (cf. Ref.\cite{BrewerPRL2015}). Note that in $d=2$ there is no non-hydrodynamic sloshing or breathing mode.}
      \label{2dnhdamp}
      \end{center}
\end{figure}

Finally, Tab.~\ref{2dfreqtab} also indicates the presence of non-hydrodynamic modes. The physics of non-hydrodynamic modes is largely unexplored (cf. Refs.~\cite{BrewerPRL2015,Bantilan:2016qos} for a brief discussion of the topic in the context of cold quantum gases). Results shown in Tab.~\ref{2dfreqtab} imply that several such non-hydrodynamic modes exist, all of which are purely damped in second-order hydrodynamics. The non-hydrodynamic mode damping rates are sensitive to $\tau_{\pi}$ and $\eta/P$. Thus the value of $\tau_\pi$ could be extracted experimentally by measuring any of the non-hydrodynamic mode damping rates in combination with a hydrodynamic mode damping rate required to determine $\frac{\eta}{P}$. In Fig.~\ref{2dnhdamp}, non-hydrodynamic damping rates are shown as a function of $\eta/P$ when setting $\tau_\pi=\eta/P$.

\section{Collective Mode Solutions in $d=3$}\label{har3dmodes}

In the case of a three-dimensional gas in a harmonic trap with trapping frequencies $\omega_z\ll\omega_x=\omega_y$, the resulting gas cloud takes on an elongated cigar shaped geometry. For $\omega_z=0$, the configuration space expansion Eqs.~\eqref{perts} can be applied because there is no dependence on the coordinate $z$ if translational invariance of the system along the $z$-axis is assumed. In this case, the collective mode structures in $d=3$ are qualitatively similar to those obtained in the two-dimensional case, cf. the discussion in Refs.~\cite{KinastPRL2005,AltmeyerPRA2007}.

\begin{figure*}[htbp]
      \includegraphics[width=1\textwidth]{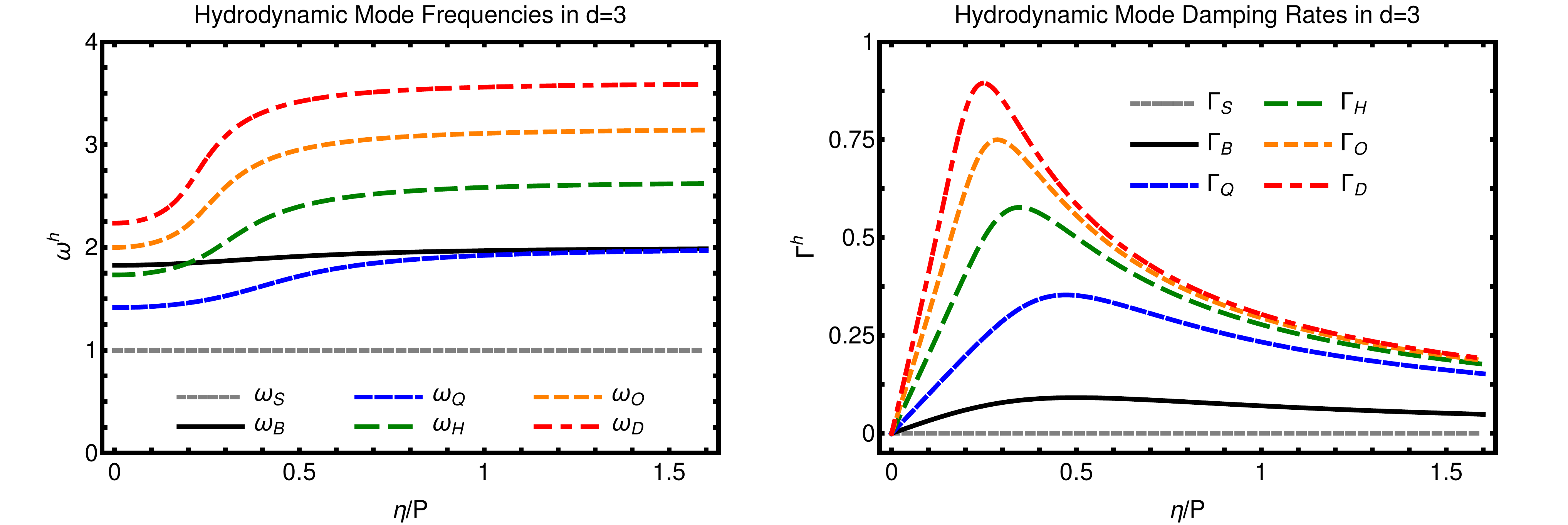}
      \caption{Three-dimensional hydrodynamic collective mode frequencies $\omega$ (left panel) and damping rates $\Gamma$ (right panel) as a function of $\eta/P$. Subscripts denote mode name (monopole ``B''; dipole ``S''; quadrupole ``Q''; hexapole ``H''; octupole ``O''; decapole ``D''). For the purpose of the figures the kinetic theory relation $\tau_\pi=\eta/P$ has been used.}
      \label{3dfreq}
\end{figure*}

\begin{figure}[htbp]
 \begin{center}
     \includegraphics[width=0.5\textwidth]{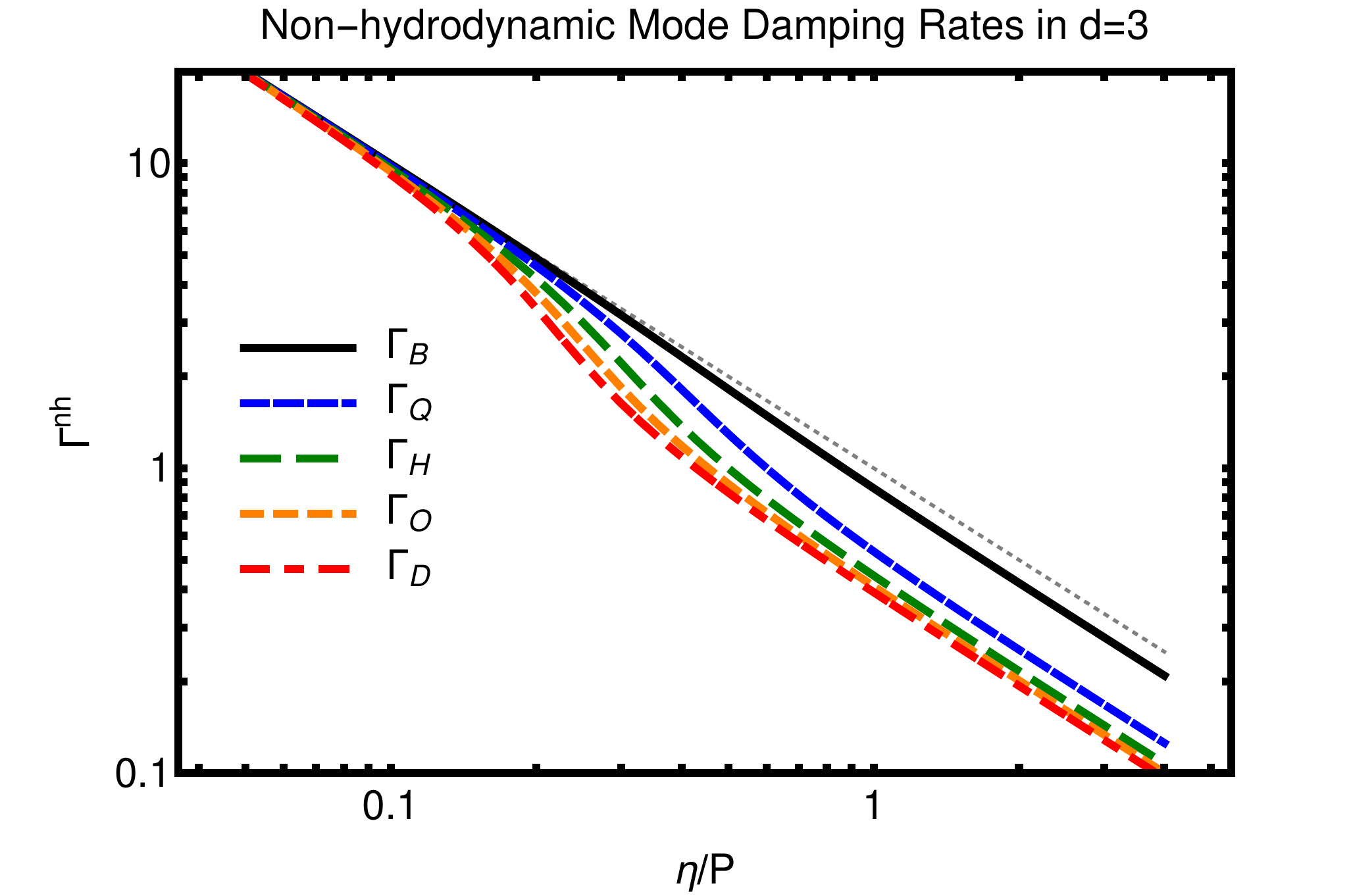}
    \caption{Three-dimensional non-hydrodynamic collective mode damping rates $\Gamma$ as a function of $\eta/P$ (using $\tau_\pi=\eta/P$). Subscripts denote mode name (monopole ``B''; quadrupole ``Q''; hexapole ``H''; octupole ``O''; decapole ``D''). Dotted line is $\Gamma^{nh}= 1/\tau_\pi$ (cf. Ref.\cite{BrewerPRL2015}).  Note that in $d=2$ there is no non-hydrodynamic sloshing mode, but there is a non-hydrodynamic breathing mode.}
      \label{3dnhdamp}
      \end{center}
\end{figure}

Results for the low-lying modes in the limit $\eta/P,\tau_\pi\ll 1$ are reported in Tab.~\ref{3dfreqtab} whereas the full dependence of frequencies and damping rates on $\frac{\eta}{P}$ is shown in Figs.~\ref{3dfreq},\ref{3dnhdamp} for the case  $\tau_\pi=\eta/P$.
The only qualitative difference with respect to the $d=2$ case is that the breathing mode in $d=3$ has a different frequency, a non-zero damping rate, and there is now a non-hydrodynamic breathing mode. See App.~\ref{App2} for more details about the spatial structure of the d=3 collective modes.
\begin{table}
\begin{center}
\begin{tabular}{| c | c | c |}
  \hline
  \phantom{...}&$\omega$ & $\Gamma$\\
  \hline			
  Temperature (Zero Mode) & $0$ & $0$\\
  \hline			
  Number (Zero Mode) & $0$ & $0$ \\
  \hline			
  Rotation (Zero Mode) & $0$ & $0$\\
  \hline			
  Breathing (Monopole) & $\sqrt{\frac{10}{3}}$  & $\frac{\eta}{3P}$ \\
  \hline
  Sloshing (Dipole) & 1 & 0 \\
  \hline
  Quadrupole & $\sqrt{2}$ & $\frac{\eta}{P}$ \\
  \hline
  Hexapole & $\sqrt{3}$ & $2\frac{\eta}{P}$  \\
  \hline
  Octupole & 2 & $3\frac{\eta}{P}$  \\
  \hline
  Decapole & $\sqrt{5}$ & $4\frac{\eta}{P}$  \\
   \hline  
  Non-hydrodynamic Breathing& 0 & $\frac{1}{\tau_\pi} - \frac{2\eta}{3P}$ \\
  \hline  
  Non-hydrodynamic Quadrupole& 0 & $\frac{1}{\tau_\pi} - 2\frac{\eta}{P}$ \\
    \hline  
  Non-hydrodynamic Hexapole & 0 & $\frac{1}{\tau_\pi} - 4\frac{\eta}{P}$ \\
    \hline  
  Non-hydrodynamic Octupole & 0 & $\frac{1}{\tau_\pi} - 6\frac{\eta}{P}$ \\
  \hline  
  Non-hydrodynamic Decapole & 0 & $\frac{1}{\tau_\pi} - 8\frac{\eta}{P}$ \\
  \hline
\end{tabular}
\end{center}
  \caption{Frequencies and damping rates in $d=3$ from linearized second-order hydrodynamics assuming $\frac{\eta}{P},\tau_\pi\ll 1$. The hydrodynamic mode damping rates depend on $\eta/P$ times a prefactor with increases with mode order. Note that there is no non-hydrodynamic sloshing mode, but, unlike for $d=2$, there is a non-hydrodynamic breathing mode for $d=3$.}
\label{3dfreqtab}
\end{table}

It should be pointed out that, while second-order hydrodynamics predicts purely damped non-hydrodynamic modes for both $d=2,3$, more general (string-theory-based, c.f. Chap.~\ref{fgd}) calculations suggest that there should be a non-vanishing frequency component in the case of $d=3$ \cite{Bantilan:2016qos}. It would be interesting to measure non-hydrodynamic mode frequencies and damping rates in order to describe transport beyond Navier-Stokes on a quantitative level.

\section{Mode Amplitudes Calculations}\label{ampcalcs}

In this section, experimentally relevant scenarios to excite the collective modes of the previous sections are discussed. The corresponding mode amplitudes are calculated. For simplicity, it is assumed that $\tau_\pi=\eta/P$ in the following. In particular, the excitation amplitudes of the non-hydrodynamic quadrupole (in $d=2,3$) and non-hydrodynamic breathing (in $d=3$) modes are considered, leaving a study of higher-order modes for future work. For simplicity, only simple trap quenches (rapid changes in trap configuration) are considered. For the analysis, the cloud is taken to start in an equilibrium configuration of a (possibly biaxial, i.e. $\omega_{x,\, init} \neq \omega_{y,\, init}$) harmonic trap. At some initial time, a rapid quench will bring the trap configuration into a final harmonic form, which is assumed to be isotropic in the x-y plane with trapping frequency $\omega_{x,\, final}=\omega_{y,\, final}=1$ in our units. 

In the case of Navier-Stokes equations, initial conditions are fully specified through the initial density $\rho_{\,init}$, velocity $\mathbf{u}_{init}$, and temperature $T_{\,init}$ or appropriate time derivatives of such quantities. However,  second-order hydrodynamics treats the stress tensor $\pi_{ij}$ as a hydrodynamic variable, so, in addition, an initial condition $\pi_{ij,\, init}$ or its time derivative needs to be specified.

For equilibrium initial conditions of a general biaxial harmonic trap with trapping force given by $\mathbf{F} = -\gamma_x \mathbf{x} - \gamma_y \mathbf{y}$ one obtains
	\begin{align}
		\rho_{init}(\mathbf{x}) = A_{i} &e^{\frac{-(\sigma_x x^2 + \sigma_y y^2)}{2 T_{\, init}}},\\
		\mathbf{u}_{init}&=0,\\
		\pi_{ij,\, init}&=0,
	\end{align} 
where $T_{\,init}$ also needs to be specified. Initial equilibrium implies the condition $\gamma_i = \omega_{i,\, init}^2 = \sigma_i /T_{\, init}$ so that the cloud width is fully specified once $\gamma_i$ for $i=x,y$ and $T_{\, init}$ are fixed. Notice that the assumption of equilibrium in the initial trap leads to the condition $\pi_{ij,\, init}=0$. The mode amplitudes can then be obtained by projecting initial conditions onto the collective modes found in the preceding sections (see App.~\ref{App3} for details of the calculation).   

\subsection{Isotropic Trap Quench in $d=2$}\label{2diso}

\begin{figure*}[htbp]
    \begin{center}
      \includegraphics[width=\textwidth]{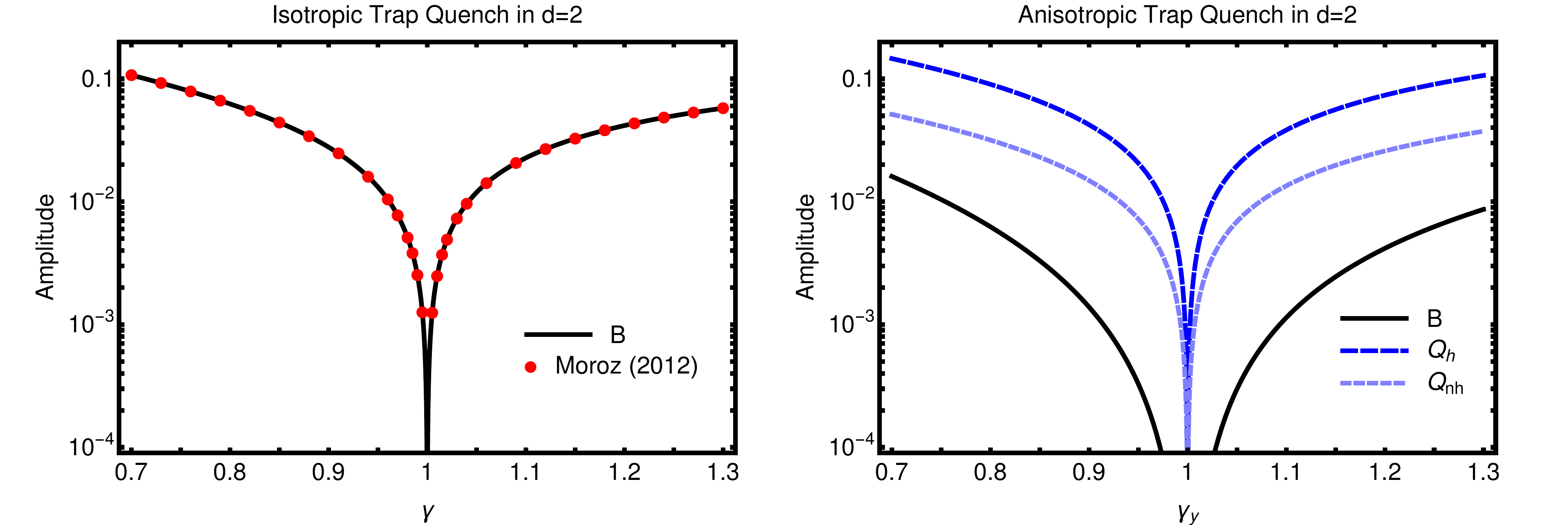}
      \caption{Left: Absolute value of the (dimensionless) breathing (``B") mode amplitude as a function the quench strength parameter $\gamma$ for an isotropic trap quench in d=2. Results are independent of $\frac{\eta}{P}$. The mode amplitude agrees perfectly with the result from an exact quantum mechanical scaling solution (``Moroz 2012") derived in Ref. \cite{Sergej2012}. Right: Absolute value of the (dimensionless) breathing (``B"),  hydrodynamic (``Q$_h$") and non-hydrodynamic (``Q$_{nh}$") quadrupole mode amplitudes as a function the quench strength parameter $\gamma_y$ for an anisotropic trap quench in d=2. Results shown are for $\frac{\eta}{P}=0.5$. Note that the temperature mode amplitude (not shown) matches the breathing mode amplitude for both the isotropic and anisotropic trap quench in d=2.}
      \label{fig:2disoamp}
    \end{center}
\end{figure*}

First the case of an isotropic trap quench $\gamma_x=\gamma_y\equiv \gamma$ in $d=2$ is considered. For simplicity it is assumed that $A_{i}/A_0=1$ and $T_{\, init}=1$. Although this case does not exhibit non-hydrodynamic or higher-order collective mode excitation, it does allow for a direct comparison to results from the literature for the breathing mode excitation amplitude. This type of initial condition corresponds to a rotationally symmetric trap quench with no initial fluid angular momentum. Symmetry then implies that only the number, temperature, and breathing modes can be excited (cf. Tab.~\ref{2dfreqtab}), and the initial amplitude for these modes are readily calculated. Fig.~\ref{fig:2disoamp} displays the (dimensionless) breathing mode amplitude as a function of the quench strength $\gamma$. (Note that the amplitude of the temperature mode is identical to the breathing mode amplitude in this case.) The number mode is not excited since the number of atoms taken in the initial condition match the number of atoms assumed in the final trap equilibrium ($A_{i}/A_0=1$).
The amplitude of the breathing mode for the isotropic trap quench is compared to the results from an exact quantum mechanical scaling solution by Moroz \cite{Sergej2012} in the left panel of Fig.~\ref{fig:2disoamp}. As can be seen from this figure, there is exact agreement between the calculations for all strength values $\gamma$.
Note that the amplitudes in this case are independent of $\eta/P$ since for $d=2$, the breathing mode does not couple to the shear stress tensor $\pi_{ij}$.

\subsection{Anisotropic Trap Quench in $d=2$}

A similar analysis to that above is performed for the case $A_i/A_0=1$, and $T_{\, init}=1$, but now taking $\gamma_x\gamma_y=1$, which corresponds to an anisotropic trap quench. The mode amplitudes in this case depend on the value of $\eta/P$.  In this case, the temperature, breathing and quadrupole modes are excited. The right panel of Fig.~\ref{fig:2disoamp} shows the absolute value of the mode amplitudes for the hydrodynamic breathing and quadrupole modes, as well as the non-hydrodynamic quadrupole mode as a function of the quench strength $\gamma_y$. Not surprisingly, Fig.~\ref{fig:2disoamp} shows that the anisotropic trap quench gives rise to a considerably larger quadrupole mode amplitude (both hydrodynamic and non-hydrodynamic) than the amplitude of the breathing mode.

\begin{figure}[htbp]
      \includegraphics[width=\textwidth]{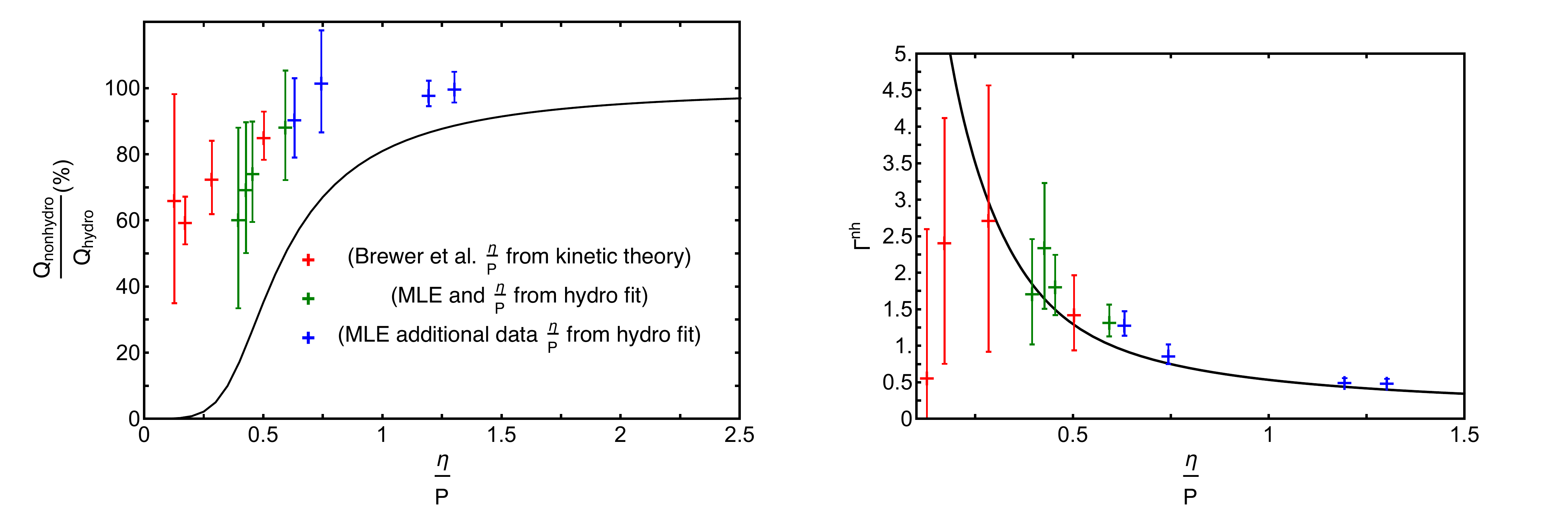}
      \caption{Comparison of second-order hydro theory (lines) and experimental (red: ``Brewer et al. 2015" using Eq.~\eqref{pauletatokfa} for $\eta/P$, green: reanalysis of data in ``Brewer et al. 2015" using Eq.~\eqref{hydrofit} and maximum likelihood estimation, blue: new analysis of data from Ref. \cite{VogtPRL2012} using Eq.~\eqref{hydrofit} and maximum likelihood estimation) results as a function of $\eta/P$ for an anisotropic trap quench (results independent of $\gamma$). Left: ratio of non-hydro to hydro quadrupole mode amplitudes. Error bars for blue and green data points are from the $1\sigma$ confidence interval for the fit parameters. This allowed for analysis of additional data (shown in blue) not studied in Ref.~\cite{BrewerPRL2015}. Right:  non-hydrodynamic quadrupole mode damping rate $\Gamma$. }
      \label{fig:anisovseta}
\end{figure}

For a potential experimental observation of the non-hydrodynamic quadrupole mode, it is interesting to consider the relative amplitude of this non-hydrodynamic mode to the (readily observable) hydrodynamic quadrupole mode. The (absolute) amplitude ratio calculated using the above anisotropic trap quench initial condition is plotted in Fig.~\ref{fig:anisovseta} as a function of $\eta/P$. One finds that the non-hydrodynamic mode amplitude is monotonically increasing as a function of $\eta/P$. This is plausible given that for small viscosities one expects the hydrodynamic mode to be dominant, whereas one expects the non-hydrodynamic mode to dominate in the ballistic $\eta/P\rightarrow \infty$ limit. 

The present calculation is compared to mode amplitude ratios extracted from experimental data \cite{VogtPRL2012} in Ref.~\cite{BrewerPRL2015}. To compare non-hydrodynamic damping rate data and theory, the procedure used in Ref.~\cite{BrewerPRL2015} of employing the approximate kinetic theory relation
\begin{equation}
	\label{pauletatokfa}
 \frac{\eta}{P} \approx K[1+\frac{4 \ln^2(k_Fa)}{\pi^2}],
\end{equation}
where $K\approx 0.12$ in order to relate the experimentally determined $k_F a$ to $\eta/P$ is followed (see discussion in Refs.~\cite{BrewerPRL2015,BrewerPRA2016} and references therein for more details on this relation). In order to get a sense of the possible errors associated with this choice, an alternate approach for extracting $\eta/P$ motivated by the discussion in Chap.~\ref{hydrointro} and the results of this chapter is used. Namely, by fitting the experimental data to the form
\begin{equation}
	\label{fitfn}
	Q(t) = A e^ {-\Gamma_H t} \cos{(\omega_H t + \phi)} + B e^{-\Gamma_{NH}t } + C,
\end{equation}
the difference between extracted values for the hydrodynamic frequency and damping ($\omega_H$ and $\Gamma_H$) and the theoretical value may be minimized by tuning $\eta/P$. That is $\eta/P$ is the value which minimizes the loss function
\begin{equation}
\label{hydrofit}
 \min_{\eta/P \in \big[0,\infty \big)} \bigg[(\omega^{exp}_H-\omega_H(\eta/P))^2 + (\Gamma^{exp}_H-\Gamma_H(\eta/P))^2 \bigg].
\end{equation}
Using these procedures, one observes qualitative agreement of the amplitude ratios between calculation and experimental data in Fig.~\ref{fig:anisovseta} (left panel). In addition, one can extract the non-hydrodynamic quadrupole mode damping rate, finding reasonable agreement with second-order hydrodynamics especially when using Eq.~\eqref{hydrofit} for the extraction of $\eta/P$ (cf. right panel of Fig.~\ref{fig:anisovseta}).

\begin{figure}[htbp]
      \includegraphics[width=\textwidth]{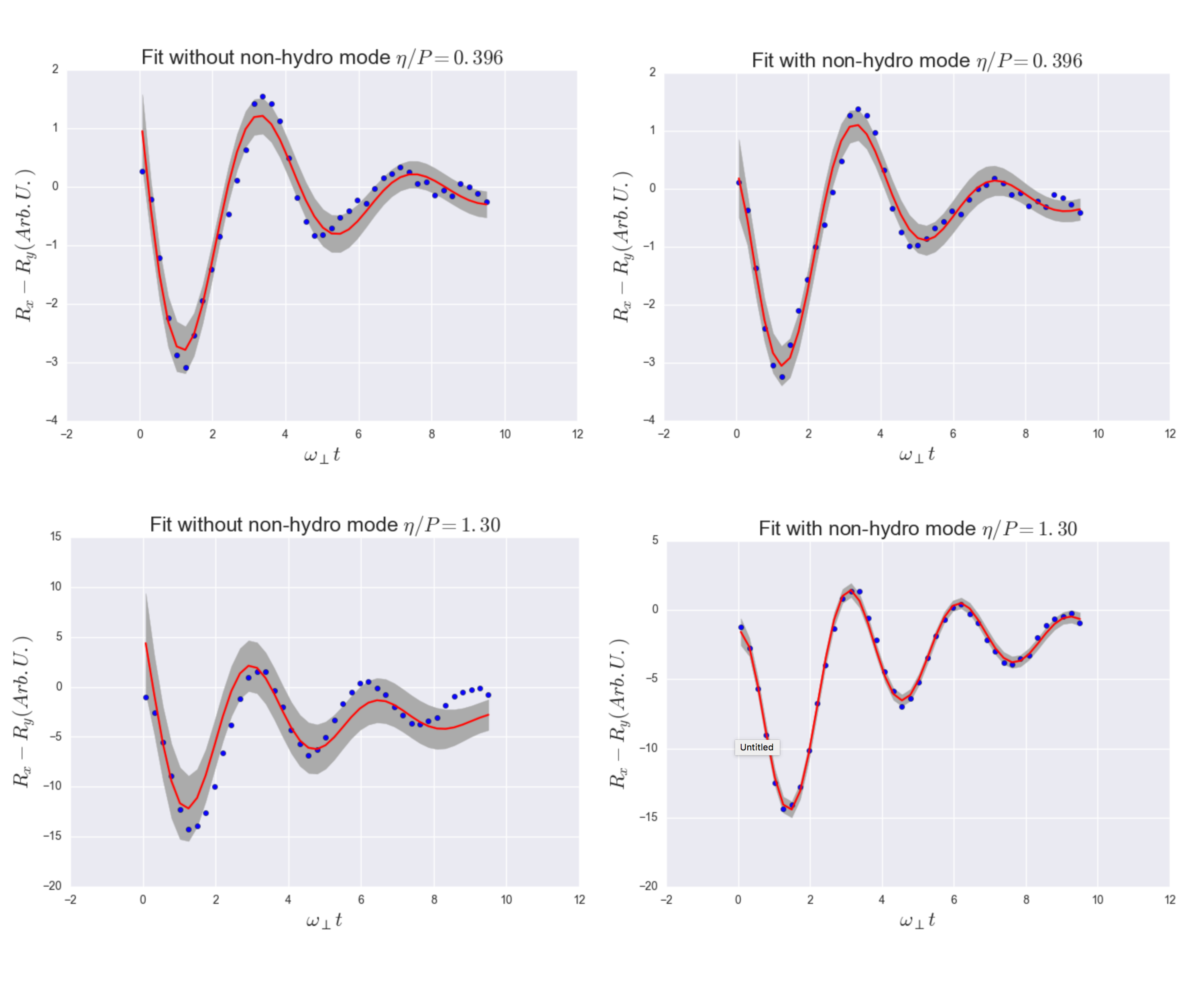}
      \caption{Top row: Fits without and with the non-hydrodynamic mode of experimental data from Ref.~\cite{VogtPRL2012} (blue points) with $\eta/P\approx 0.396$ (from Eq.~\eqref{hydrofit}). The best fit is shown in red, and the grey band is the $3\sigma$ prediction interval. Bottom row: Fits without and with the non-hydrodynamic mode of experimental data from Ref.~\cite{VogtPRL2012} (blue points) with $\eta/P\approx 1.30$ (from Eq.~\eqref{hydrofit}). The best fit is shown in red, and the grey band is the $3\sigma$ prediction interval. Notice the significantly improved performance of the fit when including the non-hydro term. The fit without the non-hydrodynamic term may be significantly improved by leaving out the early time data where the contribution from the non-hydrodynamic mode is significant.}
      \label{fig:fits}
\end{figure}

There are several possible reasons why a quantitative agreement between second-order hydro and experiment in Fig.~\ref{fig:anisovseta} should not be expected. For instance, the present theory calculations neglect the presence of a pseudogap phase
and pairing correlations (see e.g. Refs.~\cite{Baur2013,EnssPRL2014,Feld2011,Guo2011,SchaferPRA2012} on this topic). Furthermore, it is likely that the quantitative disagreement in Fig.~\ref{fig:anisovseta} is at least in part due to the assumptions discussed in Sec.~\ref{SOH}, such as small perturbations, constant $\eta/P$ and ideal equation of state. In particular, for the strongly interacting quasi-two-dimensional Fermi gas near the pseudogap temperature $T^*$, significant modification of the equation of state has been predicted and observed, cf. Ref.~\cite{EnssPRL2014}. Studies aiming for achieving a quantitative agreement most likely will have to rely on full numerical solutions, such as e.g. those discussed in Refs.~\cite{BrewerPRA2016,Bluhm2016}, which is left for future work. In addition, the framework presented here only admits a single non-hydrodynamic mode for each of the collective modes. While this may be appropriate in the kinetic theory regime, other approaches such as that of Ref. \cite{Bantilan:2016qos} (discussed in Chap.~\ref{fgd}) indicate that the model may be too simple to capture quantitative features of early time dynamics. Finally, note that the data shown in Fig.~\ref{fig:anisovseta} was extracted from experiments that were not designed with the purpose of considering early time dynamics. This may contribute to the large uncertainty of the existing data, as well as possibly introducing significant systematic error. It is particularly interesting to note that for smaller values of $\eta/P$ the error bars in Fig.~\ref{fig:anisovseta} tend to be larger. Simultaneously, the damping rate for the non-hydrodynamic mode is increased meaning the dynamics of the non-hydrodynamic modes are expected to occur on a shorter timescale. For reference, experimental data and fits of the dynamics with and without the non-hydrodynamic mode (second term in Eq.~\eqref{fitfn}) for $\eta/P\approx 0.396$ (leftmost green data point in Fig~\ref{fig:anisovseta}) and $\eta/P\approx1.30$ (rightmost blue point in  in Fig~\ref{fig:anisovseta}) are shown in Fig.~\ref{fig:fits}. One notices that the best fit for the case with $\eta/P \approx 1.30$ is significantly better when including the non-hydrodynamic mode, while for smaller $\eta/P$ the improvement is less noticeable. This is likely because the non-hydrodynamic mode amplitude is expected to be somewhat smaller, and the damping rate is higher, meaning better time resolution could improve accuracy of extraction.

\subsection{Isotropic Trap Quench in $d=3$}

\begin{figure*}[htbp]
      \includegraphics[width=1\textwidth]{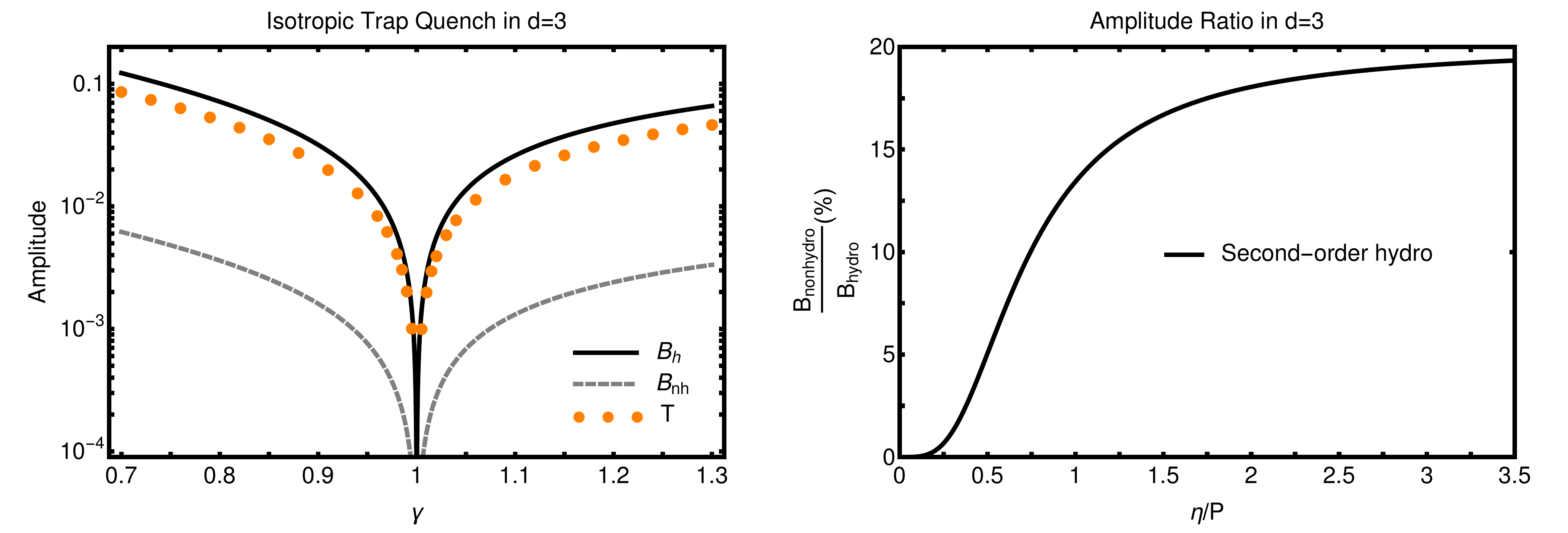}
     \caption{Left: Absolute value of the (dimensionless) hydrodynamic breathing (``B$_h$"),  non-hydrodynamic breathing (``B$_{nh}$") and temperature (``T") modes as a function the quench strength parameter $\gamma$ for an isotropic trap quench in d=3. Results shown are for $\frac{\eta}{P}=0.5$. Right: Ratio of the non-hydrodynamic to hydrodynamic breathing mode amplitudes as a function of $\eta/P$ for an isotropic trap quench in d=3 (result independent of $\gamma$). The maximal ratio of about $20\%$ is roughly consistent with the results of Ref. \cite{BrewerPRL2015}, and suggests the possibility of experimental observation.}
      \label{fig:3diso}
\end{figure*}

For the case of an isotropic trap quench ($A_i/A_0=1,T_{\, init}=1$) in d=3, results for the (hydrodynamic and non-hydrodynamic) breathing mode and temperature mode amplitudes are shown in Fig.~\ref{fig:3diso}. Unlike the case of d=2, the three-dimensional geometry is capable of supporting a non-hydrodynamic breathing mode; furthermore, the temperature mode amplitude is found to differ from the (hydrodynamic) breathing mode amplitude. 

The right panel of Fig.~\ref{fig:3diso} shows the ratio of non-hydrodynamic to hydrodynamic breathing mode amplitudes, which reaches up to about $20\%$ for large enough $\eta/P$. It is interesting to note that the apparent saturation of this ratio at about $20\%$ is consistent with the amplitude ratio from Ref.~\cite{BrewerPRL2015}, extracted from experimental data in Ref.~\cite{KinastPRL2005}. (Note that in the experiment of Ref.~\cite{KinastPRL2005}, the gas was released from a symmetric trap, allowed to expand for a short period, and then recaptured in a symmetric trapping potential, which is a different protocol than the trap quench considered here. For this reason a direct comparison to mode amplitudes of Ref.~\cite{BrewerPRL2015} is not attempted in this case.)

\chapter{Anisotropically Trapped Fermi Gas in Second-Order Hydrodynamics}
\label{sohaniso}
The assumption of an isotropic harmonic trap in the previous chapter can be relaxed. One reason to pursue this avenue is that, due to technical limitations, real experiments are often only able to achieve traps which are anisotropic even though residual anharmonicity can be minimal. In other cases the goal may be to study behavior of the gas in this geometry \cite{Wright2007}. 

In this chapter, the assumption of an isotropic trap is relaxed while retaining the assumption that the trap is harmonic. Note that all other assumptions used in the second-order hydrodynamic formalism are the same. The second-order mode spatial structure, frequencies, and damping rates are investigated. It is proposed that by controlling anharmonicity it may be possible to separate out the non-hydrodynamic modes due to the relative independence of their damping rates from the trap geometry.

\section{Equilibrium}
\label{confexp}
\begin{figure}[htbp]
	\centering
     \includegraphics[width=0.25\textwidth]{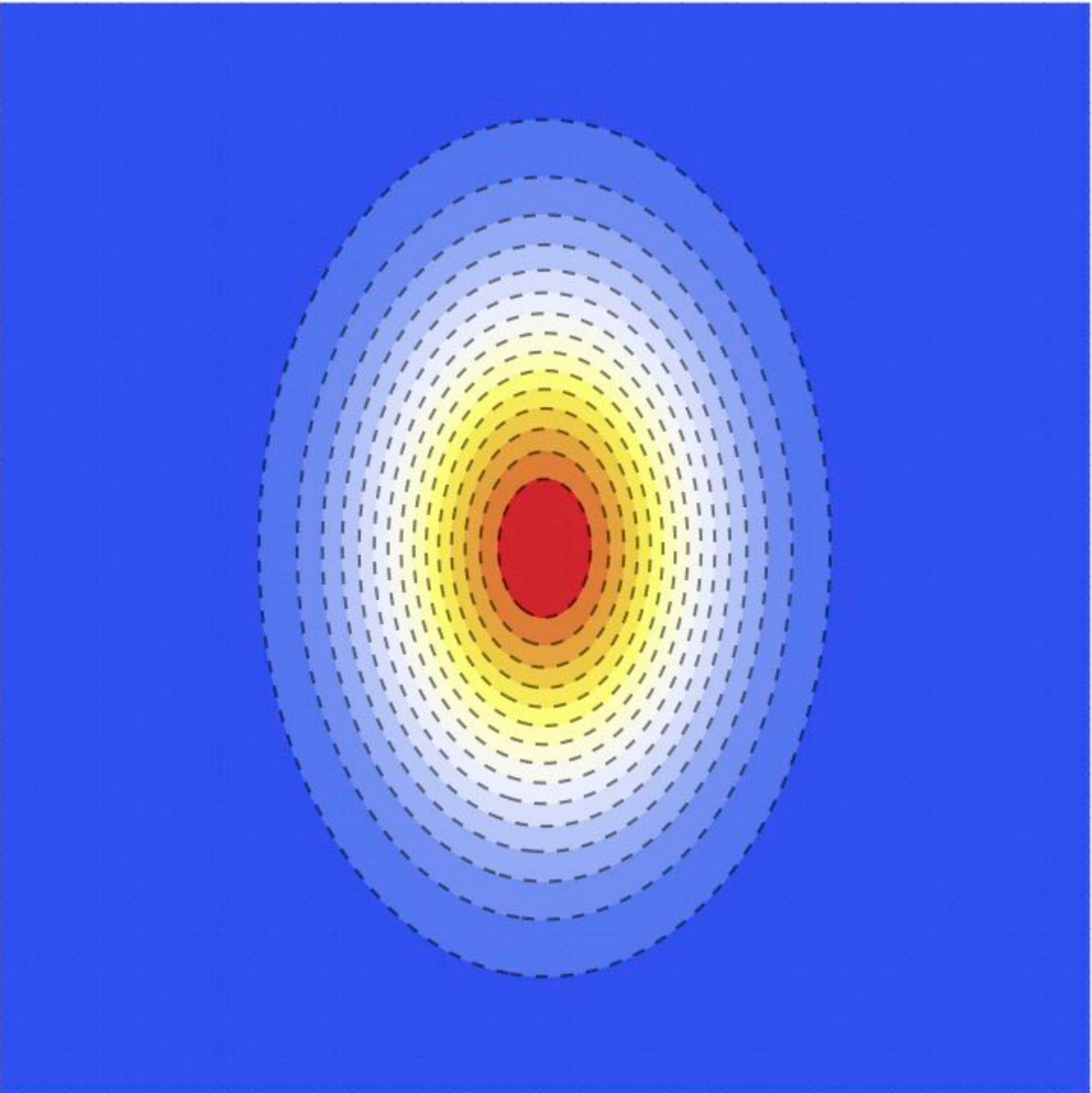}
    \caption{Spatial snapshot of the equilibrium distribution for an anisotropic trap with $\omega_x=\lambda \omega_\perp$ and $\omega_y= \omega_\perp$ and $\lambda=1.5$.}
      \label{fig:AnisEq}
\end{figure}

For an anisotropic harmonic trapping potential with trapping frequencies $\omega_x=\lambda \omega_\perp$ and $\omega_y=\omega_\perp$, the solution for the equilibrium density configuration is given by $\rho_0(\mathbf{x}) =\rho_0 \exp{[{\frac{-(\lambda^2 x^2+y^2)m \omega_\perp^2}{2 T_0}}]}$. In dimensionless units the equilibrium solution is given by
\begin{align}
	\nonumber \rho_0(\mathbf{x}) =A_0 &e^{\frac{-(\lambda^2 x^2+y^2)}{2 }}, \quad
	\mathbf{u}_0(\mathbf{x})=0,\quad T_0(\mathbf{x})= 1,
\end{align}
where $A_0$ is a dimensionless positive number setting the number of particles. For reference, the equilibrium configuration for $\lambda=1.5$ is shown in Fig.~\ref{fig:AnisEq}. A linear perturbation analysis about this equilibrium configuration can then be conducted exactly as in Chap.~\ref{sohiso}. It should be noted here that sometimes in the literature, the unit normalization $\omega_\perp = \sqrt{\omega_x \omega_y}$ is used.

\section{Collective Mode Solutions in $d=2$}\label{anishar2dmodes}
\begin{figure}[htbp]
	\centering
     \includegraphics[width=\textwidth]{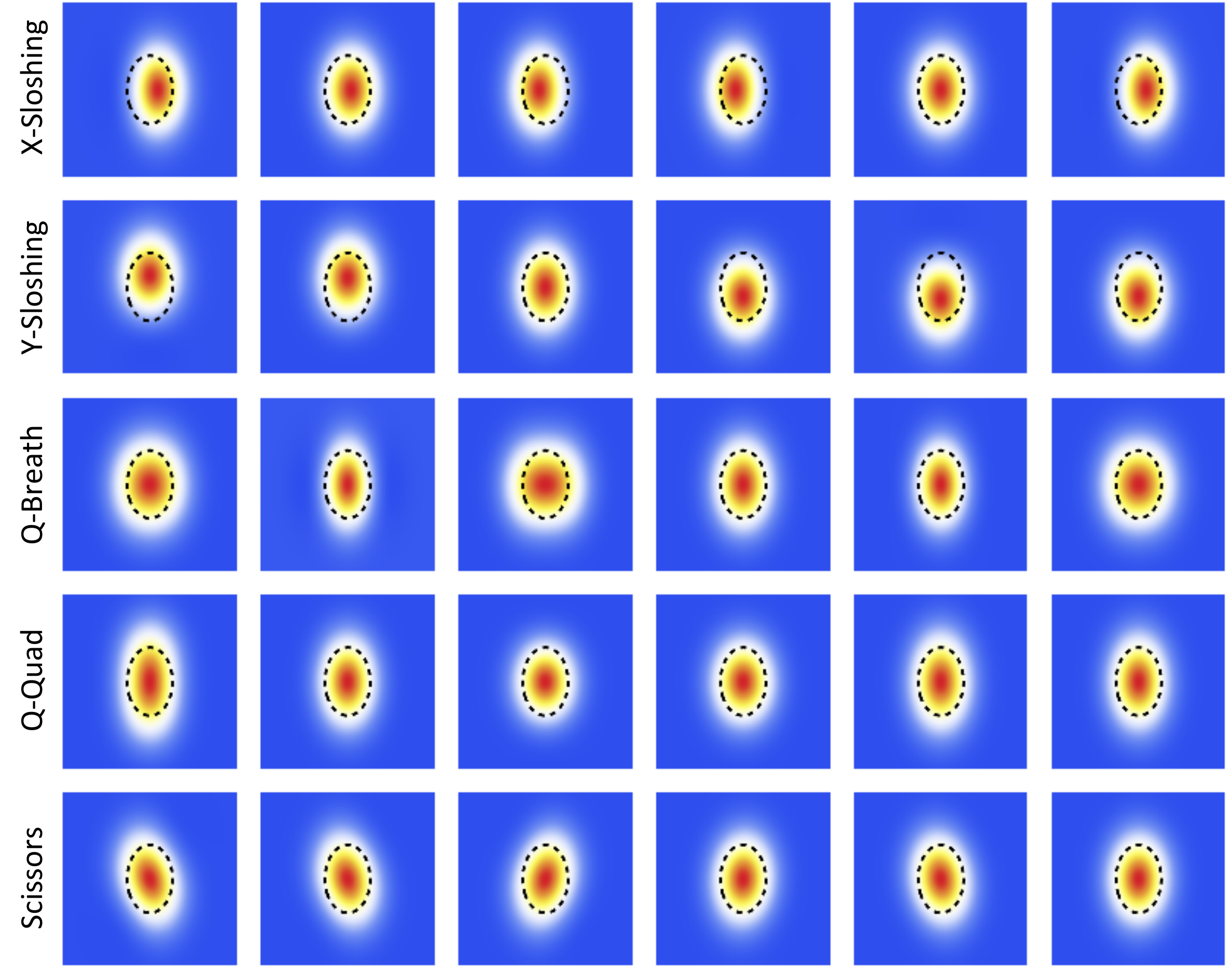}
    \caption{Each row is the time evolution (from $\omega_\perp t=0$ to $\omega_\perp t = 2\pi$ in equal steps) of a mode in an anisotropic trap with $\omega_x=\lambda \omega_\perp$ and $\omega_y=\omega_\perp$ and $\lambda=1.5$. The black dashed ellipse denotes the fixed trapping potential geometry to aid the eye in seeing the time dependence. Note that the non-hydrodynamic modes and the damped scissors mode are not shown since they have the same general spatial behavior but different time dependence.}
      \label{fig:anisev}
\end{figure}

Note that in the isotropic case at second order in the configuration space expansion, there were $8$ hydrodynamic modes and $2$ non-hydrodynamic modes. Thus in the anisotropic trap case, there should be the same number of modes. This is indeed the case. However, results for the density and velocity of the second-order collective modes in $d=2$ for an anisotropic trap are relatively complicated. This is because when introducing trap anisotropy, the diagonal-quadrupole mixes with the zero-frequency rotation mode to produce two scissors modes, and the quadrupole mode aligned with the coordinate axes mixes with the breathing mode producing two modes which share some characteristics with both the breathing and quadrupole modes (see subsequent descriptions of the modes for more detail). Thus rather than showing a single snapshot of the density profiles, the time-evolution of these modes is shown in Fig.~\ref{fig:anisev}. 

To elaborate, the hydrodynamic modes found include a zero-frequency number mode, a zero-frequency temperature mode, two different sloshing (or dipole) modes where the center of mass of the cloud oscillates about the trap center at $\omega_x (\omega_y)$ if the gas sloshes along the x-(y-)axis, a quasi-breathing mode which in the limit of no anisotropy maps back to the breathing mode,  a quasi-quadrupole which in the limit of no anisotropy maps back to the quadrupole mode, and two scissors modes which tilt back and forth making an angle with the long axis of the trap and limit to the tilted-quadrupole and rotation modes in the limit of no anisotropy. Note that the scissors mode has been discussed under different frameworks such as kinetic theory in Ref.~\cite{Bruun2007} and using the Navier-Stokes equations in Ref.~\cite{LuSciss2012}. It is interesting to note that the results discussed here for the scissors mode frequency and damping rate match exactly those from the second-order moment expansion of kinetic theory in Ref.~\cite{Bruun2007}.

\begin{table}[h!]
\begin{center}
\begin{tabular}{| c | c | c |}
  \hline
  \phantom{...}&$\omega$ & $\Gamma$\\
  \hline			
  Number (Zero Mode) & $0$ & $0$ \\
  \hline			
  Temperature (Zero Mode) & $0$ & $0$\\
  \hline				
  X-Sloshing (Dipole) & 1 & 0 \\
  \hline
  Y-Sloshing (Dipole) & $\lambda$ & 0 \\
  \hline
  Quasi-breathing& $2+\epsilon$ & $4 \epsilon^2 \frac{\eta}{P} $\\
  \hline
  Quasi-quadrupole & $\sqrt{2}+ \frac{\epsilon}{\sqrt{2}}$ & $\frac{\eta}{P}(1+\epsilon)$ \\
  \hline
  Scissors & $\sqrt{2}+ \frac{\epsilon}{\sqrt{2}}$ & $\frac{\eta}{P}(1+\epsilon)$ \\
  \hline
 Damped-Scissors & $0$ & $2\frac{\eta}{P}(1+\epsilon)$ \\
  \hline
  Non-hydrodynamic Quasi-Mode & 0 & $\frac{1}{\tau_\pi} - 2\frac{\eta}{P}(1+\epsilon)$ \\
    \hline  
  Non-hydrodynamic Scissors & 0 & $\frac{1}{\tau_\pi } - 2\frac{\eta}{P}(1+\epsilon)$ \\
    \hline  
\end{tabular}
\end{center}
  \caption{Frequencies and damping rates in $d=2$ from linearized second-order hydrodynamics letting $\epsilon=\lambda- 1 \ll 1$ and assuming $\frac{\eta}{P},\tau_\pi\ll 1$. Note that at this order the quasi-quadrupole and scissors modes are the same, but at the next order results for the frequencies and damping differ.}
\label{2dfreqtabanis}
\end{table}

\begin{figure}[htbp]
	\centering
     \includegraphics[width=\textwidth]{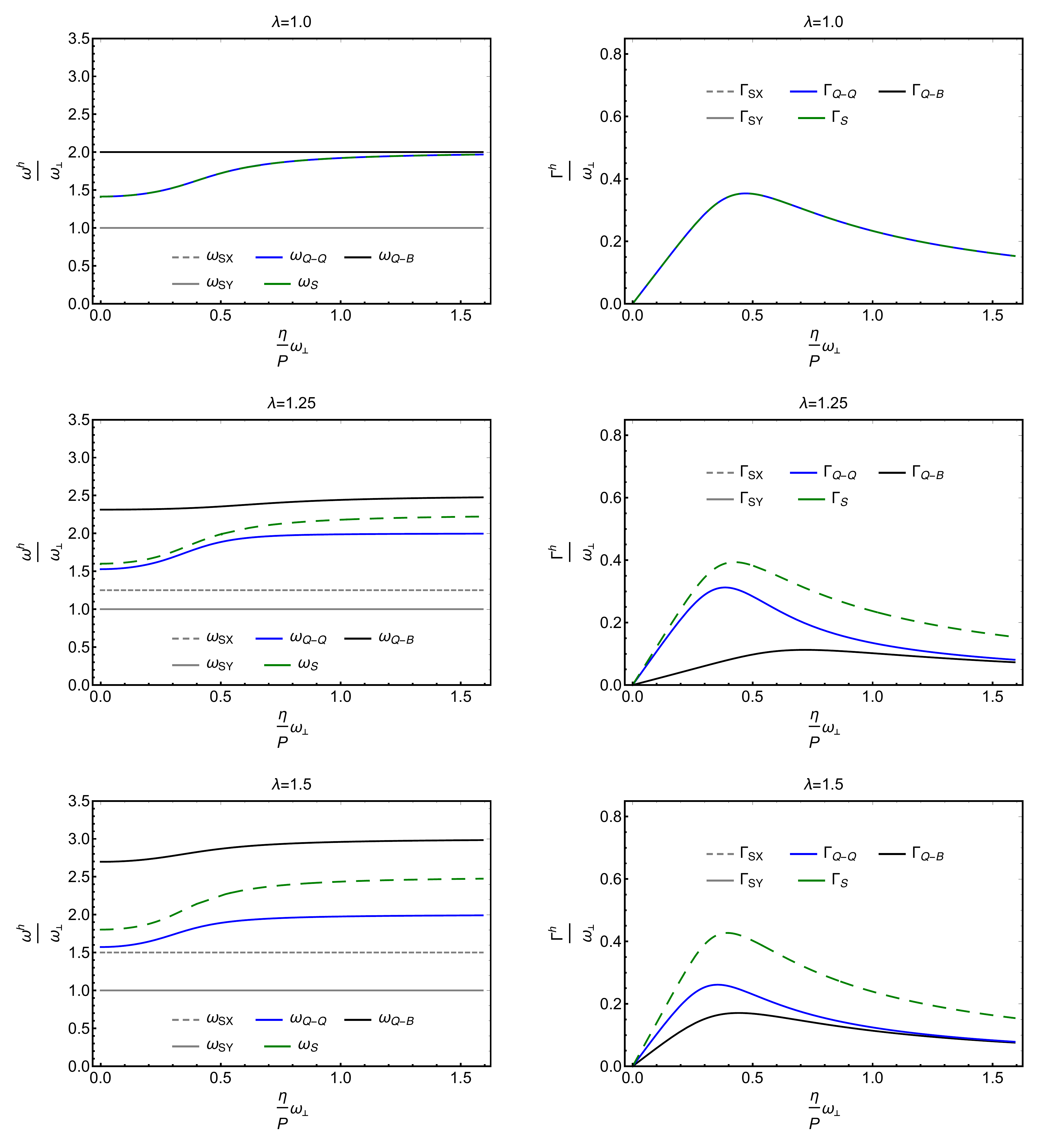}
    \caption{Two-dimensional frequencies (left panel) and damping rates (right panel) for all hydrodynamic quasi-breathing (Q-B), quasi-quadrupole (Q-Q), scissors (S), and two sloshing (SX and SY) modes vs $\eta/P$ for $\lambda = 1, 1.25, 1.5$. Note that at $\lambda=1$ the frequencies and damping rates match the isotropic trap case as expected.}
      \label{fig:2danishyd}
\end{figure}

\begin{figure}[htbp]
	\centering
     \includegraphics[width=\textwidth]{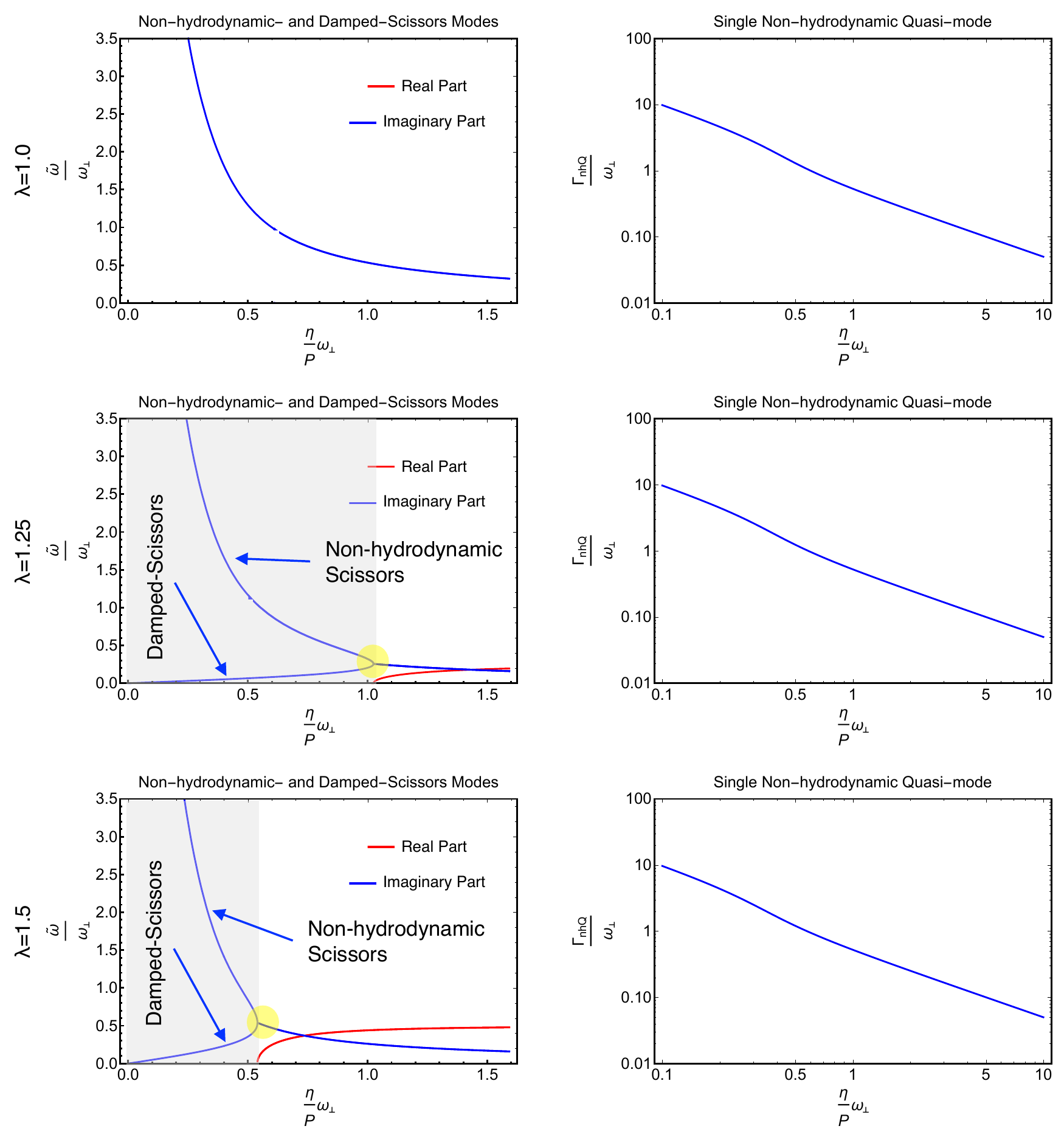}
    \caption{All plots are for $d=2$. Left panels: Frequencies (red) and damping rates (blue) for the hydrodynamic damped scissors mode and non-hydrodynamic scissors modes vs $\eta/P$ for $\lambda = 1, 1.25, 1.5$. Note that in the $\lambda=1$ case, the damped-scissors mode is no longer damped and has zero frequency as it is mapped to the rotation mode for the isotropic trap. Also note that for $\lambda \neq1$ the non-hydrodynamic scissors and damped scissors modes merge at some value of $\eta/P$ indicated by the yellow circle producing two damped modes, one with real part shown in red, and the other with the same damping but negative real part. The gray region shows where the modes are purely damped. Right panels: Damping rate for the non-hydrodynamic quasi-mode.}
      \label{fig:2danissciss}
\end{figure}

As in the previous chapter, Tab.~\ref{2dfreqtabanis} gives expressions for the complex frequencies and spatial mode structure from second-order hydrodynamics for the second-order modes for a nearly isotropic trap ($\lambda = 1+\epsilon$ with $0 \leq \epsilon \ll 1$) in the hydrodynamic limit $\eta/P \ll 1$ and $\tau_\pi \ll 1$ (assuming that $\tau_\pi$ and $\eta/P$ are of the same order of magnitude). In addition to the modes shown in Fig.~\ref{fig:anisev} there are two modes in Tab.~\ref{2dfreqtabanis} which have zero complex frequency. The first corresponds to a change in total particle number, the second corresponds to a change in temperature and width of the cloud. These modes are exactly analogous to their counterparts in Chap.~\ref{sohiso} with a modified spatial structure corresponding to the new trap equilibrium configuration. The mode frequencies and damping rates are shown in Figs.~\ref{fig:2danishyd} and \ref{fig:2danissciss} for several values of the trap anisotropy parameter $\lambda$. Fig. ~\ref{fig:2danishyd} shows the frequencies and damping rates for the sloshing modes, quasi-quadrupole, quasi-breathing, as well as the scissors mode (but not the ``damped scissors mode"). As expected when $\lambda=1$ results match the isotropic trap case. As the anisotropy increases, the scissors and quadrupole mode split with the scissors mode taking on a higher frequency. The breathing mode shifts to higher frequency as well.  In addition, the quasi-breathing mode becomes damped. 

The remaining hydrodynamic mode, the ``damped scissors mode" is one of the modes shown in the left panels of Fig.~\ref{fig:2danissciss}. Note that for $\lambda=1$, the damped scissors mode has no damping, aero frequency, and is identical to the rotation mode in the isotropic trap. The damping rate shown is that of the non-hydrodynamic quadrupole mode to which the non-hydrodynamic scissors mode limits to. However as soon as anisotropy is introduced, the rotation mode mixes with the tilted quadrupole mode leading to the damped scissors mode. Interestingly, for $\lambda \neq 1$, the damping rates of the damped scissors mode and the non-hydrodynamic scissors mode are seen to merge at some value of $\eta/P$ (see yellow circle in Fig~\ref{fig:2danissciss}) which decreases with increasing $\lambda$. When this happens, the two modes pickup an non-zero oscillation frequencies (the red curve in Fig~\ref{fig:2danissciss}). This behavior is also predicted in kinetic theory (see e.g. Refs.~\cite{Bruun2007,Urban2008}), and match exactly the frequencies and damping rates predicted here. Note that there are two such modes, corresponding to left- and right-movers. It is not clear if this behavior is just a consequence of the second-order hydrodynamic treatment breaking down, or if this could be observed experimentally. To understand the right panels, it is important to note that the quasi-breathing and quasi-quadrupole modes share the same null vector up to replacement of the mode frequency appearing in the spatial mode expressions. Consequently, there is a single non-hydrodynamic mode, termed here the ``non-hydrodynamic quasi-mode" for lack of a better descriptor, which is obtained by substituting the expression for this mode's frequency into the spatial structure of the quasi-modes. The right panels of Fig.~\ref{fig:2danissciss} show the non-hydrodynamic quasi-mode damping rate for different values of $\lambda$.

\newpage
\section{Collective Mode Solutions in $d=3$}\label{anishar3dmodes}

The spatial mode structure for the case $d=3$ with translational invariance along the z-axis is similar to the $d=2$ case. However, several of the frequencies and damping rates, while they behave in a qualitatively similar fashion, are quantitatively different. Particularly, the hydrodynamic quasi-breathing and quasi-quadrupole (to higher order than shown in Tab.~\ref{3dfreqtabanis}) modes behave differently with respect to trap anisotropy. This can be physically understood from the fact that with trap anisotropy the breathing and quadrupole modes couple, and for $d=3$ the breathing mode now has a different frequency and damping rate in the isotropic trap case. Therefore, when the modes are coupled, their behavior will be different than for the $d=2$ case. Furthermore, there are now two distinct non-hydrodynamic quasi-modes, one which becomes to the non-hydrodynamic quadrupole mode when $\lambda=1$ and the other which limits to the non-hydrodynamic breathing mode. On the other hand, the number, temperature, sloshing, scissors, damped scissors, and non-hydrodynamic scissors modes have the same frequencies and damping rates as their $d=2$ counterparts. Tab.~\ref{3dfreqtabanis} gives expressions for the complex frequencies and spatial mode structure for ($\lambda = 1+ \epsilon $ with $0 \leq \epsilon \ll1$) in the hydrodynamic limit $\eta/P \ll 1$ and $\tau_\pi \ll 1$.

\begin{table}[h!]
\begin{center}
\begin{tabular}{| c | c | c |}
  \hline
  \phantom{...}&$\omega$ & $\Gamma$\\
  \hline			
  Number (Zero Mode) & $0$ & $0$ \\
  \hline			
  Temperature (Zero Mode) & $0$ & $0$\\
  \hline				
  X-Sloshing (Dipole) & 1 & 0 \\
  \hline
  Y-Sloshing (Dipole) & $\lambda$ & 0 \\
  \hline
  Quasi-breathing& $\sqrt{\frac{10}{3}}+\epsilon \sqrt{\frac{5}{6}}$ & $\frac{\eta}{3P}(1+\epsilon) $\\
  \hline
  Quasi-quadrupole & $\sqrt{2}+ \frac{\epsilon}{\sqrt{2}}$ & $\frac{\eta}{P}(1+\epsilon)$ \\
  \hline
  Scissors & $\sqrt{2}+ \frac{\epsilon}{\sqrt{2}}$ & $\frac{\eta}{P}(1+\epsilon)$ \\
  \hline
 Damped-Scissors & $0$ & $2\frac{\eta}{P}(1+\epsilon)$ \\
  \hline
  Non-hydrodynamic Quasi-quadrupole & 0 & $\frac{1}{\tau_\pi} - 2\frac{\eta}{P}(1+\epsilon/4)$ \\
    \hline  
  Non-hydrodynamic Quasi-breathing & 0 & $\frac{1}{\tau_\pi} - (\frac{2\eta}{3P}+\frac{13\eta}{6P}\epsilon)$ \\
  \hline  
  Non-hydrodynamic Scissors & 0 & $\frac{1}{\tau_\pi } - 2\frac{\eta}{P}(1+\epsilon)$ \\
    \hline  
\end{tabular}
\end{center}
  \caption{Frequencies and damping rates in $d=3$ from linearized second-order hydrodynamics letting $\epsilon=\lambda - 1$ and assuming $\frac{\eta}{P},\tau_\pi\ll 1$. Note that at this order the quasi-quadrupole and scissors modes are the same, but at the next order results for the frequencies and damping differ.}
\label{3dfreqtabanis}
\end{table}

\begin{figure}[htbp]
	\centering
     \includegraphics[width=\textwidth]{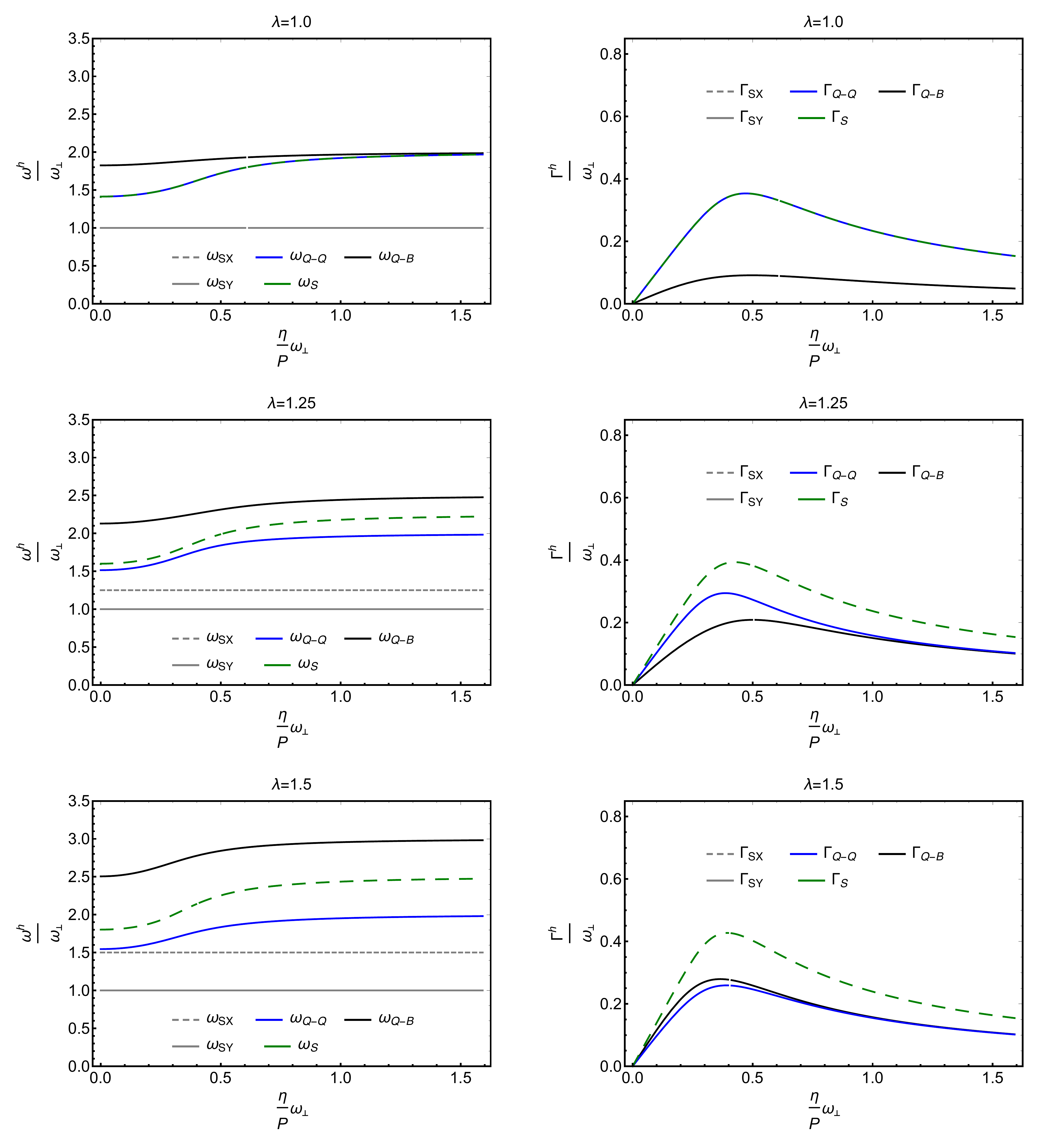}
     \caption{Three-dimensional frequencies (left panel) and damping rates (right panel) for all hydrodynamic quasi-breathing (Q-B), quasi-quadrupole (Q-Q), scissors (S), and two sloshing (SX and SY) modes vs $\eta/P$ for $\lambda = 1, 1.25, 1.5$. Note that at $\lambda=1$ the frequencies and damping rates match the isotropic trap case as expected.}
      \label{fig:3danishyd}
\end{figure}

\begin{figure}[htbp]
	\centering
     \includegraphics[width=\textwidth]{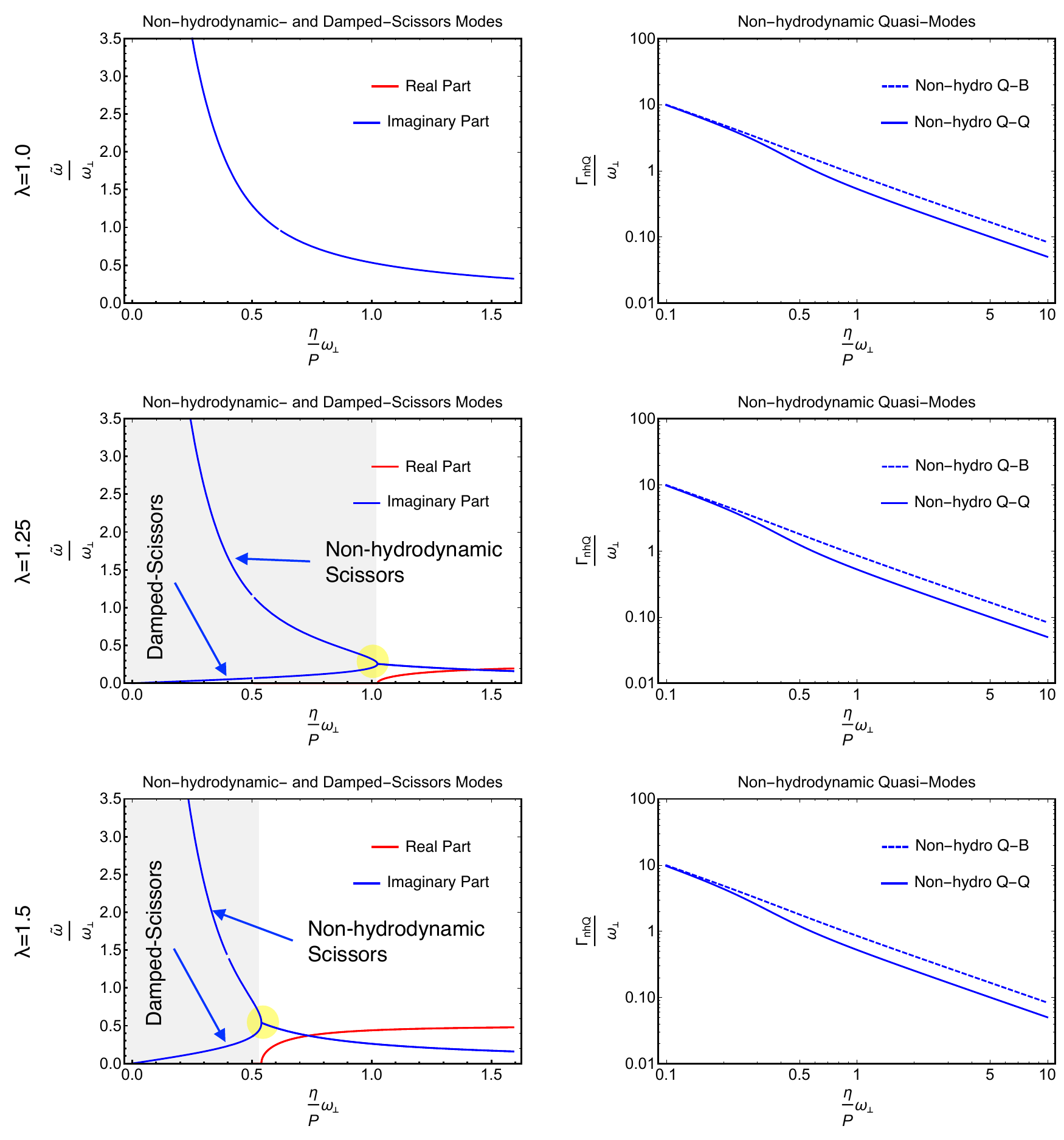}
   \caption{ All plots are for $d=3$. Left panels: Frequencies (red) and damping rates (blue) for the hydrodynamic damped scissors mode and non-hydrodynamic scissors modes vs $\eta/P$ for $\lambda = 1, 1.25, 1.5$. Note that in the $\lambda=1$ case, the damped-scissors mode is no longer damped and has zero frequency as it is mapped to the rotation mode for the isotropic trap. Also note that for $\lambda \neq1$ the non-hydrodynamic scissors and damped scissors modes merge at some value of $\eta/P$ indicated by the yellow circle producing two damped modes, one with real part shown in red, and the other with the same damping but negative real part. Right panels: Damping rate for the non-hydrodynamic quasi-mode.}      
   \label{fig:3danissciss}
\end{figure}

\section{Relative Frequency and Damping Rate Shifts}\label{freqshift}

\begin{figure}[htbp]
	\centering
     \includegraphics[width=\textwidth]{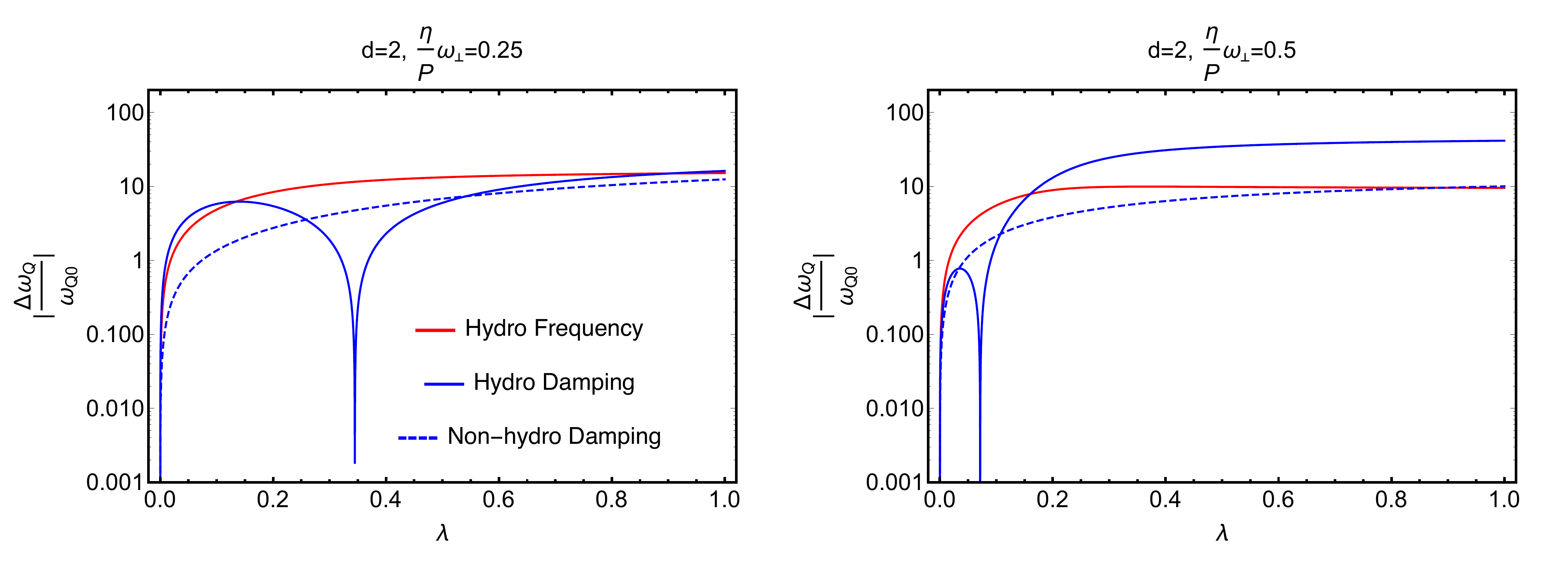}
    \caption{Plot of the percent-difference in frequency and damping of the hydrodynamic and non-hydrodynamic quasi-quadrupole modes compared to the isotropic trap case vs. trap anisotropy $\epsilon$ for $\eta/P\omega_\perp=0.25,0.5$.}
      \label{fig:aniseff2d}
\end{figure}

\begin{figure}[htbp]
	\centering
     \includegraphics[width=\textwidth]{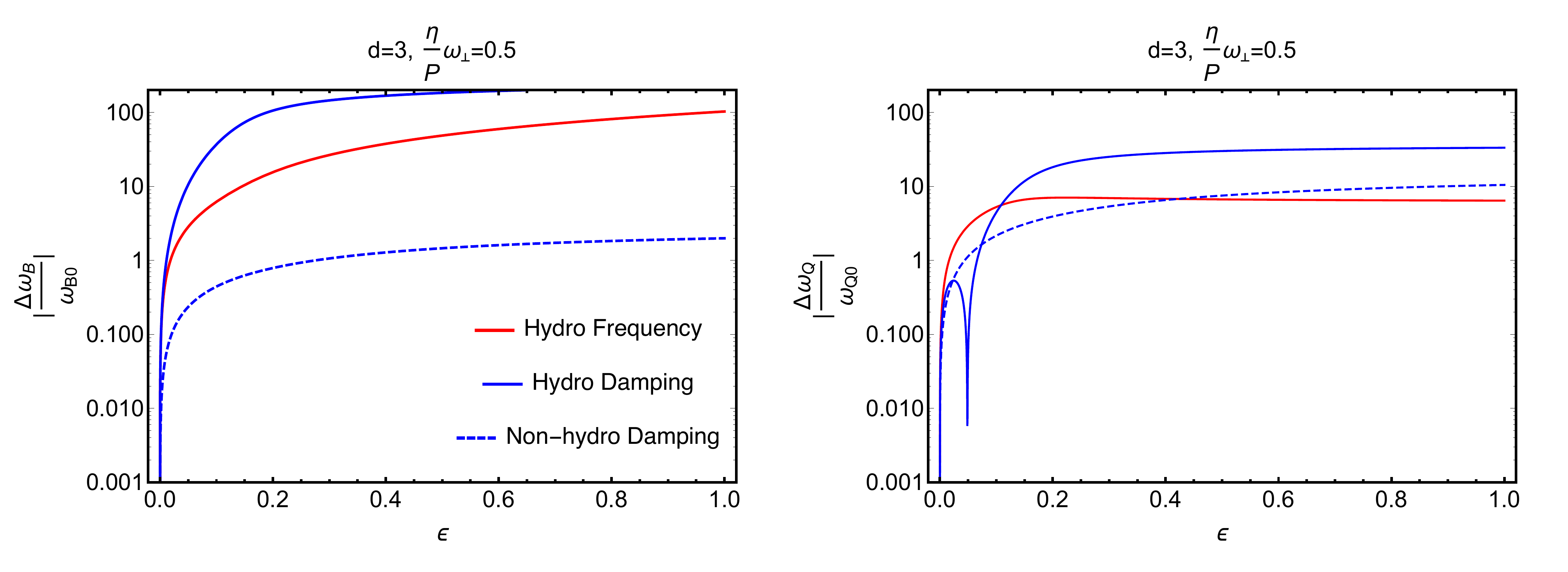}
    \caption{Left panel: Plot of the percent-difference in frequency and damping of the hydrodynamic and non-hydrodynamic quasi-breathing modes compared to the isotropic trap case vs. trap anisotropy $\epsilon$ for two values of $\eta/P\omega_\perp = 0.5$. Notice that for all values of $\epsilon$, the non-hydrodynamic damping rate is far less sensitive to trap anisotropy than the hydrodynamic damping, suggesting that anisotropy could possibly be use to extract the non-hydrodynamic mode. Right panel: Plot of the percent-difference in frequency and damping of the hydrodynamic and non-hydrodynamic quasi-quadrupole modes compared to the isotropic trap case vs. trap anisotropy $\epsilon$ for $\eta/P\omega_\perp = 0.5$ showing only slight difference from the $d=2$ case.}
      \label{fig:aniseff3d}
\end{figure}

In this section, the relative shifts  with respect to anisotropy in frequency and damping of the quasi-modes with both hydrodynamic and non-hydrodynamic terms are highlighted. For $d=2$, results are shown in Fig.~\ref{fig:aniseff2d} for the quasi-quadrupole mode for $\eta/P\omega_\perp=0.25,0.5$ with the definitions
	 \begin{equation}
	 	\Delta \omega_Q = \frac{\bigg(Re\big[\omega(\lambda)\big]-Re\big[\omega(\lambda=1)\big]\bigg)}{Re\big[\omega(\lambda=1)\big]},
	\end{equation}
	\begin{equation}
		\Delta \Gamma_Q = \frac{\bigg(Im\big[\omega(\lambda)\big]-Im\big[\omega(\lambda=1)\big]\bigg)}{Im\big[\omega(\lambda=1)\big]}.
	\end{equation}
For $d=3$ the results are shown in Fig.~\ref{fig:aniseff3d} using $\eta/P\omega_\perp=0.5$ for both the quasi-breathing and quasi-quadrupole modes which in this case both have an associated non-hydrodynamic mode. Perhaps the most interesting feature is that the non-hydrodynamic quasi-breathing mode is quite insensitive to changes in trap anisotropy as compared to the frequency and damping for the hydrodynamic quasi-breathing mode. It would be interesting to consider for future work whether this property could be utilized to isolate the non-hydrodynamic mode by tuning trap anisotropy since the hydrodynamic mode is quite sensitive to anisotropy, while the non-hydrodynamic mode is not.

\chapter{Uniform Fermi Gas in Second-Order Hydrodynamics}
\label{sohunif}
Recent experiments by the group of M. Zwierlein at MIT have created a spatially homogenous strongly interacting Fermi gas \cite{Zwierlein2017}. This may prove a fruitful ground for exploring non-hydrodynamic behavior in strongly interacting Fermi gases. Towards this end, the hydrodynamic and non-hydrodynamic sound and shear excitations in a uniform density and temperature fluid in $d=3$ are studied in second-order hydrodynamics.

\section{Solving Second-Order Hydrodynamics}\label{SOH}

\subsection{Fourier Expansion}\label{confexp}

Unlike the previous two cases in a harmonic trapping potential, here the equilibrium solution exhibits translational invariance. Specifically, for a uniform gas with no trapping potential the equilibrium configuration is given by 
\begin{align}
	\nonumber \rho_0(\mathbf{x}) =\rho_0, \quad
	\mathbf{u}_0(\mathbf{x})=0,\quad T_0(\mathbf{x})= T_0,
\end{align}
where $\rho_0$ is a dimensionless positive number setting the number of particles per unit volume and $T_0$ sets the gas temperature. In this chapter, dimensionless units such that
\begin{align}
	\nonumber \rho_0(\mathbf{x}) =1, \quad
	\mathbf{u}_0(\mathbf{x})=0,\quad T_0(\mathbf{x})= 1,
\end{align}
are employed. 

For the uniform gas, a Fourier decomposition of the dynamics is useful. In particular it is convenient to study excitations with wave-vector $\mathbf{k} \| \mathbf{u}$ (sound channel) and $\mathbf{k} \perp \mathbf{u}$(shear channel). To approach this problem, first one may make the ansatz $\rho = (1+ \delta \rho)$, $\mathbf{u}= \delta \mathbf{u}$, and $T= 1+\delta T$ with $\delta \rho, \delta \mathbf{u}, \delta T$ assumed to be small. Then working in the spatial and time frequency domains one has $\delta \rho(t,\mathbf{x}) = e^{i(\mathbf{k}\mathbf{x}-\omega t)} \delta \rho_0$, with similar expressions holding for $\delta \mathbf{u}$ and $\delta T$.

\section{Shear Mode}\label{2dmodes}

\begin{figure*}[ht!]
\includegraphics[width=\textwidth]{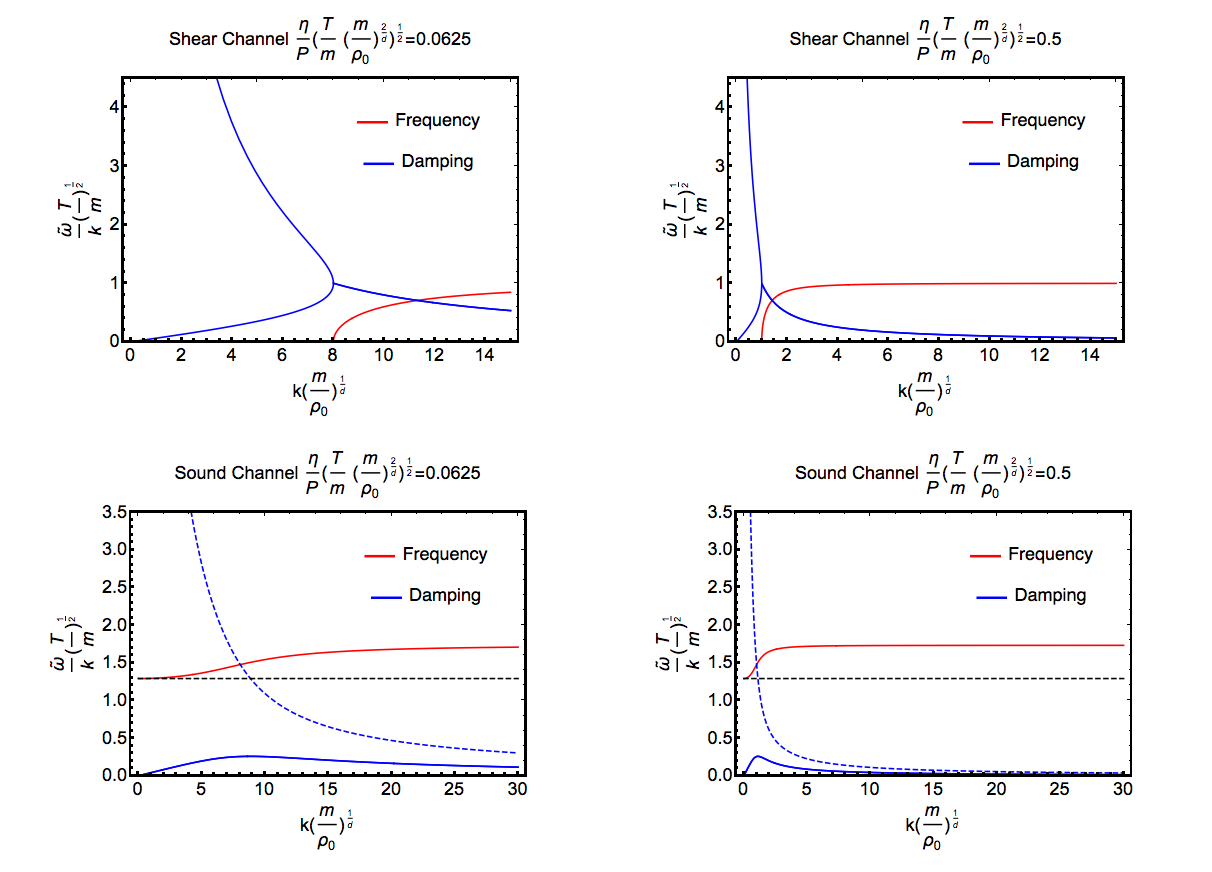}
  \caption{Top row: Normalized shear channel frequencies and damping vs. momentum for two values of $\eta/P$. Notice that for small $k$ there is a mode whose frequency diverges as $k\to0$. This is the non-hydrodynamic shear mode. Also note that similar to the scissors mode in the previous chapter, the purely damped non-hydrodynamic and hydrodynamic modes merge and split into two oscillatory modes with the same damping and opposite sign frequency. Bottom row: Normalized sound channel frequencies and damping vs. momentum for two values of $\eta/P$. Notice that for small $k$ there is a mode whose frequency diverges as $k\to0$ (dashed blue line). This is the non-hydrodynamic shear mode. Also note $\lim_{k \to 0}\frac{d\omega_{sd H \pm }}{dk} = c_s$, where $c_s$ is the speed of sound (dashed black line).}
  \label{fig:shearunif}
\end{figure*}

For the shear channel, the analytic expressions for the complex frequencies may be written as 
\begin{equation}
	\omega_{sh \pm} = \frac{-i \pm \sqrt{-1 + 4 k^2 \tau \frac{\eta}{P}}}{2 \tau}.
\end{equation}
The shear channel frequencies and damping rates are shown in the top row of Fig.~\ref{fig:shearunif} for $\tau =\eta/P$. There are a number of interesting features to notice. First, in the limit of small $k$, the frequencies may be expanded as
\begin{equation}
\label{shearchannel}
 \omega_{sh \pm}  \approx
  \begin{cases}
                                  - i \frac{\eta}{P} k^2 & \text{for the + branch} \\
                                   \frac{-i}{\tau} + i \frac{\eta}{P} k^2 & \text{for the - branch} \\
  \end{cases}.
\end{equation}
In this case, $\omega_+$ becomes the usual hydrodynamic result for the shear and diffusion mode, while $\omega_-$ is not present in the Navier-Stokes formalism and does not vanish as $k \to 0$. Additionally, $\omega_+ = \omega_-$ when $k=1/(2\sqrt{\tau \eta/P}) \equiv k_c$. For $k>k_c$, the two purely damped modes split into two oscillatory modes with the same damping rate but opposite signed frequency. Thus, while the applicability of hydrodynamics is normally restricted to $k\ll k_c$, one may wonder if the oscillatory character of the shear modes for large $k$ is physical. It is interesting to note that oscillatory shear modes, while not present in the Navier-Stokes formalism, have been found with a similar dispersion relation in both a different extended hydrodynamic formalism than that present here as well as from a kinetic theory formalism in Ref.~\cite{Colosqui2009}. Furthermore, the presence of oscillatory shear modes was found necessary to describe experimentally measured high frequency quality factors of nano-mechanical resonators \cite{COLOSQUI2007}. This is at least encouraging evidence that this high frequency oscillatory behavior is at least qualitatively correct.

\section{Sound Mode}\label{3dmodes}

For the sound channel, the analytic expression for the complex frequency is unilluminating. The frequencies and damping rates for the sound channel are shown in the bottom row of Fig.~\ref{fig:shearunif} for $\tau =\eta/P$. Notice that there are three distinct sound modes. The first two (left and right movers) are oscillatory with vanishing damping as $k\to0$ shown in solid lines with frequency shown in red and damping in blue. These are the usual hydrodynamic sound modes. The third is a purely damped non-hydrodynamic mode (dashed red and blue lines). For $k,\eta/P$, and $\tau \ll 1$ the complex frequencies for the hydrodynamic sound modes are given by
\begin{equation}
 \omega_{sd \phantom{.}H \pm}  \approx \pm \sqrt{\frac{5}{3}} |k|- i \frac{2\eta}{3P} k^2 ,
\end{equation}
which are exactly the Navier-Stokes results, with the factor $\sqrt{5/3}$ coming from the speed of sound of a monoatomic ideal gas. The non-hydrodynamic sound mode dispersion relation is given by 
\begin{equation}
\label{soundnh}
 \omega_{sd \phantom{.}NH}  \approx  \frac{-i}{\tau} + i \frac{4\eta}{3P} k^2 +\mathcal{O}(k^3).
\end{equation}
Notice that $\lim_{k \to 0}\frac{d\omega_{sd H \pm }}{dk} = c_s$, where $c_s$ is the speed of sound in the fluid (see the black dashed line in Fig.~\ref{fig:shearunif}.


\newcommand{\bb}{\mathbb}
\newcommand{\ph}{\phantom}
\newcommand{\mc}{\mathcal}
\newcommand{\rr}{\rightarrow}
\newcommand{\lra}{\longrightarrow}
\def\({\left(}
\def\){\right)}
\def\[{\left[}
\def\]{\right]}
\def\imp{$\Rightarrow$}
\def\sgn{{\rm sgn\ }}
\def\mod{{\rm\; mod\;}}
\def\Hom{{\rm Hom}}
\def\ker{{\rm ker\ }}
\def\cok{{\rm cok\ }}
\def\im{{\rm im\ }}
\def\ind{{\rm ind\ }}
\def\vol{{\rm vol}}
\def\Vol{{\rm Vol}}
\newcommand{\one}{\mathbf{1}}
\def\ch{{\rm ch}}
\def\td{{\rm Td}}
\def\cc{{\rm c.c.}}

\newcommand{\myatop}[2]{\genfrac{}{}{0pt}{}{#1}{#2}}
\newcommand{\detail}[1]{{~\\~\underline{\it #1}} ~}
\newcommand{\com}[2]{{ \left[ #1, #2 \right] }}
\newcommand{\acom}[2]{{ \left\{ #1, #2 \right\} }}
\newcommand{\C}[1]{$(\ref{#1})$}
\newcommand{\mapsvia}[1]{\stackrel{#1}{\longrightarrow}}
\newcommand{\pp}[1]{\frac{\p}{\p #1}}
\def\thetafunction#1#2#3#4{\vartheta\left[\matrix{ #1 \cr #2} \right] (#3 |#4)}
\newcommand{\draftnote}[1]{\marginpar{\tiny\raggedright\textsf{\hspace{0pt}#1}}}
\def\frac#1#2{{#1 \over #2}}
\def\bra#1{{\langle}#1|}
\def\ket#1{|#1\rangle}
\def\bbra#1{{\langle\langle}#1|}
\def\kket#1{|#1\rangle\rangle}
\def\vev#1{\langle{#1}\rangle}
\renewcommand{\tfrac}[2]{{\textstyle\frac{#1}{#2}}}
\newcommand{\sbinom}[2]{{\textstyle\binom{#1}{#2}}}
\newcommand{\tsum}{{\textstyle\sum}}
\newcommand{\msum}[2]{{\displaystyle\sum_{#1}^{#2}}}
\newcommand\uitem[1]{\item{\underline{#1}}}

\newcommand{\shalf}{{\scriptstyle \frac{1}{2}}}
\newcommand{\half}{\frac{1}{2}}
\newcommand{\qrt}{\frac{1}{4}}
\newcommand{\iot}{\frac{i}{2}}
\newcommand{\rt}{{\sqrt 2}}
\newcommand{\ort}{\frac{1}{\rt}}

\renewcommand{\a}{\alpha}
\renewcommand{\b}{\beta}
\renewcommand{\d}{\delta}
\newcommand{\g}{\gamma}
\newcommand{\gt}{{\tilde{\gamma}}}
\newcommand{\gtb}{{\bar{\tilde{\gamma}}}}
\newcommand{\gbar}{{\bar\gamma}}
\newcommand{\kap}{\kappa}
\newcommand{\f}{\psi}
\renewcommand{\r}{\rho}
\newcommand{\s}{\sigma}
\newcommand{\sbar}{{\bar\sigma}}
\renewcommand{\l}{\lambda}
\newcommand{\lb}{\bar{\lambda}}
\newcommand{\w}{\omega}
\newcommand{\m}{\mu}
\newcommand{\n}{\nu}
\newcommand{\ep}{\epsilon}
\newcommand{\eps}{\epsilon}
\newcommand{\eb}{\bar{\epsilon}}
\renewcommand{\t}{\theta}
\newcommand{\tb}{\bar{\theta}}
\newcommand{\psib}{{\bar\psi}}
\newcommand{\psit}{{\tilde{\psi}}}
\newcommand{\psitb}{{\bar{\tilde{\psi}}}}
\newcommand{\chib}{{\bar\chi}}
\newcommand{\phib}{\bar{\phi}}
\newcommand{\taub}{\bar{\tau}}
\newcommand{\vp}{\varphi}
\newcommand{\vt}{\vartheta}
\newcommand{\xib}{{\bar\xi}}
\newcommand{\xit}{{\tilde\xi}}
\newcommand{\z}{\xi}

\newcommand{\G}{\Gamma}
\newcommand{\T}{\Theta}
\newcommand{\U}{\Upsilon}
\newcommand{\LL}{\Lambda}
\newcommand{\LB}{\overline{\Lambda}}
\newcommand{\Om}{\Omega}
\renewcommand{\S}{\Sigma}
\newcommand{\GB}{\overline{\Gamma}}
\newcommand{\TB}{{\overline{\Theta}}}

\newcommand{\CA}{{\cal A}}
\newcommand{\CB}{{\cal B}}
\newcommand{\CC}{{\cal C}}
\newcommand{\CD}{{\cal D}}
\newcommand{\CE}{{\cal E}}
\newcommand{\CF}{{\cal F}}
\newcommand{\CG}{{\cal G}}
\newcommand{\CH}{{\cal H}}
\newcommand{\CI}{{\cal I}}
\newcommand{\CJ}{{\cal J}}
\newcommand{\CK}{{\cal K}}
\newcommand{\CL}{{\cal L}}
\newcommand{\CM}{{\cal M}}
\newcommand{\CN}{{\cal N}}
\newcommand{\CO}{{\cal O}}
\newcommand{\CP}{{\cal P}}
\newcommand{\CQ}{{\cal Q}}
\newcommand{\CS}{{\cal S}}
\newcommand{\CV}{{\cal V}}
\newcommand{\CX}{{\cal X}}
\newcommand{\CW}{{\cal W}}
\newcommand{\CZ}{{\cal Z}}

\newcommand{\IC}{{\mathbb C}}
\newcommand{\IH}{{\mathbb H}}
\newcommand{\IQ}{{\mathbb Q}}
\newcommand{\IR}{{\mathbb R}}
\newcommand{\IZ}{{\mathbb Z}}
\newcommand{\WP}{{\mathbb WP}}


\newcommand{\MB}{{\overline{M}}}
\newcommand{\bbar}{\overline}

\newcommand{\zt}{\tilde{z}}
\newcommand{\tz}{{\tilde z}}
\newcommand{\tg}{{\tilde g}}
\newcommand{\tq}{\widetilde{q}}
\newcommand{\tf}{\widetilde{\f}}
\newcommand{\tX}{\widetilde{X}}
\newcommand{\tK}{\widetilde{K}}
\def\XT{{\widetilde X}}
\newcommand{\tG}{\tilde{\Gamma}}
\newcommand{\tth}{\tilde{\theta}}

\newcommand{\hy}{\hat y}
\newcommand{\hz}{\hat z}
\newcommand{\hx}{\hat x}
\newcommand{\hxm}{\hat{x}^-}
\newcommand{\hxp}{\hat{x}^+}
\newcommand{\hpp}{\hat{p}^+}
\newcommand{\hpm}{\hat{p}^-}
\newcommand{\hp}{\hat{p}_x}
\newcommand{\hpz}{\hat{p}_z}
\newcommand{\hM}{\widehat M}
\newcommand{\hw}{\hat w}
\newcommand{\ha}{\widehat \alpha}

\def\xv{\vec{x}}
\def\yv{\vec{y}}
\def\zv{\vec{z}}

\newcommand{\ah}{~\!{\textrm{\dn a}}}
\newcommand{\uu}{~\!{\textrm{\dn u}}}
\newcommand{\ka}{~\!{\textrm{\dn k}}}
\newcommand{\ga}{~\!{\textrm{\dn g}}}
\newcommand{\ta}{~\!{\textrm{\dn t}}}
\newcommand{\na}{~\!{\textrm{\dn n}}}
\newcommand{\pa}{~\!{\textrm{\dn p}}}
\newcommand{\pha}{~\!{\textrm{\dn P}}}
\newcommand{\ma}{~\!{\textrm{\dn m}}}
\newcommand{\la}{~\!{\textrm{\dn l}}}

\newcommand{\STr}{{\rm STr}}
\def\Re{{\rm Re\hskip0.1em}}
\def\Im{{\rm Im\hskip0.1em}}
\def\even{{\rm even}}
\def\odd{{\rm odd}}
\def\lcm{{\rm lcm}}
\def\diag{{\rm diag}}
\def\gcd{{\rm gcd}}
\def\mod{{\rm mod}}
\def\rk{{\rm rk}}
\def\eg{{\it e.g.}}
\def\ie{{\it i.e.}}
\def\cf{{\it c.f.}}
\def\etal{{\it et. al.}}
\def\cg{{\rm g}}

\def\p{\partial}
\def\pb{{\overline \p}}
\def\pgp{\pb g g^{-1}}
\newcommand{\del}{\nabla}
\def\Dslash{\rlap{\hskip0.2em/}D}

\newcommand{\tp}{{\theta^+}}
\newcommand{\tbp}{{\bar{\theta}^+}}
\newcommand{\tm}{{\theta^-}}
\newcommand{\tbm}{{\bar{\theta}^-}}

\newcommand{\D}{{\rm D}}
\newcommand{\DB}{\overline{\rm D}}
\newcommand{\CDB}{\overline{\cal D}}
\newcommand{\Dp}{{\rm D}_{+}}
\newcommand{\DBp}{\overline{\rm D}_{+}}
\newcommand{\CDp}{{\cal D}_{+}}
\newcommand{\CDBp}{\overline{\cal D}_{+}}
\newcommand{\DD}{{\mathbb D}}
\newcommand{\DDp}{{{\mathbb D}_+}}
\newcommand{\DDm}{{{\mathbb D}_-}}

\newcommand{\QP}{Q_+}
\newcommand{\QPB}{\overline{Q}_+}
\newcommand{\QB}{{\overline{Q}}}

\newcommand{\xp}{x^+}
\newcommand{\xm}{x^-}
\newcommand{\zb}{\bar{z}}

\def\RV{Ro{\v c}ek-Verlinde}
\def\HV{Hori-Vafa}
\def\GLSM{gauged linear sigma model}
\def\GLSMs{gauged linear sigma models}
\def\NLSM{nonlinear sigma model}
\def\NLSMs{nonlinear sigma models}
\def\nsusy{non-supersymmetric}
\def\susy{supersymmetry}
\def\susic{supersymmetric}
\def\dt{discrete torsion}
\def\CY{Calabi-Yau}
\def\Ka{K\"{a}hler}
\def\NK{Non-K\"{a}hler}
\def\nK{non-K\"{a}hler}
\def\TT{$(2,2)$}
\def\ZT{$(0,2)$}
\def\ads{AdS}
\def\adsd{AdS$_{d}$}
\def\adsfive{AdS$_{5}$}
\def\adsfour{AdS$_{4}$}

\def\I{{\rm I}}
\def\IA{{\rm IA}}
\def\II{\relax{I\kern-.10em I}}
\def\IIa{{\II}a}
\def\IIb{{\II}b}
\def\TeV{{\rm TeV}}
\def\hk{hyperk\"ahler\  }
\def\Hk{Hyperk\"ahler\  }
\def\ls{l_s}
\def\ms{m_s}
\def\gs{g_s}
\def\mb{{m_{\rm brane}}}
\def\vb{{v_{\rm brane}}}
\def\imt{{\rm Im} \tau}
\def\mfl{ (-1)^{F_L}}
\def\mfr{ (-1)^{F_R}}

\def\Sch{Schr\"odinger}
\def\sch#1#2{ $Sch^{#1}_{#2}$ }
\def\lif#1#2{ $Lif^{#1}_{#2}$ }
\def\bDLCQ{$\b$DLCQ}

\hyphenation{Di-men-sion-al}

\newcommand{\rH}{r_{\mathrm H}}
\newcommand{\AH}{A_{\mathrm  H}}
\newcommand{\scale}{{u}}

\chapter{Introduction to Holographic Duality}
\label{ggintro}
The formalism of second-order hydrodynamics provided perhaps the simplest possible approach to pre-hydrodynamization physics, that is, evolution of the hydrodynamic quantities such as density and temperature on a timescale $t \lessapprox \tau_\pi$. This is achieved through the incorporation of a single purely damped non-hydrodynamic mode into the set of governing differential equations by treating the viscous stress tensor as an independent dynamical variable. This type of approach can be modified so that there are oscillatory non-hydrodynamics modes as discussed in Ref.~\cite{Heller2014,Florkowski:2017olj}. Another approach which, at least in part, motivated the discussions in Ref.~\cite{Heller2014,Florkowski:2017olj} is the conjectured gauge-gravity duality. This chapter aims to provide a big picture overview of gauge-gravity duality as an alternate approach for capturing pre-hydrodynamization physics as well as perhaps indicating the presence of so-called transport universality in strongly coupled quantum fluids. Particularly, emphasis will be placed on an intuitive picture of the framework, as well as understanding the types of results that may be obtained rather than a rigorous discussion of the framework itself. For a much more detailed discussion, the reader is referred to the many excellent books and reviews on the topic \cite{Adams2012,SchaferPR2009,AdSUserBook,GaugeBook}. This chapter is largely a summary of the exposition in Ref. ~\cite{Adams2012}.

Generally speaking, holographic duality is a conjectured mathematical correspondence between quantum field theories and gravity (or more generically string theory). A particularly useful specific case of the duality is that of certain strongly coupled quantum field theories which correspond to weakly coupled classical gravity theories (i.e. Einstein gravity) in one higher dimension. One intuitive picture of how such a duality might arise when considering a strongly coupled quantum field theory comes from renormalization group flow. 

Suppose one has the general Hamiltonian
\begin{equation}
	H = \sum_{x,i} J_i(x) \mathcal{O}^i(x),
\end{equation}
where $J_i(x)$ are position dependent couplings for the different position dependent operators $\mathcal{O}^i$. Given such a Hamiltonian, it is typically desirable to study properties of the ground state and low energy excitations. Unfortunately, exact diagonalization of the hamiltonian is possible only in very special cases. One approach then to understand the low energy physics is via the Kadanoff-Wilson renormalization group (RG) theory. Depending on whether one works in real space or energy space this corresponds to either i) successively coarse graining the spatial lattice by a given scale factor and replacing the coupling constants with new coupling constants $J_i(x,u)$ at the larger lattice scale $u$ which preserve the low energy physics or ii) successively integrating out high energy degrees of freedom on energy scales larger than some cutoff. By pursuing such a procedure, one may derive a differential equation for the coupling constants as a function of the lattice spacings which takes a form such as
\begin{equation}
 u \frac{\partial J_i(x,u)}{\partial u} = \beta_i(J_j(x,u),u),
\end{equation}
where the $\beta$ function describes the change of effective coupling with energy scale. 

Given the $\beta$-function, one would then be able to describe physics of the system at fixed points of the beta-function to which the couplings flow upon the renormalization procedure. Yet, one is often confronted with the fact that the $\beta$ function cannot be easily derived. However, one may think of the RG scale $u$ as labeled by a new coordinate $r$ which runs continuously from the original lattice scale $r=a$ to $r\to \infty$. Then in this higher dimensional space one may posit bulk fields $\Phi_i(x,u)$ such that $\Phi_i(x,r=a) = J_i(x,a)$, and the dynamics of $\Phi$ encode the $\beta$-function without its explicit calculation. Based on fairly general arguments (see e.g. Ref.~\cite{Adams2012}) the governing field theory for $\Phi_i(x,u)$ should be a theory of gravity which reduces to general relativity at low energy. Given this picture, the mapping from the gravity side to the quantum field theory shown in Tab.~\ref{table:holdict} taken from Ref.~\cite{Adams2012} is expected to hold. Furthermore, if the QFT is a conformal field theory (CFT), and obeys Lorentz invariance, the metric for the gravity dual must be asymptotically Anti-de-Sitter (AdS) in $d+1$ spacetime dimensions hence the term AdS/CFT correspondence. However, it is important to note that the AdS/CFT correspondence is still currently a conjectured rather than proven duality.
\begin{table}[t!]\label{table:holdict}
\begin{center}
\begin{tabular}{| l r c l r | }
  \hline
  Boundary QFT &~~~~~~~~&~~~~~~~~&~~~~&  Bulk Gravity \\
  \hline
  \hline
\vspace{-0.2cm}& & & & \\
  Operator 				&$\CO(x)$		&$\longleftrightarrow$	& $\Phi(x,r)$		&  Field \\
  Spin 					&$s_{\CO}$		&$\longleftrightarrow$	&$s_{\Phi}$ 		& Spin \\
  Global Charge 			&$q_{\CO}$		&$\longleftrightarrow$	&$q_{\Phi}$ 		& Gauge Charge \\
  Scaling dimension 			&$\Delta_{\CO}$	&$\longleftrightarrow$	&$m_{\Phi}$ 		& Mass \\
  Source  					&$J(x)$	 		&$\longleftrightarrow$	&$\Phi(x,r)|_{\p}$ 	& \!\!\!\! Boundary Value (B.V.) \\
  Expectation Value 			&$\vev{\CO(x)}$ 	&$\longleftrightarrow$	& $\Pi_{\Phi}(x,r)|_{\p}$ 	& \!\!\!\!\!\!\!\!B.V. of Radial Momentum \\
\vspace{-0.2cm}& & & & \\
  Global Symmetry Group		&$G$			&$\longleftrightarrow$	&$G$				&Gauge Symmetry Group\\
  Source for Global Current	&$\CA_{\m}(x)$		&$\longleftrightarrow$	& $A_{\m}(x,r)|_{\p}$ 	&B.V. of Gauge Field\\
  Expectation of Current		&$\vev{\CJ^{\m}(x)}$	&$\longleftrightarrow$	& $\Pi_{A}^{\m}(x,r)|_{\p}$ 	& B.V. of  Momentum\\
\vspace{-0.2cm}& & & & \\
  Stress Tensor 				&$T^{\m\n}(x)$	 	&$\longleftrightarrow$	& $g_{\m\n}(x,r)$		&Spacetime Metric\\
  Source for Stress-Energy 	&$h_{\m\n}(x)$		&$\longleftrightarrow$	& $g_{\m\n}(x,r)|_{\p}$ 	& B.V. of Metric\\
  Expected Stress-Energy	\!\!\!\!			&$\vev{T^{\m\n}(x)}$	&$\longleftrightarrow$	& $\Pi_{g}^{\m\n}(x,r)|_{\p}$ 	& B.V. of Momentum \\
\vspace{-0.2cm}& & & & \\
  \# of Degrees of Freedom 	&\multirow{2}{*}{$N^{2}$}	&\multirow{2}{*}{$\longleftrightarrow$}	&\multirow{2}{*}{$\({L\over \ell_{p}}\)^{d-1}$}	& Radius of Curvature \\
  Per Spacetime Point		&&&&  In Planck Units\\
\vspace{-0.2cm}& & & & \\
  Characteristic Strength	&\multirow{2}{*}{$\l$}	&\multirow{2}{*}{$\longleftrightarrow$}	& \multirow{2}{*}{$\({L\over\ell_{s}}\)^{d}$}	& Radius of Curvature \\
  of Interactions			&&&&  In String Units\\
\vspace{-0.2cm}& & & & \\
  \hline
  \hline
\vspace{-0.2cm}& & & & \\
  QFT Partition Function  	
  &\multirow{2}{*}{$Z_{\mathrm{QFT_{d}}}[J_{i}]$}
  &\multirow{2}{*}{$\longleftrightarrow$}
  &\multirow{2}{*}{$Z_{\mathrm{QG_{d\!+\!1}}}[\Phi_{i}[J_{i}]]$}\!\!\!\!\!\!\!\!
  & Quantum Gravity\\
   with Sources $J_{i}(x)$ &&&& (QG) Partition Function  \\
 	
  &&&&
  in \ads\ w/ $\Phi_{i}|_{\p}=J_{i}$ \\
\vspace{-0.2cm}& & & & \\
  QFT Partition Function
  &\multirow{2}{*}{$Z^{\l,N\gg1}_{\mathrm{QFT_{d}}}[J_{i}]$}
  &\multirow{2}{*}{$\longleftrightarrow$}
  &\multirow{2}{*}{$e^{-I_{\mathrm{GR_{d\!+\!1}}}[\Phi[J_{i}]]}$}\!\!\!\!\!\!\!\!
  & Classical GR Action\\
  at Strong Coupling
  &&&
  & in \ads\ w/ $\Phi_{i}|_{\p}=J_{i}$\\
\vspace{-0.2cm}& & & & \\
  QFT $n$-Point    	&	
  \!\!\!\!\!\!\!\!\!\!\!\!\!\!\!\!\!\!\!\!\!\!\!\!\!
  \multirow{3}{*}{$ \vev{\CO_{1}(x_{1}) \dots \CO_{n}(x_{n})}$}	&&
  \multirow{3}{*}{${\delta^{n}   I_{\mathrm{GR_{d\!+\!1}}}[\Phi[J_{i}]]  \over \delta J_{1}(x_{1})\dots \delta J_{n}(x_{n})}{\Big |}_{J_{i}=0}$}
  \!\!\!\!\!\!\!\!\!\!\!\!\!\!\!\!\!\!\!\!\!\!\!\!\!
  & Classical Derivatives of \\
  Functions at  	& &$\longleftrightarrow$&& the On-Shell Classical \\
  Strong Coupling	& &			 	      && Gravitational Action \\
\vspace{-0.2cm}& & & & \\
\hline
\hline
\vspace{-0.2cm}& & & & \\
  Thermodynamic State 		&				&$\longleftrightarrow$	& 				& Black Hole  \\
  Temperature 				&$T$			&$\longleftrightarrow$	&$T_{H}$			& Hawking Temperature $\sim$ Mass \\
  Chemical Potential	 		&$\m$			&$\longleftrightarrow$	&$Q$			& Charge of Black Hole \\
  Free Energy 				&$F$			&$\longleftrightarrow$	&$I_{\rm GR}|_{\rm (on-shell)}$	& On-Shell Bulk Action \\
  Entropy 					&$S$			&$\longleftrightarrow$	&$A_{H}$ 		& Area of Horizon \\
& & & & \\
\hline
\end{tabular}
\end{center}
\caption{The holographic dictionary relating quantities of the boundary field theory to the bulk gravity dual. Not all arrows are equalities. For example, the temperature depends on both the charge and mass of the black hole. Table reproduced from Ref.~\cite{Adams2012}}
\end{table}

There are a few key points to note from the dictionary in Tab.~\ref{table:holdict} for the purposes of the next chapter. First, the presence of a black hole in the spacetime determines the thermodynamic state of the boundary field theory. Second, one may calculate for example $2-$point correlation functions of the stress-energy tensor, allowing access to transport coefficients such as $\eta/s$ where $s$ is the entropy density either by applying Kubo-relations or comparing to the Navier-Stokes result for the same quantity. Finally, the ringdown or quasi-normal mode spectrum of the black hole on the gravity side using an appropriate probe field for evaluating the spectrum gives the corresponding operator spectrum on the boundary theory side. This will be the program of the next chapter, where an approximate black hole dual is used to predict the non-hydrodynamic mode spectrum of a harmonically trapped strongly interacting Fermi gas.

\chapter{Non-hydrodynamic Modes from a Lifshitz Black Hole}
\label{fgd}
As indicated in Chap.~\ref{ggintro}, AdS/CFT provides a route to calculating transport quantities in a gravitational picture. For a number of black hole duals including the one considered in this chapter, one finds $\eta/s=\frac{1}{4\pi}\approx 0.079$ (recall $\hbar = 1 = k_B$). This is quantitatively similar to $(\eta/s)_{UFG}\approx 0.2-0.4$ in the cold Fermi gases in the regime of strong interactions \cite{CaoSci2011} as well as $(\eta/s)_{QGP}\approx0.1-0.2$ for the quark-gluon plasma \cite{Gale:2012rq} despite the extreme difference in temperature and pressure between these two systems (see Fig.~\ref{fig:SIQFs}). While precision determination of $\eta/s$ in both of these systems is still an active area of research  \cite{Bluhm2015,PhysRevA.75.043612,2011AnPhy.326..770E,PhysRevLett.109.020406,Joseph2015}, the minimum values of $\eta/s$ in these systems appear similar to well within a factor of 10.

It is possible that these quantitative similarities hint at (approximately) universal transport properties in a wide range of strongly interacting quantum systems as proposed in Ref.~\cite{BrewerPRL2015}. It is interesting to note in support of this that clean graphene \cite{Muller2009} and high $T_c$ superconductors \cite{Rameau2014} are also being explored as possibly having $\eta/s \sim 1/(4 \pi)$. In order to further investigate this idea, the quasi-normal modes of a Lifshitz black hole are used in this chapter to make predictions about non-hydrodynamic modes in strongly interacting Fermi gases. It is important to note, however, that despite recent progress \cite{Son2008,Balasubramanian2008,Bekaert2012}, no exact black hole dual is known for a strongly interacting Fermi gas. As a result it will be argued that a Lifshitz black hole provides a reasonable starting point for making experimentally verifiable predictions.

To begin, a review of the modes of interest, namely the breathing and quadrupole modes is provided. Following that, properties of the Lifshitz black hole are discussed, and assumptions needed to relate the black hole dual to experimentally realizable Fermi gases are stated. Finally, by utilizing experimental input, predictions are made for the non-hydrodynamic quadrupole and breathing modes in $d=2,3$ respectively.

\section{Modes of Interest}
\label{sec:osc} 

\begin{figure}[ht!]
\centering
\includegraphics[width=0.45\textwidth]{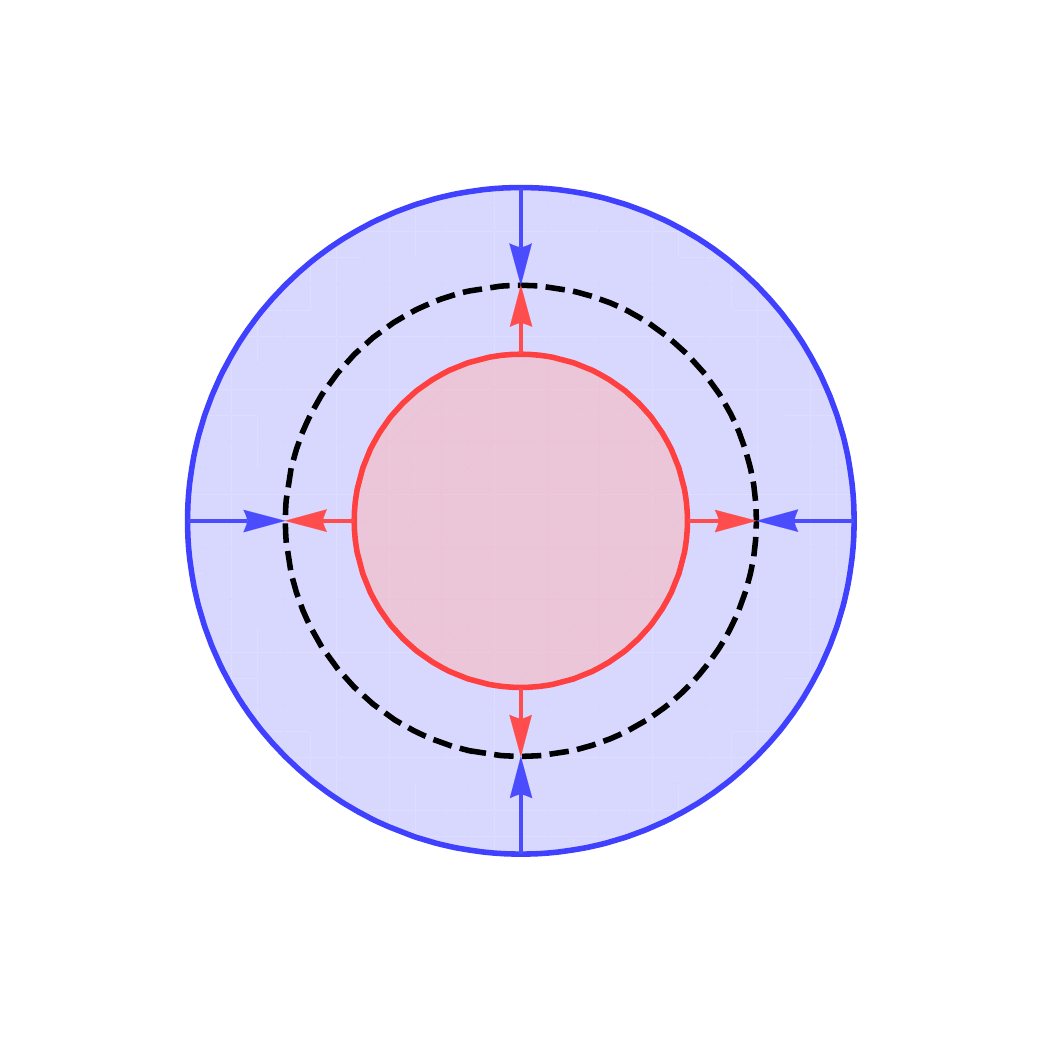}\hfill
\includegraphics[width=0.45\textwidth]{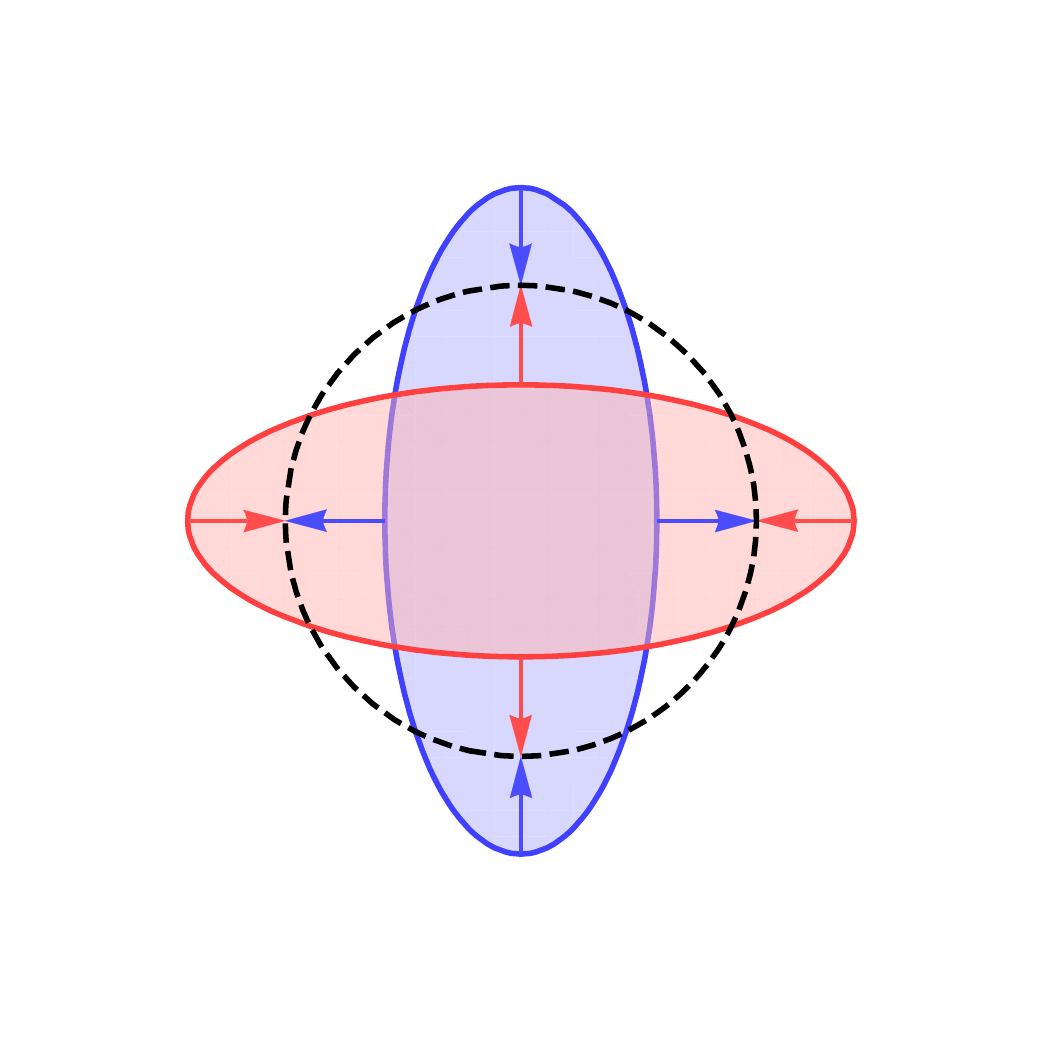}
\caption{\label{fig:one} Sketch of shape oscillations of an atomic gas cloud in the x-y plane:
breathing mode (left) and quadrupole mode (right). The breathing mode changes the overall cloud volume while the quadrupole mode corresponds to a surface deformation without volume change. The cloud's equilibrium configuration is indicated by the dashed circle.}
\end{figure}

The breathing mode is sketched in the left side of Fig. \ref{fig:one}. The frequency and damping of this mode may be extracted from experiment by considering the time dependence of the sum of the widths in along the x- and y-axes. The resulting time dependence for a gas in a harmonic potential and with small excitation amplitude may be found analytically as \cite{Book,BrewerPRA2016}:
\begin{equation}
B(t)\propto\cos(\omega_B t) e^{- \Gamma_B t},
\label{brea}
\end{equation}
where $\omega_{B}=\sqrt{10/3}\ \omega_\perp$ is the oscillation frequency and $\Gamma_{B}=\eta\, \omega_\perp^2/3 P$ is the damping rate in $d=3$ \cite{BrewerPRL2015,Massignan2005,Riedl2008} (see also the Chap.~\ref{sohiso}). In the expressions for $\omega$ and $\Gamma$, $P$ is the local equilibrium pressure and $\omega_\perp\equiv \sqrt{\omega_x \omega_y}$ is the average trap frequency in the x-y plane. The breathing mode in $d=2$ does not have a corresponding non-hydrodynamic mode, and will not be considered in this chapter. The quadrupole mode represented in the right side of Fig.~\ref{fig:one} is found by taking the difference in widths along the x- and y-directions resulting in a similar time dependence \cite{Book,BrewerPRA2016}:
\begin{equation}
Q(t)\propto\cos(\omega_Q t) e^{- \Gamma_Q t},
\label{qua}
\end{equation}
with $\omega_{Q}=\sqrt{2}\,\omega_\perp$, $\Gamma_{Q}=\eta\, \omega_\perp^2/P$. Note that, as in previous chapters of this work, a constant ratio $\eta/P$ has been assumed in order to derive these results. These hydrodynamic modes will be contrasted with their non-hydrodynamic counterparts in the coming sections.
 
\section{Black Hole in Asymptotically Lifshitz Space-Time}
\label{sec:string}
Here, a Lifshitz black hole dual is considered as it is capable of reproducing the equation of state ($\epsilon(P) = dP/2$ where $\epsilon$ is the energy density) of a strongly interacting Fermi gas. Particularly, Lifshitz black holes involve a scaling parameter $z$ which will specify the equation of state $\epsilon = d P/z$ \cite{Hoyos2013}. In order to match the strongly interacting Fermi gas, $z=2$ is chosen. The first key assumption in this chapter is that black holes in asymptotically Lifshitz spaces are capable of generally describing bulk features of strongly interacting Fermi gases. 

The quasi-normal mode spectrum may be evaluated using a probe scalar with scaling dimension $\Delta$. Specifically, the quasi-normal modes of a scalar field propagating in a fixed Lifshitz black brane background are computed as described in \cite{Sybesma:2015oha}. $\Delta$ is determined by the type of perturbation considered. For example, $\Delta=d$ corresponds to density perturbations (fermionic bilinears), while $\Delta=d+1$ is the scaling dimension for energy density perturbations (fermionic bilinears with a gradient). However, the probe field used in the above setup is scalar rather than a fermionic bilinear, and hence it is unclear what operator scaling dimension should be used. Here  $d \leq \Delta \leq d+1$ is assumed reasonable, where final results will include estimates of systematic error based on the mean and difference from $\Delta=d$ and $\Delta=d+1$. 

For $d=z$ analytic expressions for the quasi-normal modes are available \cite{Sybesma:2015oha}, and may be applied here for $d=2$ giving
\begin{equation}
\omega^{(d=2)}_n=0\,,\quad \Gamma_n^{(d=2)}= \left(n-1+\frac{\Delta}{2 z} \right)\times 4 \pi T\,,\quad n\geq 1\,.
\label{eq:one}
\end{equation}
For $d=3$, results were found numerically following the approach in Ref.~\cite{Starinets:2002br} and are collected in Tab. \ref{tab:one}. 

\begin{table}[ht]
\centering
\begin{tabular}{|c|c|c|c|c|}
\hline
& \multicolumn{2}{c|}{$\Delta=3$} & \multicolumn{2}{c|}{$\Delta=4$}\\ \hline
$n$ & $\omega_n/(4\pi T)$ & $\Gamma_n /(4\pi T)$ & $\omega_n /(4\pi  T)$ & $\Gamma_n /(4\pi T)$\\ \hline
1 & 0.2812 & 0.5282 & 0.3560 & 0.7540 \\ \hline
2 & 0.5776 & 1.437 & 0.6507 & 1.663 \\ \hline
3 & 0.8714  & 2.342 & 0.9446 & 2.568 \\ \hline
4 & 1.165  & 3.246 & 1.239 & 3.472 \\ \hline
\end{tabular}
\caption{\label{tab:one} Numerical results for frequencies and damping rates for ring-down frequencies for $d=3$ and two choices of $\Delta$. }
\end{table}

Eq.~\eqref{eq:one} and the results in Tab. \ref{tab:one} correspond to quasi-normal modes for a liquid of temperature $T$, vanishing chemical potential, and $\eta/s=1/4\pi$ \cite{Taylor2016}. However, real strongly interacting Fermi gases may have considerable chemical potential \cite{Horikoshi442} and shear viscosity over entropy ratios different from $\eta/s=\frac{1}{4 \pi}$. Kinetic theory provides guidance when attempting to connect the above results to real Fermi gases. In particular, kinetic theory exhibits a single non-hydrodynamic mode with a damping rate $\Gamma_1=1/\tau_R$ with $\tau_R$ obeys $\tau_R \propto \eta/P$ in the hydrodynamic limit \cite{BrewerPRL2015}. In order to bring the results into this form, one may perform three steps:
\begin{itemize}
\item Make the replacement $4\pi T \rightarrow s T/\eta$ which holds trivially for a Lifshitz black hole.
\item Utilize the thermodynamic identity $s T = (\epsilon+P)$ for a fluid with finite temperature and zero chemical potential to remove reference to entropy density.
\item Use the equation of state $\epsilon = dP/2$ to replace the energy density so that only shear viscosity and pressure remain.
\end{itemize}
These three steps may be summarized with the replacement
\begin{equation}
\label{eq:myst}
4\pi T \rightarrow \frac{(1+d/2)P}{\eta}\,. 
\end{equation}
In this form, experimental or theoretical values of $\eta/P$ may be used to make predictions for the non-hydrodynamic mode frequencies and damping rates.may
An important final point before making these predictions is that, the above calculations are for the case of an untrapped Fermi gas. However, most experiments utilize approximately harmonic optical trap setups. To relate the above untrapped results to the case of a trapping potential with average trapping frequency $\omega_\perp$, again guidance from kinetic theory is employed. In kinetic theory, hydrodynamic mode oscillations change qualitatively between a free system and a system placed in a trap, but the non-hydrodynamic modes do not \cite{BrewerPRL2015}. This observation also holds for the within second-order hydrodynamics. As seen in previous chapters, the non-hydrodynamic mode always scales as $1/\tau_\pi$ in the hydrodynamic limit. Based on these observations, the non-hydrodynamic mode frequencies and damping rates from the black hole dual calculation of an untrapped system above are assumed to hold for the case of a trapped Fermi gas.

\subsection{Assumptions in Applying to a Strongly Interacting Fermi Gas}

For clarity, a summary of the assumptions made in this chapter are:
\begin{itemize}
\item
Black holes in asymptotic Lifshitz spaces describe the bulk features of a strongly interacting Fermi gas. Particularly, the equation of state is known to be recovered.
\item
A probe scalar with dimension $\Delta\simeq d+\frac{1}{2}$ approximately describes density perturbations in the strongly interacting Fermi gas. This is required since scalar rather than fermionic operators were used to probe the black hole spectrum.
\item
Strongly interacting Fermi gases at non-zero density and $\frac{\eta }{s}\neq \frac{1}{4 \pi}$ are well approximated by the calculation for a Lifshitz black hole done at zero density and $\frac{\eta }{s}= \frac{1}{4 \pi}$ when performing the \hbox {replacement (\ref{eq:myst})}
\item
The frequencies and damping rates of non-hydrodynamic modes do not differ between the untrapped and trapped Fermi gas. This is well supported by both kinetic theory and second-order hydrodynamics.
\end{itemize}

These assumptions can in practice be tested and, in most cases, lifted by performing more general calculations. However, such calculations are left for future work.

\section{Quasinormal Modes of a Lifshitz Black Hole}
\label{sec:res}

Density perturbations in a trapped, strongly interacting Fermi gas are now considered with the framework discussed above. Black holes exhibit both hydrodynamic modes as well as an infinite set of non-hydrodynamic collective modes.  Provided the black hole dual calculation is applicable, Eqs. \eqref{brea} and \eqref{qua} should take the more general form 
\begin{equation}
\label{eq:qnmform}
H(t)=\alpha_H\cos(\omega_H t+\phi_H) e^{- \Gamma_H t}+\sum_{n=1}^\infty \alpha_n \cos(\omega_n t+\phi_n) e^{- \Gamma_n t}\,,
\end{equation} 
where $H=B,Q$ depending on the shape oscillation considered. Note that the first term is the usual hydrodynamic mode, $\alpha$'s are mode amplitudes, and $\phi_H,\phi_n$ are phase shifts. The results from the Lifshitz black hole in Sec. \ref{sec:string} are recast into experimentally accessible quantities in Tab. \ref{tab:two}. Note that for two dimensions, the result can be obtained analytically for all $n$, and one finds $\omega_n=0$ and $\Gamma_n=\left(2n-3/4\pm1/4\right)P/\eta$.

\begin{table}[ht]
\centering
\begin{tabular}{|c|c|c|c|c|}
\hline
& \multicolumn{2}{c|}{$d=2$} & \multicolumn{2}{c|}{$d=3$}\\ \hline
$n$ & $\omega_n\times\eta/P$ & $\Gamma_n\times \eta/P$ & $\omega_n\times \eta/P$ & $\Gamma_n\times \eta/P$\\ \hline
1 & 0 & $1.25(25) $ & $0.8(1) $ &  $1.6(3) $\\
\hline
2 & 0 & $3.25(25) $ & $1.5(1) $ &  $3.9(3) $\\
\hline
3 & 0 & $5.25(25) $ & $2.3(1) $ &  $6.1(3) $\\
\hline
4 & 0 & $7.25(25) $ & $3.0(1) $ &  $8.4(3) $\\
\hline
\end{tabular}
\caption{\label{tab:two} Numerical values for the frequencies and damping rates of the first $n\leq 4$ non-hydrodynamic modes in $d=2$ and $d=3$ dimensions, obtained from a string theory based calculation. Results are expressed in terms of the ratio of pressure $P$ to shear viscosity $\eta$. Note that $P/\eta$ can be re-expressed in terms of the  damping rates $\Gamma_Q,\Gamma_B$ of the hydrodynamic quadrupole and breathing modes for a strongly interacting Fermi gas in a trap.}
\end{table}

Note that, unlike the hydrodynamic component, the frequencies $\omega_n$ and damping rates $\Gamma_n$ of the non-hydrodynamic modes turn out to be independent of the cloud's average trapping frequency $\omega_\perp$. (Note that this is also the case for the non-hydrodynamic modes in second-order hydrodynamics where in dimensionful units $\Gamma \approx 1/\tau_\pi$ which is independent of $\omega_\perp$.) While the time-dependence of these modes are quite different from their hydrodynamic counterparts, the spatial oscillation structure of the non-hydrodynamic modes is exactly the same. Furthermore, Tab. \ref{tab:two} demonstrates that the non-hydrodynamic modes have  greater damping than oscillation frequency. Testing the novel prediction of an infinite set of non-hydrodynamic modes for strongly interacting Fermi gases would help support the conjecture of transport universality mentioned in the introduction. The different time-dependence exhibited by the non-hydrodynamic modes offers an experimental handle to distinguish these modes from hydrodynamic oscillations (c.f. Chaps.\ref{sohiso}-\ref{sohunif}).

\begin{figure}[ht]
\centering
\includegraphics[width=0.45\textwidth]{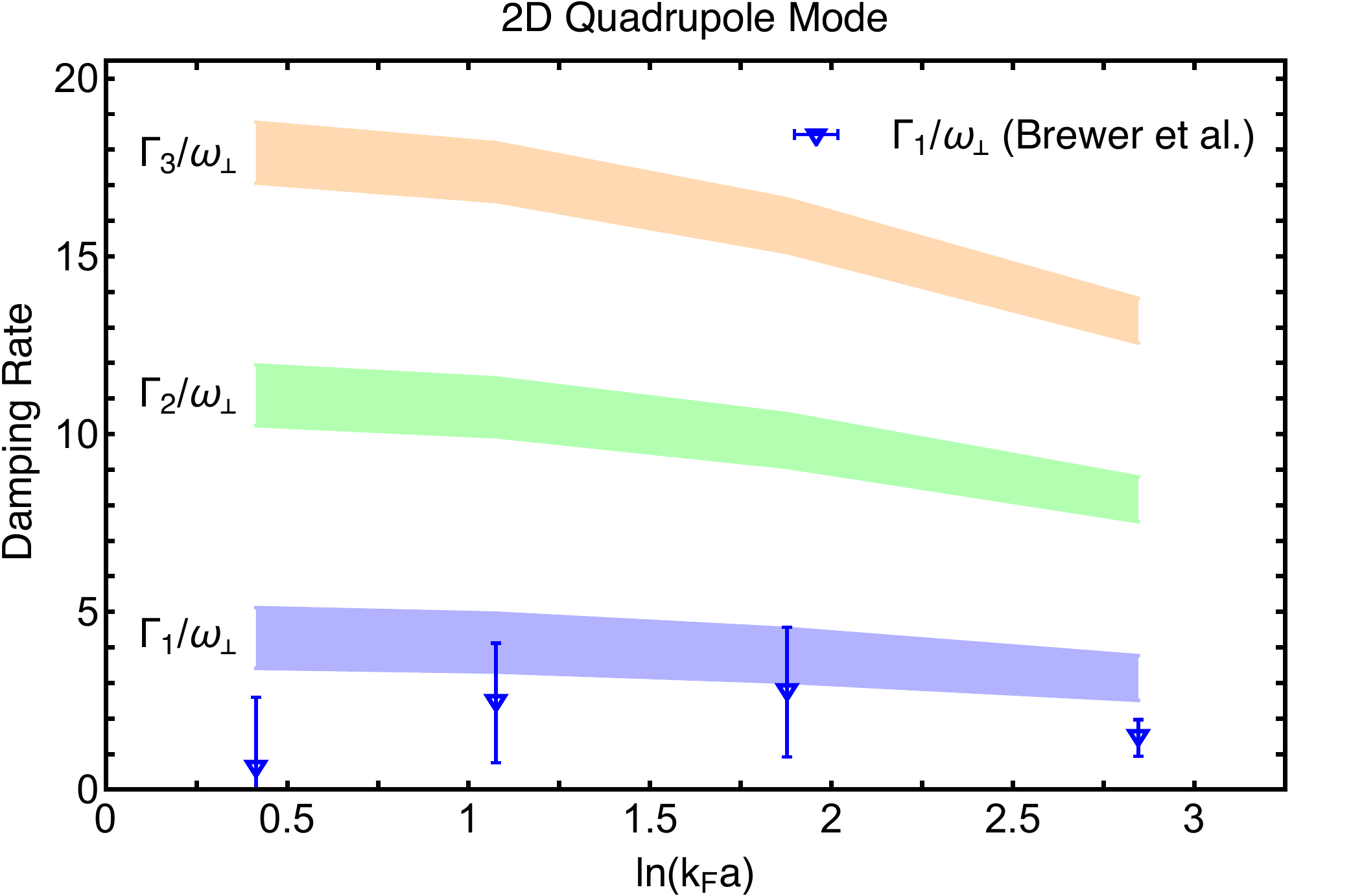}\hfill
\includegraphics[width=0.45\textwidth]{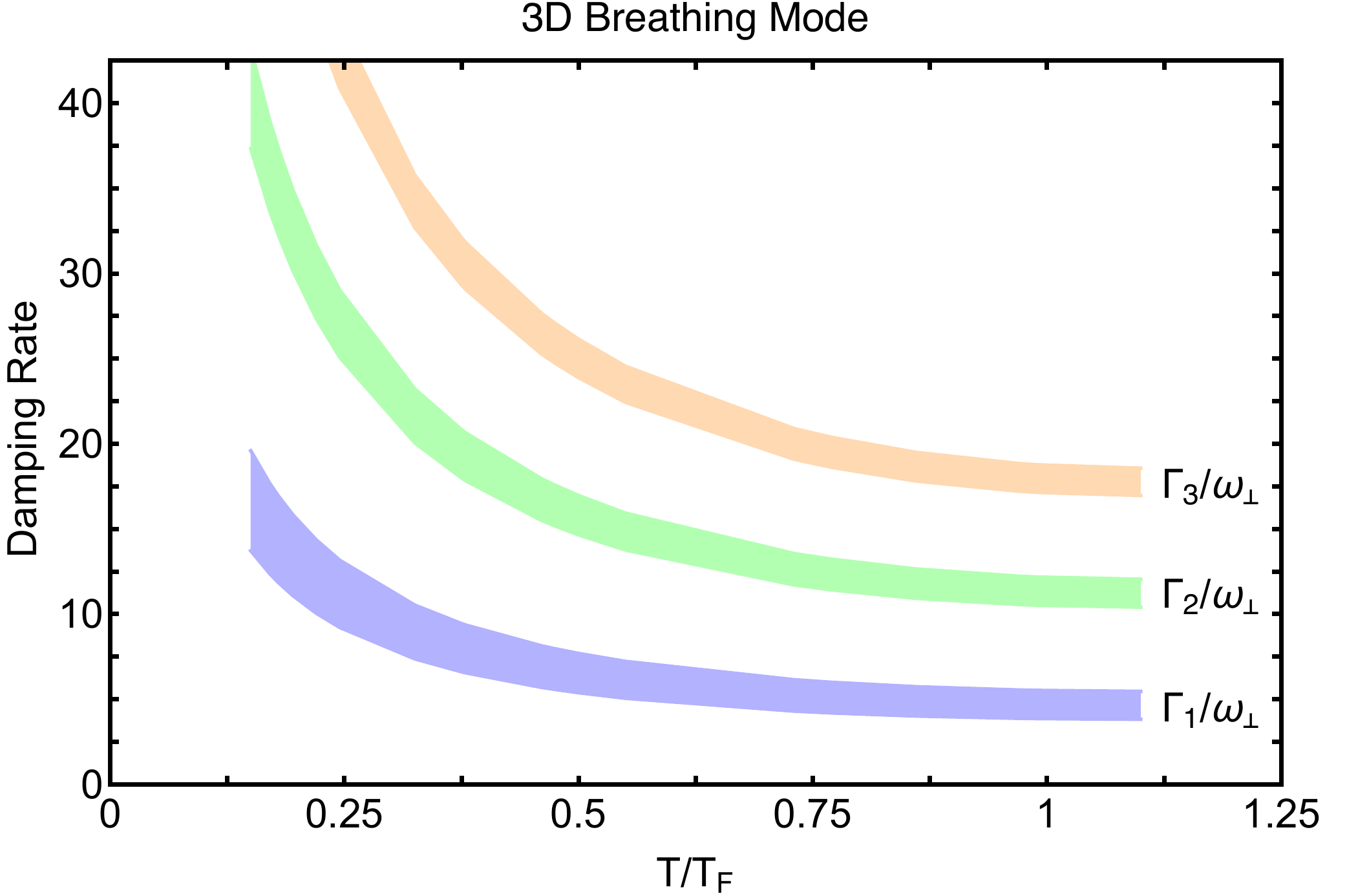}
\caption{\label{fig:two} Predicted non-hydrodynamic mode damping rates $\Gamma_{n}/\omega_\perp$ (bands).  Left: two-dimensional quadrupole mode as a function of interaction strength $\ln (k_F a)$, where $a$ is the two-dimensional s-wave scattering length and $k_F$ is the Fermi momentum  related to the density $n$ at the cloud's center as $n=k_F^2/2\pi$. Also shown is the first non-hydrodynamic mode damping rate $\Gamma_1$ extracted from experimental data \cite{BrewerPRL2015}.  Right: three-dimensional breathing mode as a function of cloud's temperature in units of the Fermi temperature $T_F=\left(3 N \omega_x \omega_y \omega_z\right)^{1/3}\hbar /k_B$ where $N\simeq 5\times 10^5$ is the number of atoms in the three-dimensional optical trap.} 
\end{figure}

For direct comparison to experiment, the results from Tab. \ref{tab:two} are converted into predictions in Fig.~\ref{fig:two} for damping rates. These predictions are calculated by employing the following two steps:
\begin{itemize}
\item First, the relations $\Gamma_{Q}=\eta\, \omega_\perp^2/P$ and $\Gamma_{B}=\eta\, \omega_\perp^2/3 P$ discussed in Sec. \ref{sec:osc} are used to rewrite the expressions in Tab. \ref{tab:two} in terms of the relevant damping rate rather than $\eta/P$.
\item Second experimentally measured damping rates $\Gamma_Q,\Gamma_B$ \cite{VogtPRL2012,KinastPRL2005,Riedl2008} are substituted in order to derive a numerical value (up to the trap frequency normalization factor).
\end{itemize}
For example, $\Gamma_1=1.25(25) \frac{P}{\eta}$ for the $d=2$ quadrupole mode from Tab. \ref{tab:two} becomes $\Gamma_1=1.25(25) \Gamma_Q^{-1} \omega_\perp^{2}$ which becomes $\Gamma_1\simeq 4.17(83)\omega_\perp$ when using the experimentally determined value of $\Gamma_Q\simeq 0.30\omega_\perp$ from Ref.~\cite{VogtPRL2012} at $\log(k_F a)\simeq 1.04$. In Fig.~\ref{fig:two}, predictions for non-hydrodynamic mode damping rates $\Gamma_n/\omega_\perp$ are shown in the case of the two-dimensional quadrupole mode  and the three dimensional breathing mode, given different choices of temperature and atom interaction strength. Fig.~\ref{fig:two} also shows the only published constraints on the two-dimensional quadrupole mode damping rate $\Gamma_1$ extracted from experimental data \cite{BrewerPRL2015}. In order to extract the non-hydrodynamic damping, Ref.~\cite{BrewerPRL2015} re-analyzed existing experimental data from Ref.~\cite{VogtPRL2012} for the quadrupole mode in $d=2$ using the form (\ref{eq:qnmform}) with $\alpha_n=0$ for $n>1$ (i.e. only a single non-hydrodynamic mode). With the time-resolution and number of data-sets obtained in the experiment, the re-analysis of Ref.~\cite{BrewerPRL2015} found evidence of the first non-hydrodynamic mode, albeit with low statistical significance. However, this is expected since the non-hydrodynamic contribution to $Q(t)$ in Eq.~(\ref{eq:qnmform}) decays quickly in comparison to the hydrodynamic mode. In Ref.~\cite{VogtPRL2012} the primary goal was the extraction of the hydrodynamic frequency and damping. Accordingly, the time-sampling rate chosen in Ref.~\cite{VogtPRL2012} was not optimal to extract early-time information with high statistical significance. The statistical significance of the evidence for a non-hydrodynamic component could likely be improved if the experiment in Ref.~\cite{VogtPRL2012} were repeated to yield 100 data sets with time-resolution increased by a factor of 20. Regardless, it is promising that the experimental constraints on $\Gamma_1$ obtained in Ref.~\cite{BrewerPRL2015} are broadly consistent with the black hole dual predictions.

Finally, while it was possible to calculate excitation amplitudes of non-hydrodynamic modes in the second-order hydrodynamics framework, the black hole dual calculation does note provide such information. As an estimate, the ratio of the first non-hydrodynamic to hydrodynamic mode amplitude $\alpha_n/\alpha_H$ can be found by assuming random phase shifts $\phi_H,\phi_n$ whereby $|\frac{\alpha_n}{\alpha_H}|\simeq (\omega_H+\Gamma_H)/(\omega_n+\Gamma_n)$. Evaluating for the first non-hydrodynamic quadrupole mode one finds $|\frac{\alpha_1}{\alpha_H}|\simeq 40 \%$ while for the first non-hydrodynamic breathing mode in $d=3$ $|\frac{\alpha_1}{\alpha_H}|\simeq 20\%$. These estimates are roughly consistent with the calculations of Chap.~\ref{sohiso} indicating that the first non-hydrodynamic modes should be within the experimental detection capabilities of present experiments.

\chapter{Conclusion}
\label{conclusion}

\section{Summary and Outlook}
In this dissertation, the applicability of hydrodynamics to cold quantum gases near a Feshbach resonance was discussed. Shortcomings of the Navier-Stokes formalism for studying transport properties such as shear viscosity were highlighted. In order to address this issue, the framework of second-order hydrodynamics was introduced. Second-order hydrodynamics was applied in the cases of a isotropic and anisotropic harmonic trapping potentials as well as for a uniform gas leading to proposals for improving precision extraction of transport coefficients as well as suggestions for observing non-hydrodynamic modes linked to pre-hydrodynamization physics on timescales $t <\tau_\pi$. Finally, the ideas of transport universality and non-hydrodynamic modes were explored within the framework of holographic duality. The key findings of this work may be summarized as:
\begin{itemize}
\item For an isotropically trapped gas, the dependence of hydrodynamic collective mode damping of the volume conserving modes is sensitive to the mode order $w$ (see App.~\ref{App1}, Figs.~\ref{fig:2dfreq}, \ref{3dfreq} and Tabs.~\ref{2dfreqtab}, \ref{3dfreqtab}). Particularly, the damping $\Gamma \approx w \eta/P$. 
\item For an isotropically trapped gas, theoretical and experimental observations indicate that the non-hydrodynamic modes may be excited with amplitudes significant enough for direct observation using a trap quench excitation (see Figs.~\ref{fig:anisovseta},\ref{fig:3diso}).
\item To leading order, non-hydrodynamic modes are insensitive to trap geometry for isotropic, anisotropic, and uniform gases (see Figs.~\ref{2dnhdamp}, \ref{3dnhdamp}, \ref{fig:2danissciss}, and \ref{fig:3danissciss}, Tabs.~\ref{2dfreqtab}, \ref{3dfreqtab}, \ref{2dfreqtabanis}, and \ref{3dfreqtabanis}, and Eqs.~\eqref{shearchannel} and \eqref{soundnh}). This fact is important, for example, in the interpretation of results in Chap.~\ref{fgd}.
\item The damping rate for the quasi-breathing mode in $d=3$ is much more sensitive to trap anisotropy than its non-hydrodynamic counterpart (see Fig.~\ref{fig:aniseff3d}). It is proposed that this fact might be utilized in experiments to tease out the non-hydrodynamic behavior.
\item Non-hydrodynamic damping of both the shear and sound channels in a uniform gas were calculated (see Fig.~\ref{fig:shearunif} as well as Eqs.~\eqref{shearchannel} and \eqref{soundnh}).
\item Calculations of non-hydrodynamic behavior from an approximate black hole dual show that while the results of second-order hydrodynamics provide good guidance, more generally an infinite tower of non-hydrodynamic modes which are not necessarily purely damped may exists (see Fig.~\ref{fig:two} and Tab.~\ref{tab:two}).
\item If experimental observations consistent with the previous point are made, this would provide strong evidence for the existence of approximately universal transport behavior in strongly interacting quantum systems.
\end{itemize}
These results indicate many interesting avenues for further theoretical or experimental investigation:
 \begin{itemize}
\item  The observation that the hydrodynamic damping increases with mode number $\Gamma \approx w \eta/P$ is similar to the higher sensitivity of $\eta/s$ to higher-order collective flows in the quark gluon plasma leading to a better constraint on $\eta/s$ \cite{Schenke2012}. Would measurements of the damping of higher-order collective modes (or expansions) in atomic gases similarly lead to better constraints on $\eta/P$?
\item Can non-hydrodynamic modes be excited more effectively by excitations other than a trap quench?
\item Can the insensitivity of non-hydrodynamic modes to trap geometry be exploited to separate out hydrodynamic and non-hydrodynamic behavior? 
\item Can the experimental techniques of Ref.~\cite{Zwierlein2017} be adapted to measure non-hydrodynamic behavior for perturbations about a fluid with uniform density and temperature?
\item Evidence for the first non-hydrodynamic modes have been found in Ref.~\cite{BrewerPRL2015}. Can the statistical significance of this result be improved by additional experiments with better time resolution at early times?
\item The random phase amplitude calculation of $|\frac{\alpha_n}{\alpha_H}|$ at the end of Chap.~\ref{fgd} give $|\frac{\alpha_2}{\alpha_H}|\simeq 15 \%$ for the second non-hydrodynamic quadrupole mode while for the second non-hydrodynamic breathing mode in $d=3$ $|\frac{\alpha_1}{\alpha_H}|\simeq 10\%$. Can these modes be experimentally detected in support of the prediction of an infinite tower of non-hydrodynamic modes provided by the Lifshitz black hole dual?
\end{itemize}
In addition to the questions listed above, it would be interesting from a theoretical standpoint to consider the effects of lifting the various approximations and assumptions made throughout this work. It is hoped that this work will inspire further progress in this field. 

\bibliographystyle{unsrt}	
\nocite{*}		
\bibliography{refs}		

\appendix
\chapter{Spatial Mode Structure for $d=2$}
\label{App1}

In this appendix the spatial mode structures for the low lying modes in $d=2$ are collected (see Tab.~\ref{tab:spatialstruct2d}). It is interesting to note that, of the modes found, only the temperature mode and breathing mode are associated with a non-zero value of $\delta T$. This is because they are the only two modes which change the volume of the cloud, and hence can lead to heating and cooling of the gas. Also note that each irrotational volume conserving mode (here: dipole, quadrupole, hexapole, octupole, and decapole modes) has two independent realizations related by an appropriate coordinate transformation. For example, the quadrupole mode and tilted quadrupole mode are related by a rotation of $45^o$. In general, we can assign a mode the winding number $w$ of the associated velocity field $\mathbf{u}$ on a circle centered on the origin. This quantity merely counts the number of full rotations made when following a vector around the prescribed circle. For example, $w_{Dip}=0$, $w_{Quad}=1$, $w_{Hex}=2$,... so that the angle of coordinate rotations to get the second independent mode for a given mode is conveniently given by
	\begin{equation}
	\label{rotang}
		\Delta \phi = \frac{\pi}{2 (w+1)}.
	\end{equation}
Of course, any rotation through an angle in the range $\Delta \theta \in (0,\pi/(w+1))$ will produce an equally valid independent mode, but the angle in Eq.~\eqref{rotang} provides a uniform approach to finding an independent mode from one already found. This provides a connection for example to the approach in Ref.~\cite{BrewerPRL2015} where inequivalent polynomials under rotation up to quadrupole mode were considered in the second-order hydrodynamics framework used here.

\begin{table*}[h]
\begin{center}
\adjustbox{max width=\textwidth}{
\begin{tabular}{| c | c | c | c | c |}
  \hline
  \phantom{...} & $\delta\rho$ & $\delta u_x$ & $\delta u_y$ & $\delta T$\\
  \hline				
  Number (Zero Mode) & $1$ & $0$ & $0$ & $0$\\
  \hline			
    Temp. (Zero Mode) & $x^2+y^2-2$ & $0$ & $0$ & $2$\\
  \hline		
  Rotation (Zero Mode) & $0$ & $y$ & $-x$ & $0$\\
  \hline			
  Breathing (Monopole) & $x^2+y^2-2$ & $-i x \widetilde{\omega}_B $ & $- i y \widetilde{\omega}_B$ & $-2$\\
  \hline
  x-axis Sloshing (Dipole) & $x$ & $-i\widetilde{\omega}_S$ & 0 & 0 \\
  \hline
  y-axis Sloshing (Dipole) & $y$ & 0 & $-i\widetilde{\omega}_S$ & 0 \\
  \hline
  Quadrupole & $y^2-x^2$ & $i x \widetilde{\omega}_Q$ & $-i y \widetilde{\omega}_Q$ & 0\\
  \hline
  Tilted Quadrupole & $xy$ & $ -i \frac{y}{2} \widetilde{\omega}_Q$ & $-i 
 \frac{x}{2}\widetilde{\omega}_Q$ & 0\\
  \hline
  Hexapole & $y^3-3x^2y$ & $2ixy\widetilde{\omega}_H$  & $-i(y^2-x^2)\widetilde{\omega}_H$ & 0\\
    \hline
  T-Hexapole & $-\frac{x^3}{3}+xy^2$ & $-i\frac{(y^2-x^2)}{3}\widetilde{\omega}_H$  & $-i\frac{2xy}{3}\widetilde{\omega}_H$ & 0\\
  \hline
  Octupole & $x^4-6x^2y^2+y^4$ & $3i(-\frac{x^3}{3}+xy^2)\widetilde{\omega}_O$  & $-i(y^3-3x^2y)\widetilde{\omega}_O$ & 0\\
  \hline
    T-Octupole & $xy^3-x^3y$ & $i\frac{(3x^2y-y^3)}{4}\widetilde{\omega}_O$  & $i\frac{(x^3-3xy^2)}{4}\widetilde{\omega}_O$ & 0\\
  \hline
  Decapole & $y^5-10x^2y^3+5x^4y$ & $4i(xy^3-x^3y)\widetilde{\omega}_D$  & $-i(y^4-6x^2y^2+x^4)\widetilde{\omega}_D$ & 0\\
  \hline  
    T-Decapole & $\frac{x^5}{5}-2x^3y^2+xy^4$ & $-i\frac{(y^4-6x^2y^2+x^4)}{5}\widetilde{\omega}_D$  & $-4i\frac{(xy^3-x^3y)}{5}\widetilde{\omega}_D$ & 0\\
  \hline 
  Non-hydro Quad.& $y^2-x^2$ & $i x \widetilde{\omega}^{nh}_Q$ & $-i y \widetilde{\omega}^{nh}_Q$ & 0\\
    \hline  
  T-Non-hydro Quad.& $xy$ & $ -i \frac{y}{2} \widetilde{\omega}^{nh}_Q$ & $-i 
 \frac{x}{2}\widetilde{\omega}^{nh}_Q$ & 0\\
   \hline 
  Non-hydro Hex.& $y^3-3x^2y$ & $2ixy\widetilde{\omega}^{nh}_H$  & $-i(y^2-x^2)\widetilde{\omega}^{nh}_H$ & 0\\
     \hline 
  Non-hydro Oct.& $x^4-6x^2y^2+y^4$ & $3i(-\frac{x^3}{3}+xy^2)\widetilde{\omega}^{nh}_O$  & $-i(y^3-3x^2y)\widetilde{\omega}^{nh}_O$ & 0\\
  \hline 
  Non-hydro Dec.& $y^5-10x^2y^3+5x^4y$ & $4i(xy^3-x^3y)\widetilde{\omega}^{nh}_D$  & $-i(y^4-6x^2y^2+x^4)\widetilde{\omega}^{nh}_D$ & 0\\
  \hline
\end{tabular}
}
\end{center}
  \caption{Spatial structure of the various modes for $d=2$ expressed in terms of the normalized complex mode frequencies. Note that the tilted modes denoted Tilted- or T- for short in the table can be found by an appropriate rotation of coordinates.
}
	\label{tab:spatialstruct2d}
\end{table*}

\chapter{Spatial Mode Structure for $d=3$}
\label{App2}

The spatial structure of modes in $d=3$ are given in Tab.~\ref{tab:spatialstruct3d}. Results are very similar to those for $d=2$ shown in App. \ref{App1}. The only differences are that for a harmonic trapping potential which is translationally invariant along one axis and isotropic along the other two in $d=3$, the breathing mode now couples to shear stresses. Hence there is now an associated non-hydrodynamic breathing mode as well as a difference in the corresponding temperature perturbation associated with the volume change of the cloud. Since there is no associated velocity field for the time independent temperature zero mode, this mode is associated with a vanishing stress tensor, and hence the mode structure is the same as in the $d=2$ case. All other modes also exhibit the same spatial and frequency structure.

\begin{table*}[h]
\begin{center}
\adjustbox{max width=\textwidth}{
\begin{tabular}{| c | c | c | c | c |}
  \hline
  \phantom{...} & $\delta\rho$ & $\delta u_x$ & $\delta u_y$ & $\delta T$\\
  \hline				
  Number (Zero Mode) & $1$ & $0$ & $0$ & $0$\\
  \hline			
    Temp. (Zero Mode) & $x^2+y^2-2$ & $0$ & $0$ & $2$\\
  \hline		
  Rotation (Zero Mode) & $0$ & $y$ & $-x$ & $0$\\
  \hline			
  Breathing (Monopole) & $x^2+y^2-2$ & $-i x \widetilde{\omega}_B $ & $- i y \widetilde{\omega}_B$ & $-\frac{4}{3}$\\
  \hline
  x-axis Sloshing (Dipole) & $x$ & $-i\widetilde{\omega}_S$ & 0 & 0 \\
  \hline
  y-axis Sloshing (Dipole) & $y$ & 0 & $-i\widetilde{\omega}_S$ & 0 \\
  \hline
  Quadrupole & $y^2-x^2$ & $i x \widetilde{\omega}_Q$ & $-i y \widetilde{\omega}_Q$ & 0\\
  \hline
  Tilted Quadrupole & $xy$ & $ -i \frac{y}{2} \widetilde{\omega}_Q$ & $-i 
 \frac{x}{2}\widetilde{\omega}_Q$ & 0\\
  \hline
  Hexapole & $y^3-3x^2y$ & $2ixy\widetilde{\omega}_H$  & $-i(y^2-x^2)\widetilde{\omega}_H$ & 0\\
    \hline
  T-Hexapole & $-\frac{x^3}{3}+xy^2$ & $-i\frac{(y^2-x^2)}{3}\widetilde{\omega}_H$  & $-i\frac{2xy}{3}\widetilde{\omega}_H$ & 0\\
  \hline
  Octupole & $x^4-6x^2y^2+y^4$ & $3i(-\frac{x^3}{3}+xy^2)\widetilde{\omega}_O$  & $-i(y^3-3x^2y)\widetilde{\omega}_O$ & 0\\
  \hline
    T-Octupole & $xy^3-x^3y$ & $i\frac{(3x^2y-y^3)}{4}\widetilde{\omega}_O$  & $i\frac{(x^3-3xy^2)}{4}\widetilde{\omega}_O$ & 0\\
  \hline
  Decapole & $y^5-10x^2y^3+5x^4y$ & $4i(xy^3-x^3y)\widetilde{\omega}_D$  & $-i(y^4-6x^2y^2+x^4)\widetilde{\omega}_D$ & 0\\
  \hline  
    T-Decapole & $\frac{x^5}{5}-2x^3y^2+xy^4$ & $-i\frac{(y^4-6x^2y^2+x^4)}{5}\widetilde{\omega}_D$  & $-4i\frac{(xy^3-x^3y)}{5}\widetilde{\omega}_D$ & 0\\
      \hline			
 Non-hydro Breath.& $x^2+y^2-2$ & $-i x \widetilde{\omega}^{nh}_B $ & $- i y \widetilde{\omega}^{nh}_B$ & $-\frac{4}{3}$\\
  \hline 
  Non-hydro Quad.& $y^2-x^2$ & $i x \widetilde{\omega}^{nh}_Q$ & $-i y \widetilde{\omega}^{nh}_Q$ & 0\\
    \hline  
  T-Non-hydro Quad.& $xy$ & $ -i \frac{y}{2} \widetilde{\omega}^{nh}_Q$ & $-i 
 \frac{x}{2}\widetilde{\omega}^{nh}_Q$ & 0\\
   \hline 
  Non-hydro Hex.& $y^3-3x^2y$ & $2ixy\widetilde{\omega}^{nh}_H$  & $-i(y^2-x^2)\widetilde{\omega}^{nh}_H$ & 0\\
     \hline 
  Non-hydro Oct.& $x^4-6x^2y^2+y^4$ & $3i(-\frac{x^3}{3}+xy^2)\widetilde{\omega}^{nh}_O$  & $-i(y^3-3x^2y)\widetilde{\omega}^{nh}_O$ & 0\\
  \hline 
  Non-hydro Dec.& $y^5-10x^2y^3+5x^4y$ & $4i(xy^3-x^3y)\widetilde{\omega}^{nh}_D$  & $-i(y^4-6x^2y^2+x^4)\widetilde{\omega}^{nh}_D$ & 0\\
  \hline
\end{tabular}
}
\end{center}
  \caption{Spatial structure of the various modes for $d=3$ expressed in terms of the normalized complex mode frequencies. Note that the tilted modes denoted Tilted- or T- for short in the table can be found by an appropriate rotation of coordinates.
}
	\label{tab:spatialstruct3d}
\end{table*}
\chapter{Details of Mode Amplitude Calculation}
\label{App3}

Given generic initial conditions on $\rho$, $\mathbf{u}$, $T$, and $\pi_{ij}$, one can derive a system of equations for complex mode amplitudes $(a_n + i b_n)$ of mode $n$ (e.g. $n=$``number'', ``temperature'', ``breathing'', etc.) by performing the following projections onto a mode $m$:
\begin{align}
	\label{proj1}&\int_{\mathbb{R}^2} d^2\mathbf{x} \bigg[\rho_{\,init}(\mathbf{x}) -\rho_0(\mathbf{x}) \bigg]\delta \rho_m(\mathbf{x})  =\\
  \nonumber &\int_{\mathbb{R}^2} d^2\mathbf{x}   \rho_0(\mathbf{x}) \sum_{modes\, n}\mathcal{R}e\bigg[(a_n + i b_n) e^{-i \omega_n t}\delta \rho_n(\mathbf{x})\bigg]\delta \rho_m(\mathbf{x}),\\
	\label{proj2} &\int_{\mathbb{R}^2} d^2\mathbf{x} \bigg[u_{i \phantom{.} \,init}(\mathbf{x}) \bigg]\rho_0(\mathbf{x}) \delta u_{im}(\mathbf{x}) =\\
\nonumber&\int_{\mathbb{R}^2} d^2\mathbf{x} \sum_{modes\, n} \mathcal{R}e\bigg[(a_n + i b_n) e^{-i \omega_n t}\delta\mathbf{u}_n(\mathbf{x})\bigg]\rho_0(\mathbf{x}) \delta u_{im}(\mathbf{x}),\\
	\label{proj3}&\int_{\mathbb{R}^2} d^2\mathbf{x} \bigg[T_{\,init}-T_{0}\bigg] \rho_0(\mathbf{x}) \delta T_{m} =\\
  \nonumber&\int_{\mathbb{R}^2} d^2\mathbf{x} \sum_{modes\, n} \mathcal{R}e\bigg[(a_n + i b_n) e^{-i \omega_n t} \delta T_{n}\bigg]\rho_0(\mathbf{x})\delta T_{m},
	\end{align}
	\begin{align}
	\label{proj4}&\int_{\mathbb{R}^2} d^2\mathbf{x} \bigg[\pi_{ij \phantom{.} \,init}(\mathbf{x}) \bigg] \delta \pi_{ij\phantom{.}m}(\mathbf{x}) =\\
\nonumber &\int_{\mathbb{R}^2} d^2\mathbf{x} \sum_{modes\, n}\mathcal{R}e\bigg[(a_n + i b_n) e^{-i \omega_n t}\delta \pi_{ij\phantom{.}n}(\mathbf{x})\bigg] \delta \pi_{ij\phantom{.}m}(\mathbf{x}),\,.
	\end{align}
It should be noted that as can be seen from Tabs.\ref{tab:spatialstruct2d} and \ref{tab:spatialstruct3d}, not all of the individual perturbations are orthogonal (the full mode structures are, however, independent). For example, while $\delta\rho_{Temp} = \delta\rho_{B}$, it is not possible to construct the full mode structure $\delta_{Temp}=\{\delta \rho_{Temp},\delta \mathbf{u}_{Temp}, \delta T_{\phantom{.}Temp} \}$ as a linear combination of the full mode structure of the other modes. Note that as a result, care should be used when obtaining the system of equations for the amplitudes give by evaluating Eqs.~\eqref{proj1}-\eqref{proj4} not to miss contributions from all the important modes. It is also pointed out that in the process for finding mode structure, mode frequencies come in pairs, one with positive and the other with negative real part. However, in allowing for a complex amplitude and taking the real part in Eqs.~\eqref{proj1}-\eqref{proj4} one need only consider modes to have positive real part of the frequency.

Consider a generic isotropic trap quench in $d=2$ for multiple initial conditions in order to demonstrate the role of the temperature and number modes. In this case, the amplitudes take on a fairly simple form 
\begin{align}
	\label{numamp}
	&a_N=\frac{A_i}{A_0}-1,\\
	&a_{T}=\frac{(1-\gamma)\frac{A_i}{A_0}-(T_{\,init}-1)\gamma}{4 \gamma},\\
	&a_{B}=\frac{(1-\gamma)\frac{A_i}{A_0}-(1-T_{\,init})\gamma}{4 \gamma},\\
	\label{bamp}	
	&b_{B}=0.
\end{align}
Note that the above expressions contain both phase and magnitude information, as they may be negative. Phase and amplitude will be plotted separately below. Additionally, one sees from Eqs.~\eqref{numamp}-\eqref{bamp} several features which should be expected. To explore this, the analysis is broken into several cases. \\

\noindent \textbf{Case 1: $\mathbf{A_i/A_0=1}$, $\mathbf{T_{\,init}=1}$}

This is the case discussed in the main text (see Sec.~\ref{2diso}). \\

\noindent \textbf{Case 2: $\mathbf{A_i/A_0\neq1}$, $\mathbf{T_{\,init}=1}$}

\begin{figure*}[htbp]
      \includegraphics[width=\textwidth]{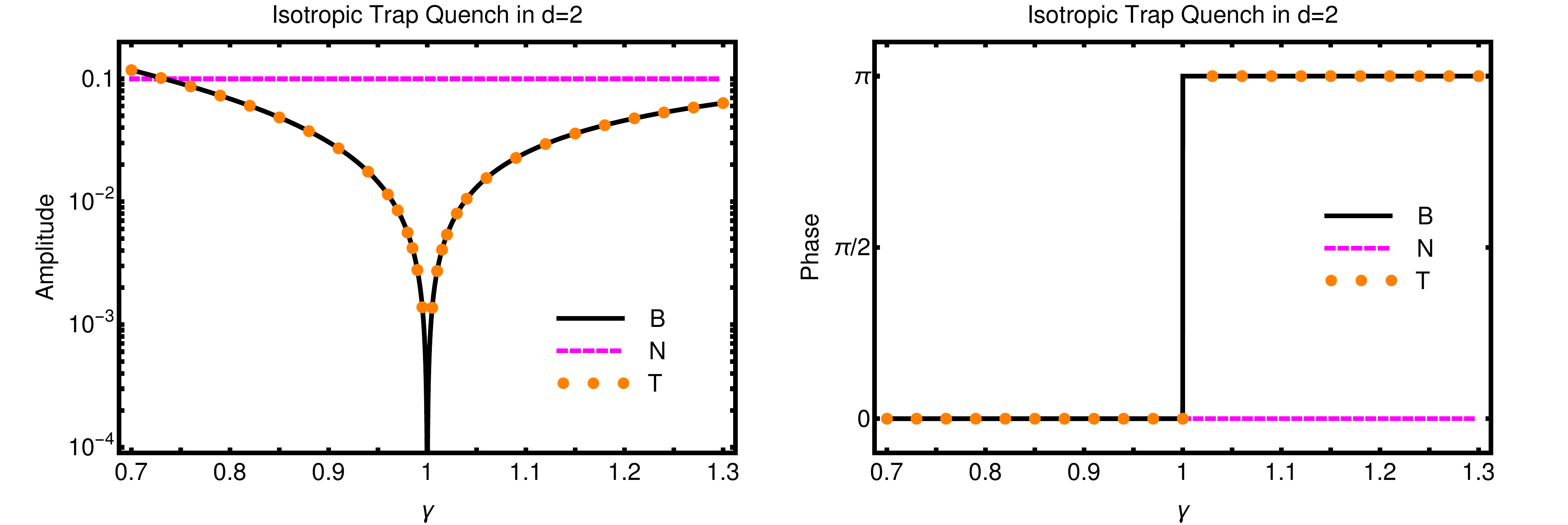}
      \caption{Left: Absolute value of the (dimensionless) number ('N'), temperature ('T') and breathing ('B') mode amplitudes as a function of the quench strength parameter $\gamma$ for an isotropic quench in d=2 assuming $A_i/A_0=1.1>1$ and $T_{\,init}=1$. Right: Phase of mode amplitude. Note the amplitude and phase of the breathing and temperature modes are identical.}
      \label{more}
\end{figure*}

For this case one may see from Eqs.~\eqref{numamp}-\eqref{bamp} that the ratio $A_i/A_0$ gives rise to a non-zero amplitude of the number mode, but leaves the location of the zero of the other two modes at $\gamma=1$. This should be expected since this merely means that at $\gamma=1$ there are more ($A_i/A_0>1$) or less ($A_i/A_0<1$) atoms in the trap than what was assumed in the equilibrium solution expanded about. This should make the role of the number mode more clear and is demonstrated for the case $A_i/A_0>$ and $T_{init}=1$ in Fig.~\ref{more}. Particularly, it is only important if for some reason one choses to expand dynamics about an equilibrium with a different number of particles than given by the initial conditions.

\noindent \textbf{Case 3: $\mathbf{A_i/A_0=1}$, $\mathbf{T_{\,init}\neq1}$}

\begin{figure*}[htbp]
      \includegraphics[width=\textwidth]{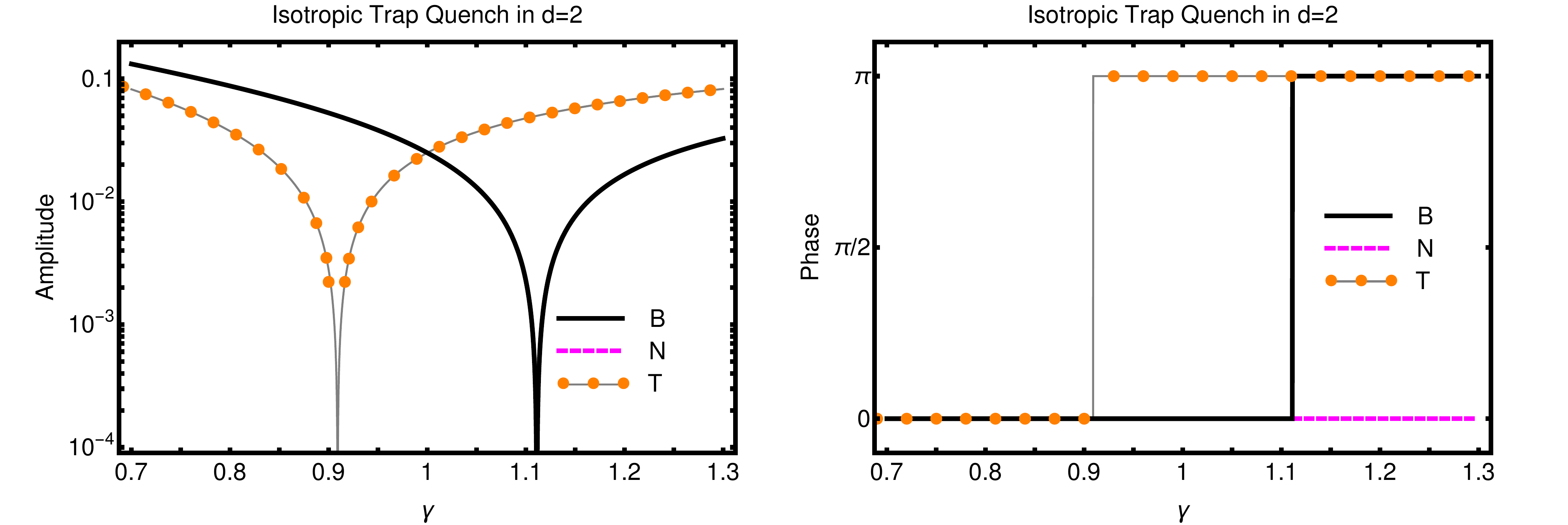}
      \caption{Left: Absolute value of the (dimensionless) temperature ('T') and breathing ('B') mode amplitudes as a function of the quench strength parameter $\gamma$ for an isotropic quench in d=2 assuming $A_i/A_0=1$ and $T_{\,init}=0.9$. Right: Phase of mode amplitude.}
      \label{lowerT}
\end{figure*}

For this case one may see from Eqs.~\eqref{numamp}-\eqref{bamp} that the number mode is not excited, while the value of $T_{\,init}$ alters the location of the zero for the temperature and breathing modes. This should be expected since at $\gamma=1$ the breathing mode is excited through the temperature difference. Fig.~\ref{lowerT} shows the case where $T_{\,init}<1$, demonstrating that the phase of the breathing mode vanishes at $\gamma=1$ while the magnitude is non-zero. A positive amplitude at $\gamma=1$  is expected since the temperature is below its equilibrium value for the given cloud radius so the cloud will reduce its size to try to reach equilibrium.


\end{document}